\documentclass[letterpaper,twocolumn]{aastex6}
\shorttitle{Optimizing Exoplanet Target Selection for \emph{JWST}}
\shortauthors{Fortenbach \& Dressing}
\usepackage{epsfig}
\usepackage{amsmath}
\usepackage{rotating}
\usepackage{natbib} 
\usepackage{enumerate}
\usepackage{hyperref}
\usepackage{url}
\usepackage{longtable}
\usepackage{color}
\begin{document}

\def\mearth{{\rm\,M_\oplus}}                                                    
\def\msun{{\rm\,M_\odot}}                                                       
\def\rsun{{\rm\,R_\odot}}                                                       
\def\rearth{{\rm\,R_\oplus}} ­­­
\def\fearth{{\rm\,F_\oplus}}
\def\searth{{\rm\,S_\oplus}}
\def\lsun{{\rm\,L_\odot}} 
\def\teffsun{{\rm\,T_{eff,\,\odot}}}                                                        
\def\kepler {{\emph{Kepler}\,}}                                              
\newcommand{\teff}{\ensuremath{T_{\mathrm{eff}}}}
\newcommand{\teq}{\ensuremath{T_{\mathrm{eq}}}}
\newcommand{\jmag}{\ensuremath{J_{\mathrm{mag}}}}
\newcommand{\logg}{\ensuremath{\log g}}

\newcommand{\tc}[2][red]{\textcolor{#1}{\emph{\textbf{#2}}}}


\title{A FRAMEWORK FOR OPTIMIZING EXOPLANET TARGET SELECTION FOR \mbox{THE JAMES WEBB SPACE TELESCOPE}}
\author{Charles D. Fortenbach\altaffilmark{1}}
\author{Courtney D. Dressing\altaffilmark{2}}
\altaffiltext{1}{Department of Physics and Astronomy, San Francisco State University, San Francisco, CA 94132, USA; \href{mailto:cfortenbach@att.net}{cfortenbach@att.net}}
\altaffiltext{2}{Astronomy Department, University of California, Berkeley, CA 94720, USA}
\date{\today}
\slugcomment{In preparation} 

\begin{abstract}
The \emph{James Webb Space Telescope (JWST)} will devote significant observing time to the study of exoplanets. It will not be serviceable as was the \emph{Hubble Space Telescope}, and therefore the spacecraft/instruments will have a relatively limited life.  It is important to get as much science as possible out of this limited observing time.  We provide an analysis framework (including publicly released computational tools) that can be used to optimize lists of exoplanet targets for atmospheric characterization. Our tools take catalogs of planet detections, either simulated, or actual; categorize the targets by planet radius and equilibrium temperature; estimate planet masses; generate model spectra and simulated instrument spectra; perform a statistical analysis to determine if the instrument spectra can confirm an atmospheric detection; and finally, rank the targets within each category by observation time required. For a catalog of simulated \emph{Transiting Exoplanet Survey Satellite} planet detections, we determine an optimal target ranking for the observing time available.  Our results are generally consistent with other recent studies of \emph{JWST} exoplanet target optimization.  We show that assumptions about target planet atmospheric metallicity, instrument performance (especially the noise floor), and statistical detection threshold, can have a significant effect on target ranking.  Over its full 10-year (fuel-limited) mission, \emph{JWST} has the potential to increase the number of atmospheres characterized by transmission spectroscopy by an order of magnitude (from about 50 currently to between 400 and 500).\\
\end{abstract}


\maketitle

\section{Introduction}\label{sec:Introduction}
\setcounter{footnote}{2}

\subsection{Background}\label{sec:Background}
The launch of the \emph{James Webb Space Telescope (JWST)} is currently scheduled for early 2021.\footnote{\raggedright\url{https://www.nasa.gov/press-release/nasa-completes-webb-telescope-review-commits-to-launch-in-early-2021}} \emph{JWST} will usher in a new era of astronomical observation, studying everything from the history of the Universe, to the formation of extrasolar planetary systems, to the evolution of our own solar system.  It will complement and extend what has been achieved with the \emph{Hubble Space Telescope (HST)} in a new wavelength regime and with much better sensitivity \citep{Gardner2006, Kalirai2018a, Madhusudhan2019}.

One of the main mission goals of \emph{JWST} is to ``measure the physical and chemical properties of planetary systems,  . . . and investigate the potential for life in those systems."\footnote{\url{https://jwst.nasa.gov/science.html}} For planets outside our own solar system (exoplanets) one of the best ways to investigate the potential for life is to study the planetary atmospheres. 

\emph{JWST} will significantly advance studies of exoplanet atmospheres by providing access to a wide variety of molecular spectral features. For example, the main features of an Earth-like atmosphere in the 0.7 to 5.0\,$\mu$m range are H$_{2}$O\,(near 2.8\,$\mu$m, and 3.2\,$\mu$m), and CO$_{2}$\,(4.2\,$\mu$m), and in the 4.0 to 20\,$\mu$m range are CO$_{2}$\,(15\,$\mu$m), O$_{3}$\,(9.6\,$\mu$m), CH$_{4}$\,(7.8\,$\mu$m), H$_{2}$O\,(5.9\,$\mu$m), and HNO$_{3}$\,(11.2\,$\mu$m) \citep{Kaltenegger2009}. \emph{JWST} has the capability to take high quality spectra over these wavelength ranges. The target exoplanets may well have atmospheres with significantly different composition, but many interesting molecular features are still expected to be observed.
\bigskip

\subsubsection{The Mission}
The spacecraft will orbit the Sun at the second Sun-Earth Lagrange point, L2.  This is a meta-stable point about 1.5 million km away from the Earth along the Earth-Sun line outward (in the opposite direction) from the Sun.  This orbital geometry allows the large sunshield to provide protective cooling for the telescope with minimal maneuvering, which would be very difficult, if not impossible, if the spacecraft were in a typical low-Earth orbit, as with \emph{HST}.

Due to this orbital location, \emph{JWST}'s lifetime will be limited. There is currently no plan for on-orbit servicing as there was with \emph{HST}.  The spacecraft will have a nominal five-year mission, but will carry ten years worth of fuel to enable an extended mission.  

We can estimate the time available for exoplanet observing over the spacecraft's ten-year fuel-limited lifetime.  NASA expects \emph{JWST} to be in routine science mode roughly six months after launch.\footnote{\url{https://jwst.nasa.gov/faq.html}}  This allows for cooling down the spacecraft and doing various calibrations and maneuvering operations.  This leaves approximately 83,220 hours available for the overall observing program over the 10-year extended mission.

The question is: out of this 83,220 hours, how much time will be available for exoplanet studies and in particular for transmission spectroscopy?

The purpose of our work is to provide a tool to rank an exoplanet target catalog (simulated or actual, delivered by \emph{TESS} or other precursor mission) by atmospheric observability.  We are not attempting to justify a particular time allocation from a bottom-up scientific perspective.  We are only trying to provide a rough estimate of the amount of time that might be available for exoplanet transmission spectroscopy based on a top down view of high-level mission priorities.  

Our ranking study is not in any way intended to be an observing proposal.  Any strategic survey program related to our ranking work would need to go through the proper reviews and would be competitively chosen by the \emph{JWST} Telescope Allocation Committee (TAC) and/or the Director.  

One of the early documents providing guidance on the allocation of \emph{JWST} mission time to various scientific programs was the \emph{JWST} Science Operations Design Reference Mission report (SODRM, released 27 Sept. 2012).\footnote{\url{http://www.stsci.edu/files/live/sites/www/files/home/jwst/about/history/science-operations-design-reference-mission-sodrm/_documents/SODRM-Revision-C.pdf}}  The 2012 SODRM was an early estimate of how observing time might be allocated for the first year of \emph{JWST} science operations. This document was built up from a diverse set of science and calibration programs.\footnote{\url{http://www.stsci.edu/jwst/about/history/science-operations-design-reference-mission-sodrm}}  It showed a notional allocation of 16.1\% (13,363 hours) of the total mission time to exoplanet study (after removing the 6-month commissioning activity).  

Since the release of the 2012 SODRM, interest in exoplanet studies and, in particular, in exoplanet atmospheric characterization has grown.  This has led to a shift in emphasis of \emph{JWST} observational priorities.  A recent white paper \citep{Greene2019} describes how guaranteed time observations (GTOs) and early release science (ERS) will ``advance understanding of exoplanet atmospheres and provide a glimpse into what transiting exoplanet science will be done with \emph{JWST} during its first year of operations. . . . Approximately 3,700 hours of GTO and an additional $\sim$ 500 hours of Director's Discretionary Early Release Science (ERS) observations have been accepted for \emph{JWST} Cycle 1 [for general science, including exoplanet studies]. This is $\sim$ 50\% of the time available in the first year of science operations." 

\citet{Greene2019} go on to say that ``The transiting planet observations in the Cycle 1 GTO and ERS (Bean et al. 2018) programs will enable a large step forward in the characterization of exoplanet atmospheres . . . these programs [which include multi-wavelength transmission, emission, and phase curve observations of 27 transiting planets] sum up to 816 hours, 19\% of the scheduled GTO + ERS observing time."  If we extrapolate this to a full year program we might see as much as 1600 hours dedicated to transit spectroscopy; and if extrapolated to the full 10-year (less commissioning) fuel-limited mission, we might see a total program of as much as 15,000 hours for exoplanet transit spectroscopy.

Additionally, a significant GTO + ERS time allocation will be made for coronagraphic imaging and direct spectroscopy of young planets \citep{Beichman2019}, but that is not our focus.

\citet{Greene2019} also mentioned that, ``\emph{JWST} may well characterize the atmospheres of over 50 transiting planets in its first year of science operations." Extrapolating, this suggests that \emph{JWST} could possibly characterize the atmospheres of as many as 475 transiting planets over the course of its full mission.  In some cases this would include multiple-transit observations and revisits of the same planet with multiple instruments for broader wavelength coverage.

It has been noted  that transmission spectroscopy ``is expected to be the prime mode for exoplanet atmospheric observations and provides the best sensitivity to a wide range of planets" \citep{Kempton2018a}. It would seem likely that a somewhat greater emphasis will be placed on transmission spectroscopy than emission (eclipse) spectroscopy.

Of course all of the \emph{JWST} instruments will be involved in the observing program; however, for our baseline ranking we will focus exclusively on transmission spectroscopy with the NIRSpec G395M, operating in Bright Object Time Series (BOTS) mode (the basis for selection of this instrument/mode is discussed further in Section \ref{sec:gensimspectra}).  This should give us a reasonable ranking of the atmospheric observability of our target planet catalog at least for the NIRSpec G395M wavelength range and capabilities.  Future work could rank the targets with NIRISS SOSS, or other instrument/modes.  An optimal target ranking using a mix of instruments would be a complex problem; but one that could potentially be studied with our code.  Such a study is beyond the scope of our current work.  

It is important to emphasize that the available observing time does not affect the target rankings; it only comes into play in establishing the cut-off line for the list of ranked targets.  In order to set this cut-off line for our study, we will assume an overall program of 8300 hours for \underline{transmission} spectroscopy.  This is a little over half ($\sim$ 55\%) of the total 15,000 hour transit spectroscopy program extrapolated from the Cycle 1 GTO and ERS program allocation.  

\subsection{Aim of This Work}
Given the limited lifetime of \emph{JWST}, the scarce resource of observing time must be allocated as efficiently as possible.  In particular, for the study of exoplanet atmospheres we want to prioritize targets where the exoplanet atmospheres have the best chance of successful detection and characterization.  

The scientific community needs a tool to implement this target prioritization in a robust and efficient way.  We provide a framework for analysis and an associated computer program to assist with ranking targets of interest for \emph{JWST} detected by the \emph{Transiting Exoplanet Survey Satellite (TESS)} and other precursor efforts.  

In addition, our ranking tools will help to prioritize targets for high precision radial velocity (RV) follow-up to further constrain the masses of the target planets.

\subsection{Related Work}\label{sec:Other related work}
A number of studies have addressed the range of issues associated with optimizing \emph{JWST} targets for atmospheric characterization.  These issues include: precursor missions, synthetic target surveys, exoplanet mass-radius relationships, model atmosphere spectra, simulated instrument spectra, and other target ranking schemes.
\bigskip

\subsubsection{Synthetic Target Surveys}\label{sec:syntargetsurveys}
\indent The earliest simulation of exoplanet yield from a space-based all-sky survey of bright stars was described in \citet{Deming2009}.  This was specifically aimed at simulating the \emph{TESS} survey yield. It was a particularly notable effort given that the \emph{Kepler} mission had not even been launched at the time of their work.  This was an exercise in scaling, and extrapolating from the limited data available. 

We have chosen to use the target catalog developed by \citet{Sullivan2015} (the ``Sullivan" catalog) as the basis of our analysis and code development.  This catalog is a simulation of expected \emph{TESS} planet detection results based on existing occurrence rate information.  The \emph{TESS} exoplanet statistics have been studied recently by several groups, including \citet{Bouma2017}, \citet{Muirhead2018}, \citet{Barclay2018}, and \citet{Ballard2019}.  These studies considered \emph{TESS} mission extension, yields of planets around M-dwarf hosts, and updates of planet yields using the actual \emph{TESS} catalog of target stars.  

These updated studies do show some significant differences with the Sullivan work.  Specifically, by accounting for co-planarity of planets in multi-planet systems, \citet{Ballard2019} found that \emph{TESS} would detect up to 50\,\% more M-dwarf planets than predicted by Sullivan. \citet{Bouma2017} estimated 30\,\% fewer Earths and 22\,\% fewer super-Earths for the primary 2\,x\,$10^5$ target stars that will be sampled at a 2-minute cadence, and \citet{Barclay2018} predicted 36\,\% fewer Earths and 60\,\% fewer super-Earths for the primary target stars.  

Our study was too advanced to incorporate the results of these new simulated catalogs into our baseline target ranking.  The primary goal of our project was to develop a framework for ranking planets for study with \emph{JWST}. With modest effort, our code can be adapted to work with any input planet sample including the actual \emph{TESS} detections.

\subsubsection{Target Ranking Studies}\label{sec:Targetranking}
\indent A number of studies have considered the issue of finding an optimal target set for atmospheric characterization by \emph{JWST}.  We have previously mentioned the \citet{Deming2009} study.  In addition to modeling the performance of the Mid-Infrared Instrument (MIRI) and the Near-Infrared Spectrograph (NIRSpec), they coupled their simulated \emph{TESS} target yield to their sensitivity model. They found that \emph{JWST} should be able to characterize dozens of \emph{TESS} super-Earths with temperatures above the habitable range.  They also found that \emph{JWST} should be able to measure temperature and identify absorption features in one to four habitable Earth-like planets orbiting lower{-}main-sequence stars.

They asserted that, ``although the number of habitable planets capable of being characterized by \emph{JWST} will be small, large numbers of warm to hot super Earths and exo-Neptunes will be readily characterized by \emph{JWST}, and their aggregate properties will shed considerable light on the nature of icy and rocky planets in the solar neighborhood."

More recently \citet{Crossfield2017} determined that the prominence of features in the transmission spectrum for a warm-Neptune exoplanet is related to its equilibrium temperature and its bulk H/He mass fraction.  They were able to construct an analytical relation to estimate the overall observing time needed to distinguish a gas giant's transmission spectrum from a flat line. They suggested that the atmospheric trends they describe could result in a reduction in the number of \emph{TESS} targets with atmospheres that could be detected by \emph{JWST} by as much as a factor of eight. 

\citet{Howe2017} explored the optimization of observations of transiting hot Jupiters with \emph{JWST} to characterize their atmospheres.  They constructed forward model sets for hot Jupiters, exploring parameters such as equilibrium temperature and metallicity, as well as considering host stars with a wide brightness range. They computed posterior distributions of model parameters for each planet with all of the available \emph{JWST} instrument modes and various programs of combined observations.  From these simulations, trends emerged that provide guidelines for designing a \emph{JWST} observing program.

\citet{Morley2017} offered a study of atmospheric detection for the seven TRAPPIST-1 planets, GJ 1132b, and LHS 1140b.  These are some of the smallest planets discovered to date that might have atmospheres within the detection capability of \emph{JWST}.  This was not strictly a target ranking study, but it did involve generating model atmospheres and simulating \emph{JWST} instrument performance.  They considered the observability of each planet's atmosphere in both transmission and emission.  GJ 1132b and TRAPPIST-1b are excellent targets for emission spectroscopy with MIRI, requiring less than 10 eclipse observations.  Seven of the nine planets are good candidates for transmission spectroscopy.  Using estimated planet masses they determined that less than 20 transits would be required for a $5\, \sigma$ detection of a transmission spectrum.

Another recent study \citep{Louie2018} was aimed at understanding the suitability of expected \emph{TESS} planet discoveries for atmospheric characterization using \emph{JWST}'s Near-Infrared Imager and Slitless Spectrograph (NIRISS) by employing a simulation tool to estimate the signal-to-noise ratio (S/N) achievable in transmission spectroscopy.  The tool was applied to predictions of the \emph{TESS} planet yield and then the S/N for anticipated \emph{TESS} discoveries was compared to estimates of S/N for 18 known exoplanets. They analyzed the sensitivity of the results to planetary composition, cloud cover, and the presence of an observational noise floor. Several hundred anticipated \emph{TESS} discoveries with radii $1.5\rearth <R_{p} \le 2.5\rearth$ will produce S/N higher than currently known exoplanets in this radius regime. In the terrestrial planet regime, only a few expected \emph{TESS} discoveries will result in higher S/N than currently known exoplanets.

A study, \citet{Kempton2018a} (the ``Kempton study" or ``Kempton"), was published recently that is similar to our work in its general goal of finding an optimal target set for \emph{JWST}.  The authors use a set of two analytic metrics, quantifying S/N values in transmission and thermal emission spectroscopy to rank the target planets in the Sullivan catalog.  They use the S/N predictions from the \emph{JWST}/NIRISS simulation performed by \citet{Louie2018} as the basis for their transmission metric. They determine a sample of roughly 300 transmission spectroscopy targets that meet the threshold values of their metrics for observation.\medskip

We have organized the remainder of this paper as follows:  In Section \ref{sec:analysisframework}, we describe our analysis framework and the high level structure of our ranking code. We then validate the analysis and code in Section \ref{sec:validating}.  Next, we discuss our results including a baseline target ranking run in Section \ref{sec:resultsdiscuss}.  Finally, we present our conclusions and opportunities for further study in \mbox{Section \ref{sec:conclusions}}.

\section{Analysis Framework and the \texttt{JET} Code}\label{sec:analysisframework}
We have developed an analysis framework that takes a planet detection target catalog (simulated or actual) and processes it to result in a list of prioritized exoplanet targets for atmospheric characterization by \emph{JWST}. 

Our analysis of a target catalog proceeds in a straightforward manner.  We first set target selection parameters and other values (e.g., \emph{JWST} instrument/mode, target list limits, atmosphere model equations of state, etc.). Next, we take the target catalog data, and for each of the targets determine various planetary system parameters: orbital semi-major axis, planet equilibrium temperature, planet mass, etc. We then make an estimate of the number of transits observable in a 10-year mission, given target position on the sky and spacecraft pointing constraints, for each target. Then we divide the parameter space into seven demographic categories, by planet radius and equilibrium temperature, and categorize the targets.

We then generate model transmission spectra for each target.  In this paper we focus exclusively on transmission spectroscopy because, as discussed in \citet{Kempton2018a}, this is expected to be the best approach for observations of exoplanet atmospheres and provides the best sensitivity to a wide range of planets.  We consider two bounding cases for model atmospheres for each target: low metallicity with no clouds, and high metallicity with a mid-level cloud deck. This should span the range of atmospheres that we might encounter, and their potential detectability (see Section \ref{sec:modtranspec}). 

Next, we make sure that the host star is not so bright as to saturate the detector of the particular instrument under consideration.  If not, then for each of the targets and for each model atmosphere, we run an instrument simulation which includes noise effects.  We then determine the number of transits required for a high confidence detection.  

Next, we check to see if the number of transits required for detection is available in a ten-year mission.  If so, we can then determine the total observing time for the target, with all of the overheads included, for each atmosphere case.  From this we can determine an average total observing time for that particular target. 

Finally, we sort the targets by category and then rank them within each category by the average total observing time needed for detection.  No attempt is made to prioritize one demographic category over another, but future users could modify the code according to their own priorities

We will describe each of these analysis steps in more detail in the following subsections.  In addition, we have developed a suite of computer tools: the \underline{J}WST \underline{E}xoplanet \underline{T}argeting code (\texttt{JET}) that embodies our analysis procedure.

\subsection{Top Level Analysis Framework and \texttt{JET} Architecture}
\begin{figure*}[!htbp]
	\centering
	\includegraphics[width=0.9\textwidth]{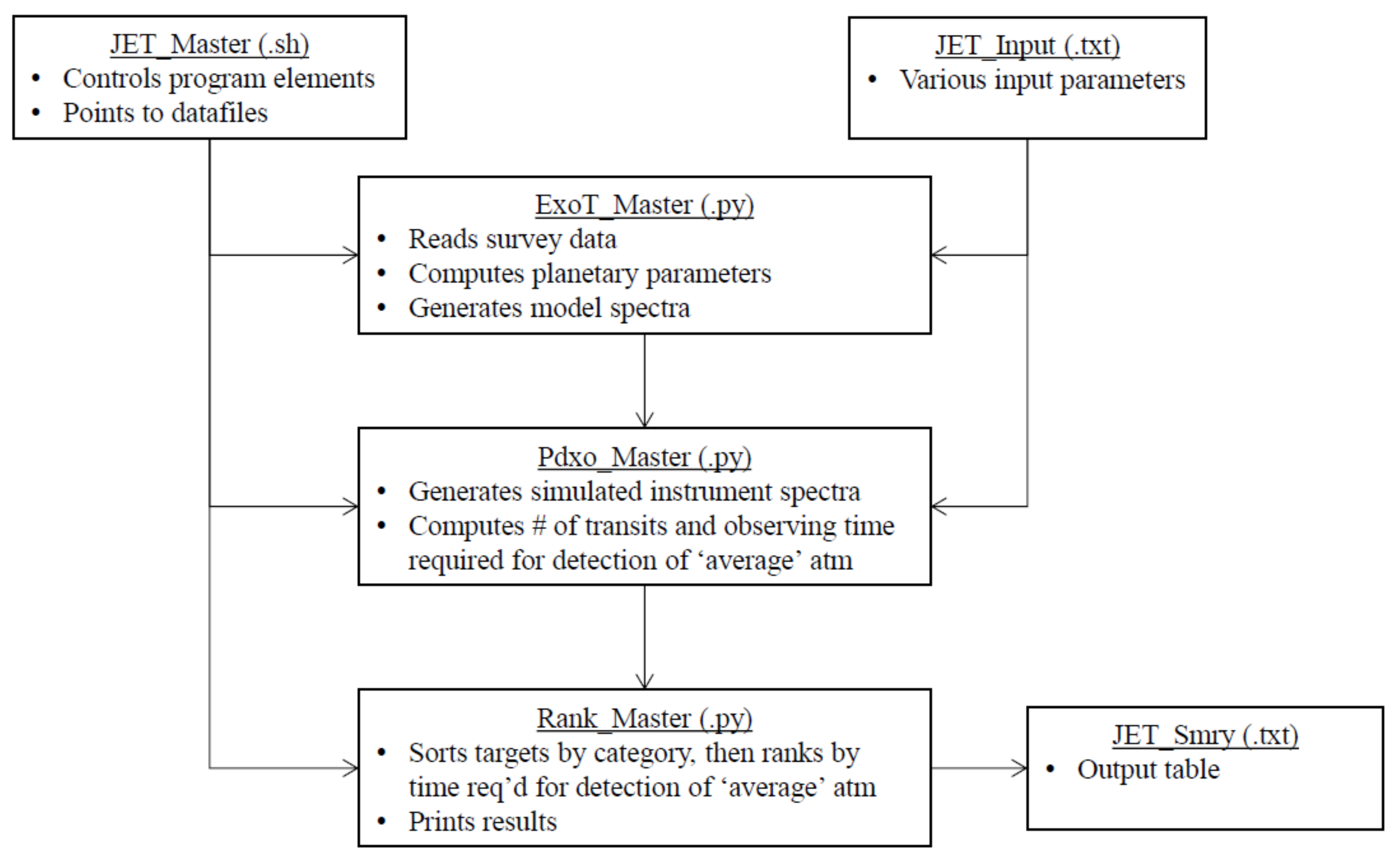}
	\setlength{\abovecaptionskip}{5pt}
	\caption{Top level \texttt{JET} program architecture.\label{fig:JETflow}}
\end{figure*}

The analysis (and code) can be divided logically into five main parts: (1) General input, (2) Generating model spectra, (3) Detecting the spectra with \emph{JWST}, (4) Sorting and ranking the target list, and (5) Controlling the program execution. Figure \ref{fig:JETflow} presents a flowchart of this top level analysis framework and program architecture.

The complete source code, detailed installation instructions, and sample input files for \texttt{JET} are available via Github (\url{https://github.com/cdfortenbach/JET}).  \texttt{JET} has been written in \texttt{Python}, but it incorporates a fully compiled executable of the \texttt{Exo-Transmit} atmospheric modeling code which was written in C (see Section \ref{sec:modtranspec}).  The \texttt{JET} code is designated as open source under the GNU General Public License.

\subsection{Generating Model Transmission Spectra (ExoT\_Master)}
\begin{figure*}[!htbp]
	\centering
	\includegraphics[width=1.0\textwidth]{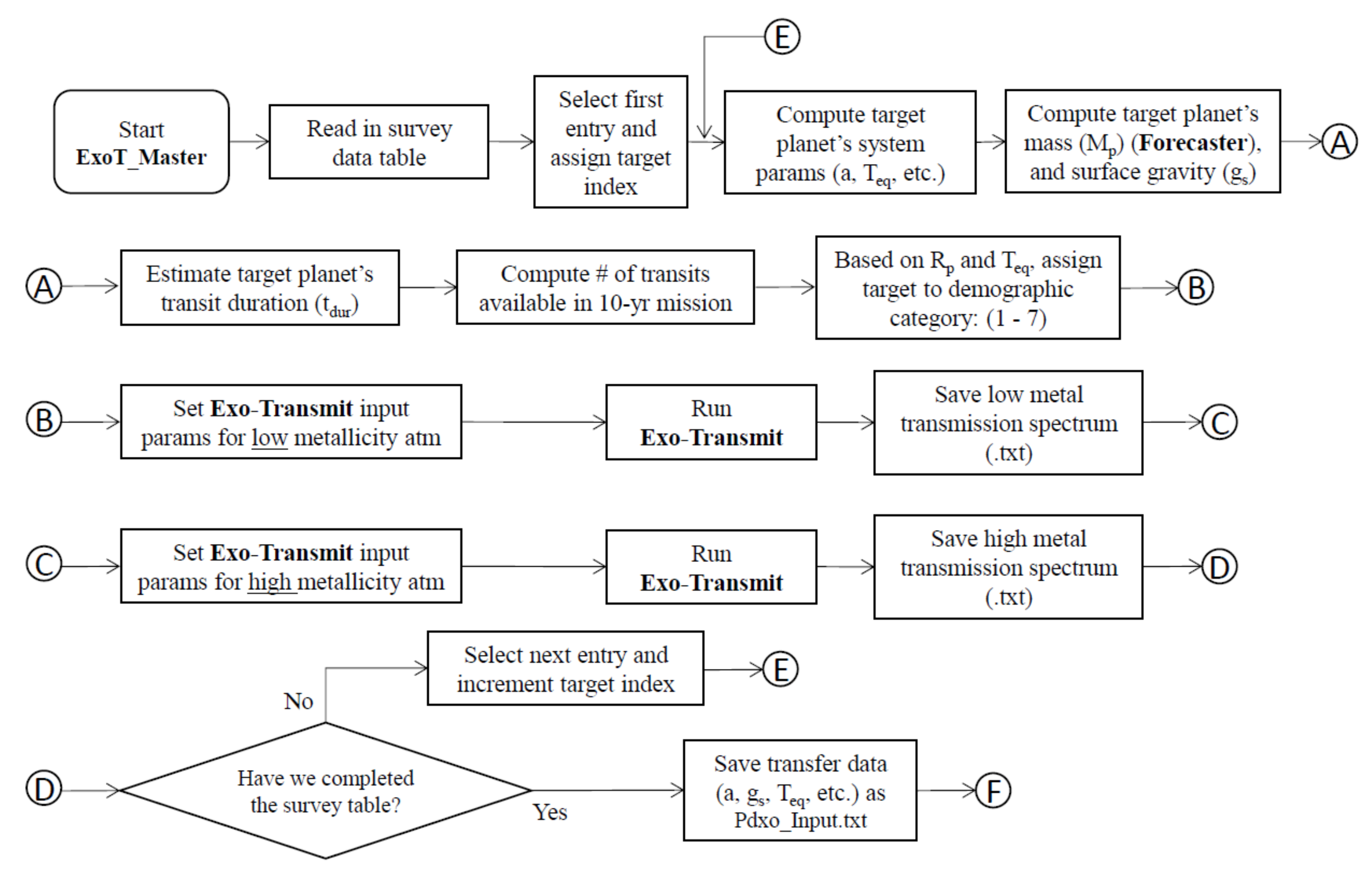}
	\setlength{\abovecaptionskip}{5pt}
	\caption{Analysis and code flow for \texttt{JET} program element \texttt{ExoT{\_}Master}.\label{fig:ExoTflow}}
\end{figure*}

The first major step in the analysis process is to generate model transmission spectra for the various targets.  This task is carried out by the \texttt{JET} program element: \texttt{ExoT{\_}Master}, shown in flowchart form in Figure \ref{fig:ExoTflow}.
\bigskip

\subsubsection{Reading the Survey Data}
Our analysis begins by reading in a planet detection target catalog.  As we discussed in Section \ref{sec:syntargetsurveys}, there have been several attempts at developing a synthetic detection catalog that would emulate a full \emph{TESS} (or other mission) program catalog.  As previously mentioned, we have adopted the format of the catalog described in \citet{Sullivan2015}.  Of course, when an actual \emph{TESS} (or other mission) catalog is available, that would be used for a \texttt{JET} production/science run.  

\subsubsection{Estimating Planet Masses}\label{sec:estplanetmasses}
Given that the precursor missions are transit surveys, they will not directly provide planet mass data. \texttt{JET} estimates planet masses using an approach implemented in the \texttt{Forecaster} package described by \citet{0004-637X-834-1-17}. To estimate the median values of planet mass, we use two power laws that can be derived from this work.  We note that our power law derivation here is consistent with the analysis of \citet{Louie2018}.\bigskip 

For $R_{p} < \text {1.23}\rearth$:
\begin{equation}\label{masseq1}
M_{p} = 0.9718\, (R_{p})^{ 3.58}
\end{equation}\smallskip

For $\text {1.23}\rearth \leq R_{p} \leq \text {10}\rearth$:
\begin{equation}\label{masseq2}
M_{p} = 1.436\, (R_{p})^{ 1.70}
\end{equation}

\noindent where $M_{p}$ is the planet mass relative to the mass of the Earth ($\!\mearth \cong 5.9736$\,x\,$10^{24}$\,kg) and, $R_{p}$ is the planet radius relative to the radius of the Earth \mbox{($\!\rearth \cong 6.37814$\,x\,$10^{6}$\,m)}.

We have set an upper limit for $R_{p}$ of $10\rearth$.  The mass-radius relationship described in the \citet{0004-637X-834-1-17} study is ambiguous for radii above this level. This is due to the very wide range of mass for planets with radii similar to that of Jupiter.  While there are target planets in the Sullivan catalog (and there will be in actual catalog data) that have larger radii, we will flag them as outside the range of our analysis. This is equivalent to imposing a planet mass upper limit of roughly 72\,$\mearth$ (roughly 3/4 the mass of Saturn) for our analysis.

\subsubsection{Deriving Other Planetary System Parameters}
In addition to estimating planet mass, we need to determine certain other planetary system parameters for each of the target planets.  These parameters include the semi-major axis of the planet's orbit, the planet's equilibrium temperature, the planet's surface gravity, and the transit duration.  These parameters are derived directly from the catalog data, given certain assumptions.\\

\noindent \textbf{Semi-major axis ($a_{AU}$)}:\newline
\indent The relative insolation of the planet, $S$, is given by Equation (13) of \citet{Sullivan2015}.  By rearranging this equation we can determine the semi-major axis of each of the target planets as follows:	

\begin{equation}\label{semimajoraxaueq}
a_{AU} = \bigg(\frac{R_*}{\sqrt{S}}\bigg)\bigg(\frac{\teff}{\teffsun}\bigg)^2
\end{equation}
where $S$ is the stellar insolation at the top of the planet's atmosphere in $\searth$ units, $\teff$ is the effective temperature of the host star in K, while $R_{*}$ is the radius of the host star in solar radii, $\teffsun$ is the Sun's effective temperature $\cong 5777\text{ K}$, and $a_{AU}$ is the semi-major axis of the planet's orbit in AU.\\

\noindent \textbf{Surface gravity ($g_{s}$)}:\newline
\indent The target planet's surface gravity is needed to calculate the pressure scale height, and is given by:

\begin{equation}\label{surfgraveq}
g_{s} = \frac{\text{G} M_{p}}{(R_{p})^2}
\end{equation} 
where G is the constant of universal gravitation, 6.67428\,x\,$10^{-11}$ N$\text{m}^{2}$/$\text{kg}^{2}$; and $g_{s}$ is the surface gravity in $\text{m s}^{-2}$.\\   

\noindent \textbf{Equilibrium temperature ($\teq$)}:\newline
\indent The planet's equilibrium temperature is also needed in the generation of spectra of model atmospheres. We are assuming circular obits, zero albedo, and full day-night heat distribution through the atmosphere.  We can estimate equilibrium temperature as follows (see Equation (12) of \mbox{\citet{Sullivan2015}}):

\begin{equation}\label{teqeq}
\teq = \teff\sqrt\frac{R_{*}}{2a}                 
\end{equation}
where $\teq$ is the planet's equilibrium temperature in K, and $a$ is the orbital semi-major axis in solar radii. \,\,This, of course, ignores any potential greenhouse effect.\\

\noindent \textbf{Transit duration ($t_{\rm dur}$)}:\newline
\indent The transit duration for each target planet is a key element in our analysis.  For our purposes, we define the transit duration to be the interval between the first ($t_{\rm I}$) and fourth ($t_{\rm IV}$) contacts as shown in Figure 2 of \citet{Winn2010}.  This interval is otherwise known as $t_{14}$.  	

If we assume a circular orbit, with an inclination of 90 deg., we can estimate the total transit time by the method described in Section 2.4 of \citet{Winn2010}.  The full transit duration is given by:
\begin{equation}\label{tdureq4}
T_{\rm tot} \cong \bigg(1 + \bigg(\frac {R_{p}}{R_{*}}\bigg)\bigg(\frac {\rearth}{\rsun}\bigg)\bigg)\frac{R_{*}(P*24)}{\pi a}             
\end{equation}

\noindent and,
\begin{equation}\label{tdureq5}
t_{\rm dur} = t_{14} = T_{\rm tot}            
\end{equation}
where $P$ is the planet's orbital period in days, $t_{\rm dur}$ is the transit duration in hours, and again $a$ is the orbital semi-major axis in solar radii.

\subsubsection{Categorizing Targets}
The next step in our analysis process is to categorize the target planets.  We have chosen to divide the catalog into seven categories along planet radius and equilibrium temperature dimensions with breaks at planet radii of 1.7 and 4.0$\rearth$, and at equilibrium temperatures of 400 and 800\,K (as shown in Table \ref{tab:categories}). In this work, as mentioned in Section \ref{sec:estplanetmasses}, we are focusing our attention on planets with generally smaller radii $(R_{p} < \text {10}\rearth)$. 

The category divisions are somewhat arbitrary; however, the break at an $R_{p}$ of 1.7 $\rearth$ is based on recent exoplanet population studies \citep{Fulton2017, Fulton2018} that indicate that a gap exists in the exoplanet population around this radius.

\begin{table*}[!htbp]
	\centering
	\caption{Target Planet Categories\label{tab:categories}}                                          
	\begin{center}						                                                              
		\setlength{\tabcolsep}{25pt}                                                                  
		\renewcommand{\arraystretch}{1.5}                                                             
		\begin{tabular}{c l l l}                                                                      
			\hline\hline
			Category & Description        & $\teq$ range                                 & $R_{p}$ range  \\ [0.5ex]  
			\hline   
			1        & Cool Terrestrials   & $\teq < \text {400 K}$                       & $R_{p} < \text {1.7}\rearth$\\
			2        & Warm Terrestrials   & $\text {400 K}\leq \teq \leq \text {800 K}$  & $R_{p} < \text {1.7}\rearth$\\
			3        & Hot Terrestrials    & $\teq > \text {800 K}$                       & $R_{p} < \text {1.7}\rearth$\\
			4        & Cool sub-Neptunes   & $\teq < \text {400 K}$                       & $\text {1.7}\rearth\leq R_{p}\leq \text {4.0}\rearth$\\
			5        & Warm sub-Neptunes   & $\text {400 K}\leq \teq \leq \text {800 K}$  & $\text {1.7}\rearth\leq R_{p}\leq \text {4.0}\rearth$\\
			6        & Hot sub-Neptunes    & $\teq > \text {800 K}$                       & $\text {1.7}\rearth\leq R_{p}\leq \text {4.0}\rearth$\\
			7        & sub-Jovians         & $\cdots$                                     & $R_{p} > \text {4.0}\rearth$\\ [1ex] 
			\hline   
			\multicolumn{4}{l}{%
				\begin{minipage}{16cm}~\\[1.5ex]%
					Notes. --- Category 7 captures all catalog targets of large radii. Targets with $R_{p}>\text{10.0}\rearth$ are flagged and no transit detection calculations are made on them. These targets drop to the bottom of the ranking for Category 7.%
				\end{minipage}%
			}\\
		\end{tabular}
	\end{center}		
\end{table*}

\subsubsection{Determining \emph{JWST} Observing Constraints}
\emph{JWST} has viewing constraints that are defined by its orbit and configuration.  The spacecraft will orbit the Sun (in the ecliptic plane) at the L2 position with a one year period.  At all times the large sunshield must be positioned between the Sun and the science instruments to maintain the very cold environment necessary for infrared observations.  The sunshield geometry creates limitations on where the spacecraft can point at any particular time and for how long.  \emph{JWST} can point to solar elongations between 85 deg.\ and 135 deg. It can also point to any location in the 360 deg.\ circle perpendicular to the sun line.  This defines a large annulus where \emph{JWST} can observe at any given time (the field of regard).  Targets at low ecliptic latitudes are particularly impacted by these constraints.  This observing geometry is well described in Figure 1 of the Space Telescope Science Institute's (STScI) on-line User Documentation for \emph{JWST} Target Viewing Constraints.\footnote{\raggedright{\url{https://jwst-docs.stsci.edu/jwst-observatory-hardware/jwst-target-viewing-constraints}}}

For a target at a particular ecliptic latitude there are only a certain number of days per year that \emph{JWST} can observe it.  This means that depending on the position and transit period of the target system, not all transits may be observable.  This could be a problem for low S/N targets (meaning many transit observations necessary) that have long periods, and that are at low ecliptic latitudes.

For each target we compute the number of transits that would be observable over the full 10-year fuel life of the spacecraft.  If the transit requirement for detection exceeds the number of transits available we raise a flag in the output.

Within \texttt{ExoT{\_}Master}, our algorithm for computation of transits available over the 10-year mission is only approximate.  We would need to include the absolute transit timing and absolute \emph{JWST} orbital position to more accurately determine this parameter for a given target.  However, our estimate should be acceptable for our purposes.

We first compute the number of mission days available in the 10-year (fuel limited) mission, less a six month commissioning period:
\begin{equation}\label{tmission}
t_{\rm mission} = (10 - 0.5)*365, \text{ days}             
\end{equation}

Then we compute the number of transits available (but not necessarily observable) in a 10 yr mission:
\begin{equation}\label{ntavail}
nt_{10yr,\text{ available}} = (1/P)*t_{\rm mission}                
\end{equation}
where $P$ is the planet's orbital period in days.

Next, we convert the target's decimal coordinates into hours-minutes-seconds coordinates.  This is done using the \texttt{Python} \texttt{astropy} \texttt{SkyCoord} routines.\footnote{\raggedright\url{http://docs.astropy.org/en/stable/api/astropy.coordinates.SkyCoord.html}} Then we convert to Ecliptic coordinates using the \texttt{Python} \texttt{PyEphem} package.\footnote{\url{https://rhodesmill.org/pyephem/index.html}}

We then take advantage of the STScI analysis that produced an estimate of observable days vs ecliptic latitude for \emph{JWST} shown in Figure 2 of the previously referenced STScI on-line User Documentation for \emph{JWST} Target Viewing Constraints. For the target's particular ecliptic latitude we can then estimate the number of days out of 365 that the target will be observable: the observable days factor ($f_{\rm obsdays}$).  Below 45 deg.\ this observability comes as two shorter time periods separated by six months. Above 45 deg.\ one longer viewing period is available, increasing until the continuous viewing zone is reached at approximately 85 deg.\ ecliptic latitude. We are not considering the impact of the six month split for observations below an ecliptic latitude of 45 deg.  We only consider aggregate observable days available.

Finally, we can estimate the rough upper bound on the total number of transits that would be observable in the 10-year mission as:
\begin{equation}\label{nt10yr}
nt_{\rm 10yr} = nt_{\rm 10yr,\text{ available}}*f_{\rm obsdays}               
\end{equation}

For further information on target visibility we recommend the \emph{JWST} General Target Visibility Tool (\texttt{GTVT}).  This is a command-line \texttt{Python} tool that provides quick-look assessments of target visibilities and position angles for all \emph{JWST} instruments.\footnote{\url{https://jwst-docs.stsci.edu/jwst-other-tools/target-visibility-tools/jwst-general-target-visibility-tool-help}}
\smallskip

\subsubsection{Generating Model Transmission Spectra with Exo-Transmit}\label{sec:modtranspec} 
We have chosen to implement the \texttt{Exo-Transmit} code for our study.  \citet{1538-3873-129-974-044402} presents a detailed description of \texttt{Exo-Transmit}, an open source code for generating model exoplanet transmission spectra.  This is an extension of a super-Earth radiative transfer code originally described by \citet{Miller-Ricci2009}, and \citet{Miller-Ricci2010}.    

\texttt{Exo-Transmit} is a flexible tool aimed at calculating transmission spectra for a wide range of exoplanet size, surface gravity, equilibrium temperature, and atmospheric composition.  The code essentially solves the equation of radiative transfer for absorption of the host star's light as it travels through the planet's atmosphere during a transit.  It was originally developed to work with low-mass exoplanets, but it can be used to model giant planet transmission spectra as well.  The models can be set up with or without clouds, and can include the effects of Rayleigh scattering, and collision induced absorption.

In general we will not know the characteristics of the target planet's atmosphere.  We might be able to make an educated guess, but our approach is to assume that we do not know.  

Figure 5 of \citet{1538-3873-129-974-044402} showed that spectral line strength is related to atmospheric metallicity and cloud levels. High metallicity (related to high mean molecular weight) and the presence of higher altitude (lower pressure) level clouds reduce the strength of spectral features.

We set up two bounding cases of observability: (1) a relatively easy to detect low-metallicity atmosphere (5xSolar) with no clouds, and (2) a more difficult to detect high-metallicity (1000xSolar) atmosphere with a low (100\,mbar) cloud deck. In each case the condensation and removal via rain-out of molecules (excluding graphite) from the gas phase is included. The specific \texttt{Exo-Transmit} Equation of State (EOS) input file names and other input parameters for the two atmosphere cases are given in Table \ref{tab:inputparams}. 

\begin{table*}[!htbp]
	\centering
	\caption{\texttt{JET} Baseline Run - Input Parameters\label{tab:inputparams}}      
	\begin{center}                                                                     
		\setlength{\tabcolsep}{10pt}                                                   
		\renewcommand{\arraystretch}{1.1}                                              
		\begin{tabular}{l r}                                                           
			\hline\hline
			Parameter & Value  \\[0.5ex]                                               
			\hline                                                                     
			Starting target from catalog for this run:        & 1 \\
			Ending target from catalog:                       & 1984 \\
			&  \\
			\emph{JWST} instrument:                           & NIRSpec G395M \\
			Wavelength short limit (microns):                 & 2.87 \\
			Wavelength long limit (microns):                  & 5.18 \\
			$\jmag$ limit ($\teff$ = 10000K):                    & 6.2 \\
			$\jmag$ limit ($\teff$ = 5000K):                     & 6.8 \\
			$\jmag$ limit ($\teff$ = 2500K):                     & 7.4 \\
			Detector linear response limit (\% FW):           & 80 \\
			Noise floor (nfloor) (ppm):                       & 25 \\
			R value of sim (Res):                             & 100 \\
			dBIC samples for each ntr grid pt.:               & 2000 \\
			Detection threshold (dBIC):                       & 10 \\
			Free model spectrum BIC parameters:				  & 5 \\
			&  \\
			Eq. of State (lo metal atm):                      & eos\_5Xsolar\_cond \\
			Cloud\_lo (Pa):                                   & 0 \\
			Eq. of State (hi metal atm):                      & eos\_1000Xsolar\_cond \\
			Cloud\_hi (Pa):                                   & 10000 \\
			&  \\
			Out of transit factor (\% tdur):                  & 100 \\
			+ Out of transit ``timing tax" (sec):              & 3600 \\
			Slew duration avg. (sec):                         & 1108 \\
			SAMs: small angle maneuvers (sec):                & 0 \\
			GS Acq: guide star acquisition(s)(sec):           & 284 \\
			Targ Acq: target acquisition (sec):               & 492 \\
			Exposure Ovhd: factor 1:                          & 0.0393 \\
			Exposure Ovhd: factor 2 (sec):                    & 26 \\
			Mech: mechanism movements (sec):                  & 110 \\
			OSS: Onbd Script Sys. compilation (sec):          & 65 \\
			MSA: NIRSpec MSA config. (sec):                   & 0 \\
			IRS2: NIRSpec IRS2 Detector setup (sec):          & 0 \\
			Visit Ovhd: visit cleanup (sec):                  & 110 \\
			Obs Ovhd factor (\%):                             & 16 \\
			DS Ovhd (sec):                                    & 0 \\
			&  \\
			RunExoT (Y/N):                                    & Y \\
			RunPdxo (Y/N):                                    & Y \\
			RunRank (Y/N):                                    & Y \\
			&  \\
			\hline
			\multicolumn{2}{l}{
				\begin{minipage}{10cm}~\\[1.5ex]
					Notes. --- The ``EOS (Equation of State)" file nomenclature is defined in the \texttt{Exo-Transmit} user manual. The detection threshold ($dBIC$) is discussed in detail in Section \ref{sec:detecting}.
				\end{minipage}
			}
		\end{tabular}
	\end{center}
\end{table*}

\texttt{Exo-Transmit} provides the user with over 60 different EOS file choices, modeling various conditions of metallicity, condensation, etc.  We have chosen files with solar constituents, and with high and low levels of metal abundance.  These particular files also allow for modeling condensation and removal via rain-out of molecules from the gas phase \text{-} but they do not include condensation and rainout of graphite.  This was an arbitrary choice.  Alternative EOS files are available that would allow consideration of the rain-out of graphite, which would deplete the model atmosphere of all carbon-bearing species at low temperature.

As described in \citet{1538-3873-129-974-044402}, \texttt{Exo-Transmit} ``allows the user to incorporate aerosols into the transmission spectrum calculation following one of two ad-hoc procedures. The first is to insert a fully optically thick gray cloud deck at a user specified pressure. The second is to increase the nominal Rayleigh scattering by a user-specified factor."  We are using the first procedure.
  
For a given target, and for each of the atmosphere cases, our \texttt{ExoT{\_}Master} subprogram will write the appropriate set of planetary parameters to a transfer input file (\texttt{userInput.in}).  The parameters written include the equilibrium temperature (we assume an isothermal atmosphere); the equation of state (defines the metallicity and other atmospheric characteristics, e.g, rain-out of condensates); planet surface gravity; planet radius; host star radius; and in the case of clouds, the pressure at the cloud top.  We have assumed the default Rayleigh scattering factor (1.0), and default collision induced absorption factor (1.0).
 
\begin{figure*}[!htbp]
	\centering
	\includegraphics[width=1.005\textwidth]{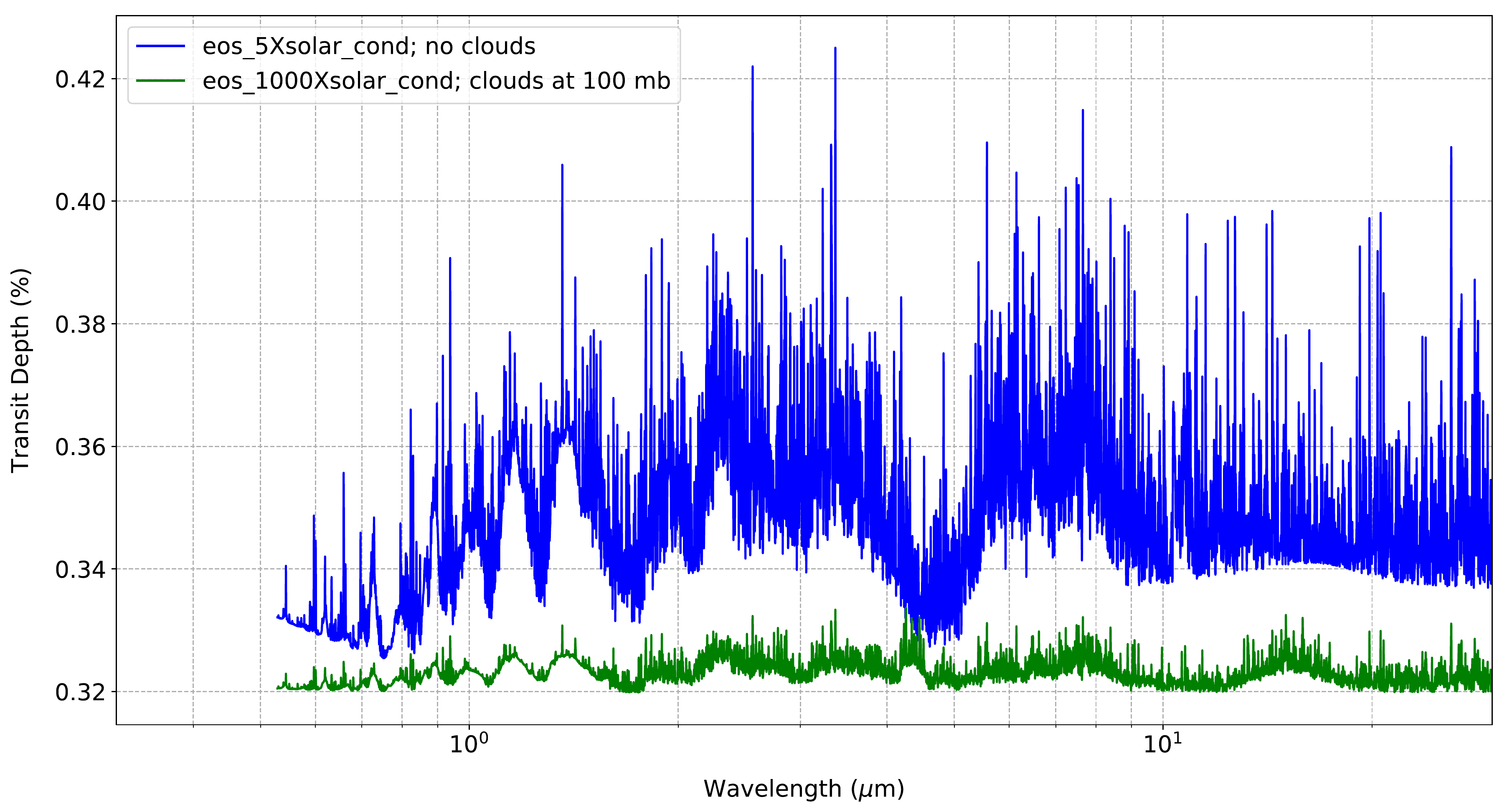}
	\setlength{\abovecaptionskip}{-5pt}
	\caption{Model transmission spectra for Sullivan Target 1292.  The spectra show the full resolution computed by \texttt{Exo-Transmit}.  Note the substantial difference (on the order of 100 to 300\,ppm) between the low and high metallicity atmospheric spectra.\label{fig:modelatmospheres1292}}
\end{figure*}

For each target/atmosphere case our \texttt{Python} program \texttt{ExoT{\_}Master} calls \texttt{Exo-Transmit} to generate model spectra.  For each case, \texttt{Exo-Transmit} generates a moderate resolution (R $\sim$ 1000) spectrum across a wide wavelength range ($0.3\text{  - 30 } \mu \text{m}$).  Figure \ref{fig:modelatmospheres1292} presents an example of the model spectra generated for Sullivan Target 1292, a Category 1 planet (Cool Terrestrial) with \mbox{$R_{p}$ $\sim$ 1.52$\rearth$}, and $\teq$ $\sim$ 300 K.

\begin{figure*}[!htbp]
	\centering
	\includegraphics[width=1.0\textwidth]{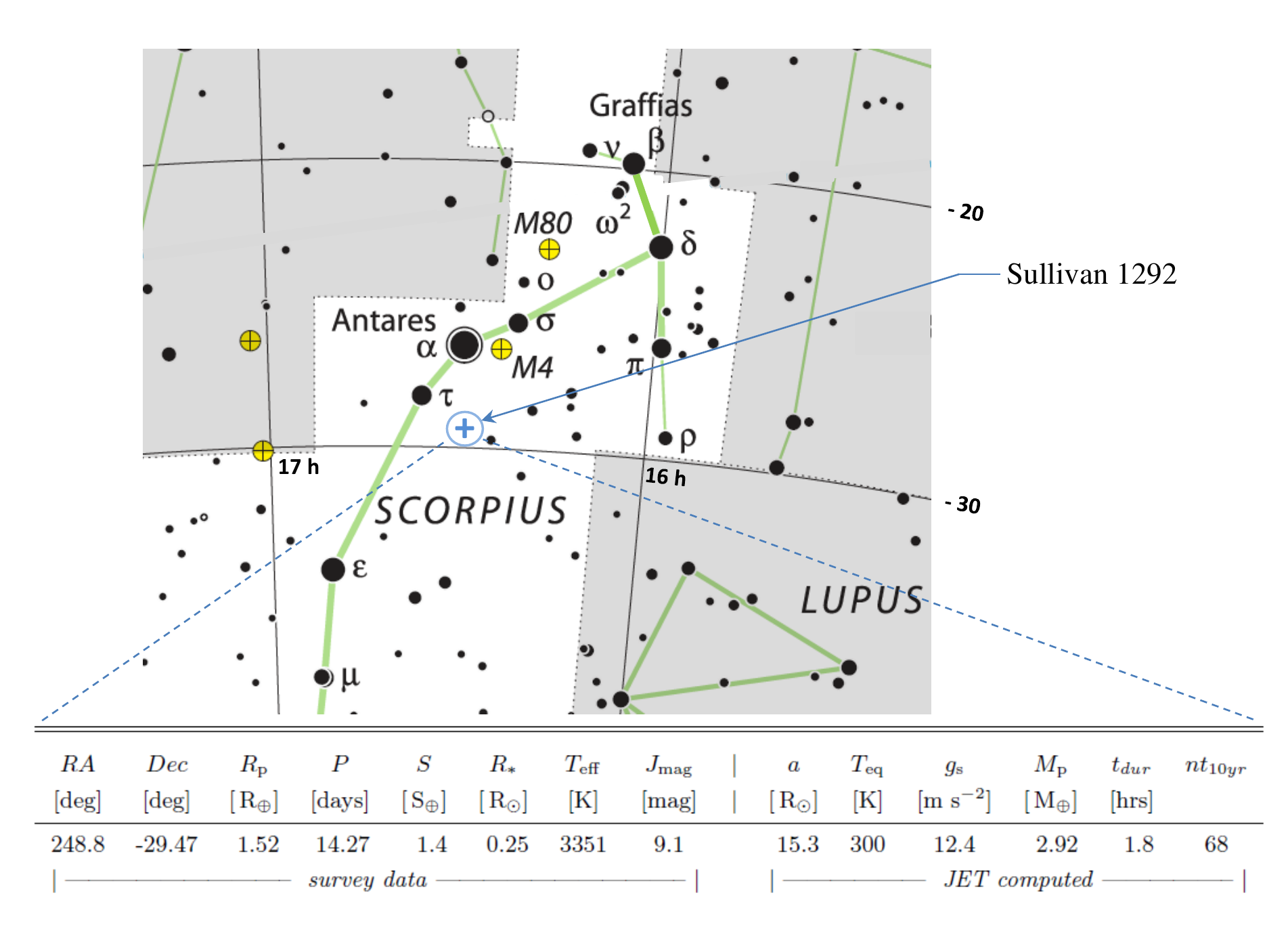}
	\setlength{\abovecaptionskip}{-15pt}
	\caption{We have arbitrarily chosen Sullivan Target 1292 as a reasonable example case. We show the catalog data and the \texttt{JET} computed parameters for the target, as well as its location (as simulated) on the sky. The chart shows relative $V_{\rm mag}$.  For reference, Antares has $V_{\rm mag} = 0.9$, and Sullivan 1292 has $V_{\rm mag} = 12.7$.  Attribution: IAU and Sky \& Telescope magazine (Roger Sinnott \& Rick Fienberg) [CC BY (\url{https://creativecommons.org/licenses/by/3.0})] The figure has been modified from the original.}\label{fig:Sullivan1292}	
\end{figure*}

We have chosen Target 1292 arbitrarily, but it is a reasonable example case.  The catalog data and \texttt{JET}-computed parameters for this target as well as its location (as simulated) on the sky are shown in \mbox{Figure \ref{fig:Sullivan1292}}.

\subsection{Detecting Transmission Spectra with \emph{JWST} (Pdxo\_Master)}
\begin{figure*}[!htbp]
	\centering
	\includegraphics[width=1.0\textwidth]{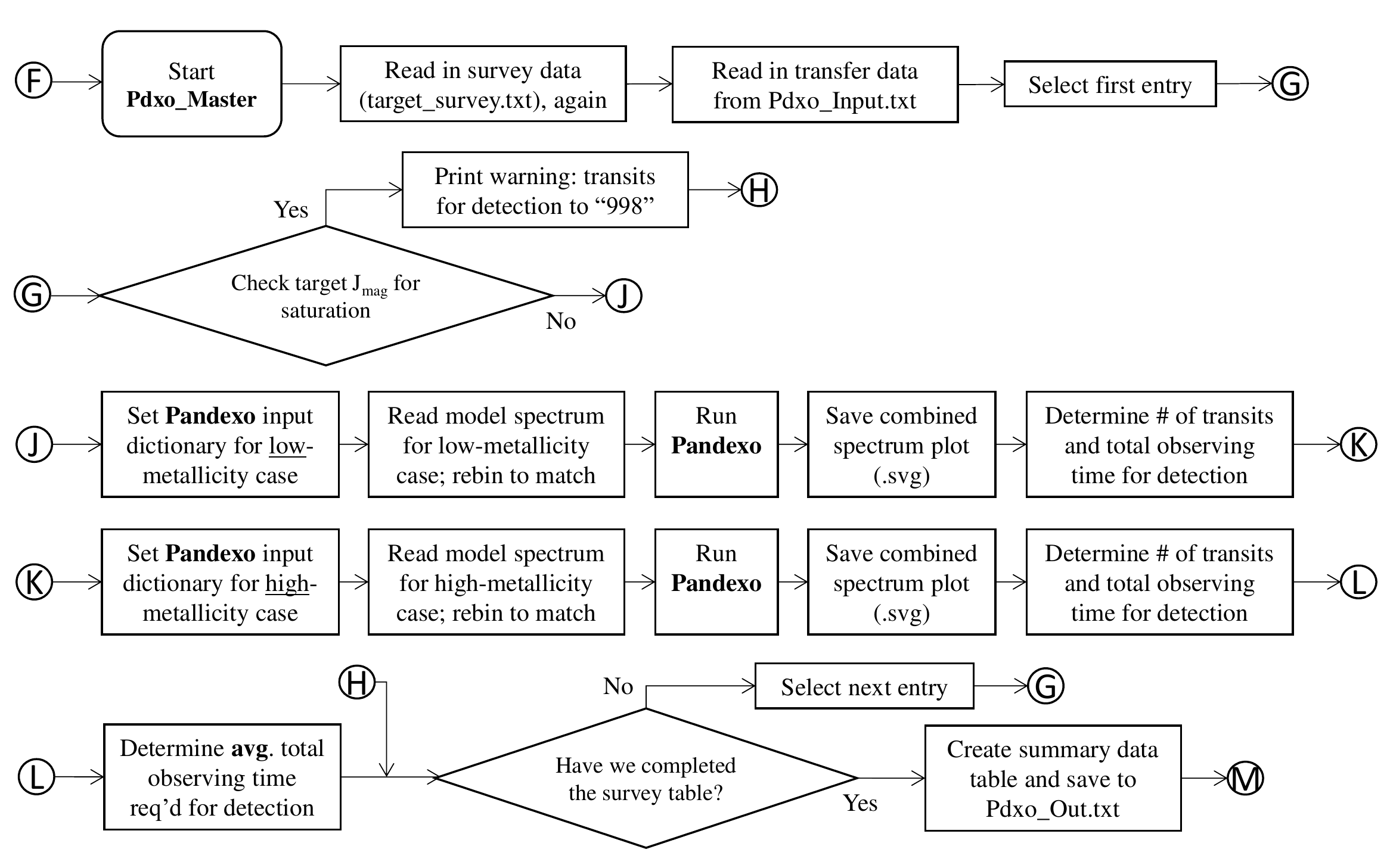}
	\setlength{\abovecaptionskip}{-5pt}
	\caption{Analysis and code flow for \texttt{JET} program element \texttt{Pdxo{\_}Master}.\label{fig:Pdxoflow}}
\end{figure*}

The next major step in the analysis process is to determine if the \emph{JWST} instruments can detect the target transmission spectra. Where ``detect" indicates that the simulated spectrum fits the model spectrum better than a flat line to a statistically significant level.  This task is carried out by the \texttt{JET} program element: \texttt{Pdxo{\_}Master}.  Figure \ref{fig:Pdxoflow} presents a flowchart for the \texttt{Pdxo{\_}Master} program element.\\

\subsubsection{Generating Simulated Transmission Spectra with PandExo}\label{sec:gensimspectra}
For each target/atmosphere case, \texttt{Pdxo{\_}Master} first sets up a \texttt{Python} input dictionary for the instrument simulator, \texttt{PandExo}.

\citet{1538-3873-129-976-064501} presented an open-source \texttt{Python} package, \texttt{PandExo}, to provide community access to \emph{JWST} instrument noise simulations.  \texttt{PandExo} relies on the Space Telescope Science Institute's Exposure Time Calculator, \texttt{Pandeia} \citep{Pickering2016}. It can be used as both an online tool\footnote{\url{https://exoctk.stsci.edu/pandexo/}} and a \texttt{Python} package for generating instrument simulations of \emph{JWST}'s NIRSpec, NIRCam (Near-Infrared Camera), NIRISS, MIRI, and \emph{HST}'s WFC3. \texttt{PandExo} has been shown to be within 10\,\% of the \emph{JWST} instrument team's noise simulations and has become a trusted tool of the scientific community.

\texttt{PandExo} takes as input the target catalog data, the system parameter data, and the model spectrum for the particular target/atmosphere in question.  It also takes input of the top level settings that define the particular instrument considered, its wavelength range, noise floor value, and brightness limits, among other things.

The brightness limits for NIRSpec are given in Table 3 of the STScI's on-line User Documentation for NIRSpec Bright Object Time-Series Spectroscopy.\footnote{\url{https://jwst-docs.stsci.edu/near-infrared-spectrograph/nirspec-observing-modes/nirspec-bright-object-time-series-spectroscopy}}  The full-well J-magnitude brightness limits as a function of a target's $\teff$  are given for each mode/filter.  The SUB2048 subarray is assumed in all cases except for PRISM/CLEAR values that were determined using SUB512.  These values are for gain = 2, and a full well depth of 65,000.  The \texttt{JET} user manually inputs the magnitude limits for $\teff$ values of 2500\,K, 5000\,K, and 10,000\,K for the chosen disperser.  \texttt{JET} then uses an interpolation routine to determine the value of the full-well magnitude limit for any particular target $\teff$ over the full range of targets considered in the survey/catalog.  For targets with $\teff$ greater than 10,000\,K, or less than 2500\,K, the \texttt{JET} code will extrapolate the full-well magnitude limit vs $\teff$  curve, holding a constant endpoint slope. The detector percent-full-well limit (e.g., 80\%) is a user specified value.  We use a simple correction formula to estimate the J-magnitude limit for a particular percent-full-well condition: 

\begin{equation}\label{maglim}
\begin{split}
J_{\mathrm{mag, limit}}\text{(X\%\,FW)}  & = J_{\mathrm{mag, limit}}\text{(100\%\,FW)}\\
& \hspace{8ex} - 2.5*log(\text{X\%\,FW}) 
\end{split}          
\end{equation}

In order to provide a margin for linearity we are using 80\% full-well in our baseline run with NIRSpec.
  
Similarly, the brightness limits for NIRISS SOSS are shown in Table 1 and Figure 1 of the STScI's on-line User Documentation for NIRISS Bright Limits.\footnote{\url{https://jwst-docs.stsci.edu/near-infrared-imager-and-slitless-spectrograph/niriss-predicted-performance/niriss-bright-limits}}   Currently \texttt{JET} can simulate the performance of the SUBSTRIP96 subarray, with Order 1.  The full-well J-magnitude limit is listed as 7.5 for a G2V star.  This bright limit varies by +0.15 magnitudes for an A0V star to -0.10 magnitudes for an M5V star.  The \texttt{JET} user manually inputs the three data points (for $\teff$ = 2500\,K, $\jmag$ $\sim$ 7.4; for 5000\,K, $\jmag$ $\sim$ 7.5 and for 10,000\,K, $\jmag$ $\sim$ 7.65) for the full-well limits.  The same correction formula described in Equation \ref{maglim} above is used to estimate the J-magnitude limit for a particular percent-full-well condition for NIRISS SOSS. Again, we are using an 80\% full-well cap on the brightness limit in our variation studies with NIRISS.  

\begin{figure*}[!htbp]
	\centering
	\begin{picture}(800, 240)
	\includegraphics[width=1.0\textwidth]{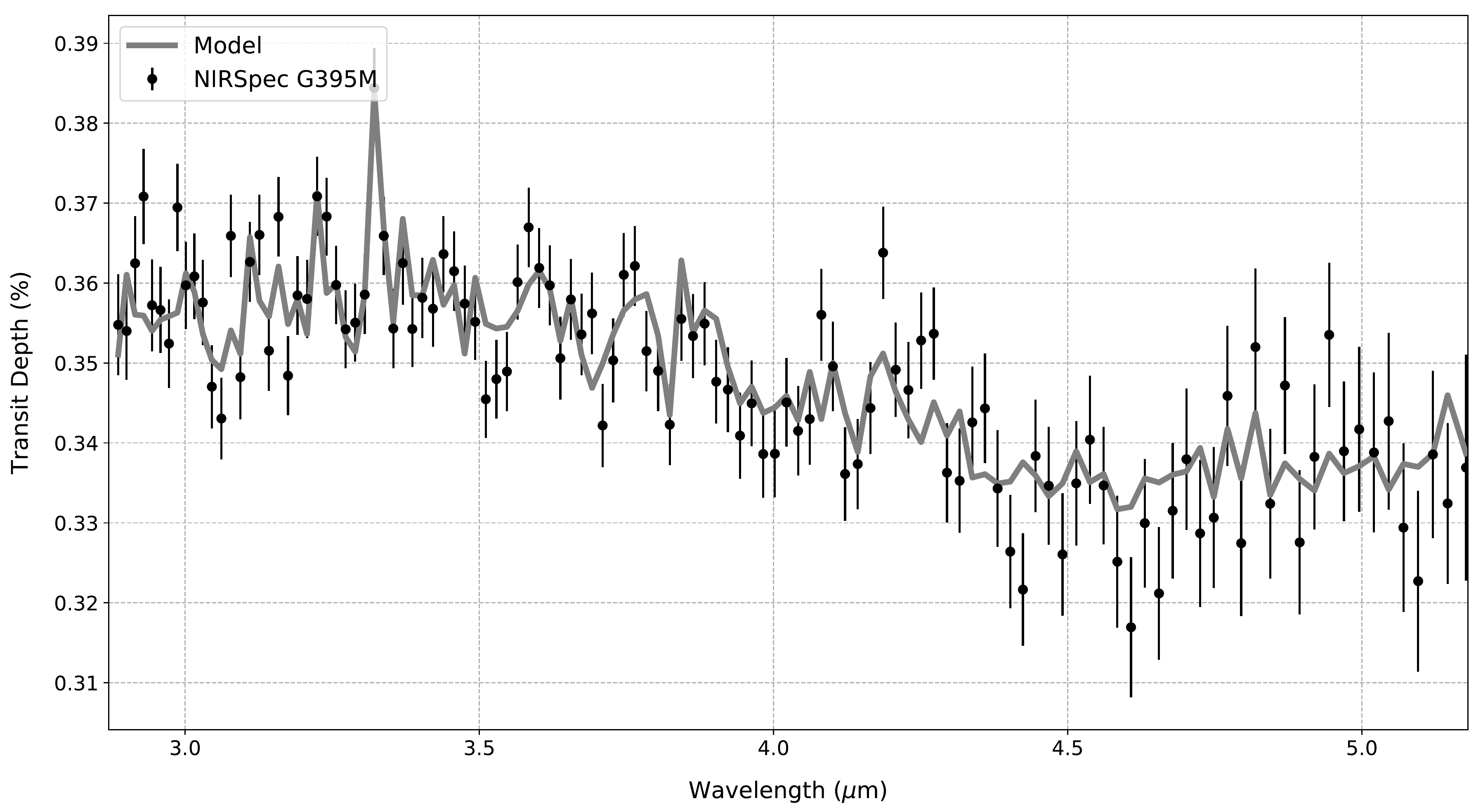}
	\end{picture}
	\begin{picture}(800, 290)
	\includegraphics[width=1.0\textwidth]{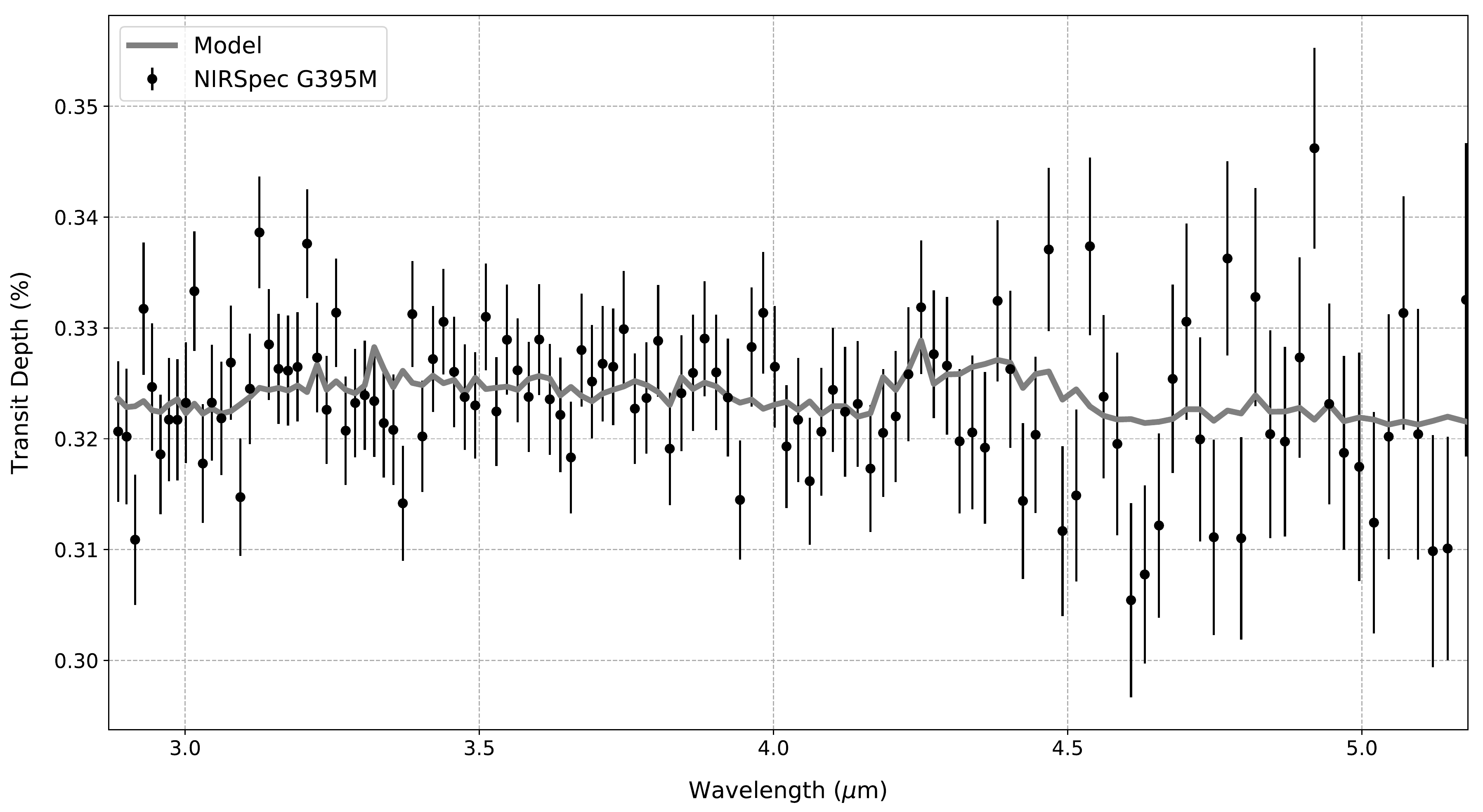}
	\end{picture}
	\setlength{\abovecaptionskip}{0pt}
	\caption{Simulated spectra for a single transit of Sullivan Target 1292 using NIRSpec G395M. The model spectra shown as gray background lines have been binned down to a resolution consistent with the simulated data (R $\sim$ 100). \emph{Top:} A low-metallicity (5xSolar, no clouds) atmosphere. For the lo-metal case the mean dBIC (less 1 $\sigma$) is $> 10$ for a single transit. \emph{Bottom:} A high-metallicity (1000xSolar, clouds at 100\,mbar) atmosphere. The mean dBIC (less 1 $\sigma$) for a single transit of the Sullivan 1292 hi-metal case is $\sim$ -18.  It takes 17 transits to reach the dBIC detection threshold of 10.\label{fig:tsp1292lonirspecg395m}}
\end{figure*}

\texttt{PandExo} is first run for a single transit.  This yields a baseline simulated instrument spectrum with 1$\sigma$ error bars for each data point (at $n$ wavelength locations, depending on the instrument and spectral resolution chosen).  Figure \ref{fig:tsp1292lonirspecg395m} presents simulated transmission spectra for Sullivan Target 1292 using NIRSpec G395M, with a low-metallicity atmosphere (5xSolar, no clouds), and a high metallicity atmosphere (1000xSolar, clouds at 100\,mbar).  The NIRSpec G395M disperser covers the wavelength range from 2.87 to 5.18 $\mu$m, with a noise floor estimated to be 25\,ppm.

The noise floor assumption is an important one.  Unfortunately, in the literature it is not always clear if a given noise floor value is associated with a single transit observation, or is a residual value after multiple transits.  We use the term in the latter sense.  This is consistent with our detection algorithm and the way that \texttt{PandExo} is structured.  In \texttt{PandExo} the propagated error generally falls with the square root of the observed transit time and number of transits.  When the error reaches the value of the noise floor it will go no lower for additional observation time or number of transits.

First, we will address the NIRISS noise floor assumption.  For the variation studies that we present in Section \ref{sec:noisefloorvar}, we assume a noise floor of 20 ppm for NIRISS SOSS.

\citet{Beichman2014} presented ``a transmission spectrum simulation of what we would expect with NIRISS for GJ1214b, a super-Earth around a star of magnitude J = 9.75. The simulation assumes 12 hours of clock time spread over 4 transits, and a 20 ppm noise floor."  Here the noise floor is given in a multi-transit residual sense.

\citet{Greene2016} suggested 20 ppm for the NIRISS SOSS noise floor.  They pointed out that,``The best \emph{HST} WFC3 G141 observations of transiting systems to date have noise of the order of 30 ppm (Kreidberg et al. 2014a) , . . . .   We adopt reasonably optimistic systematic noise floor values of 20 ppm for NIRISS SOSS, . . . . These are less than or equal to the values estimated by Deming et al. (2009) for the \emph{JWST} NIRSpec . . . . The excellent spatial sampling of the NIRISS GR700XD SOSS grism approaches that of the \emph{HST} WFC3 G141 spatial scanning mode, and both instruments have reasonably similar HgCdTe detectors. We anticipate that decorrelation techniques will continue to improve, so we assign a 20 ppm noise floor value to NIRISS even though \emph{HST} has not yet [as of early 2016] done quite this well."

More recently, \emph{HST} WFC3 has demonstrated even better performance.  \citet{Line2016} reported a precision for \emph{HST} WFC3 (at 1.1 to 1.6 $\mu$m) of $\sim$ 17 ppm for multiple secondary eclipse observations of HD209458b.  In spite of this we choose to remain conservative with a noise floor estimate of 20 ppm for NIRISS SOSS.  There are still many unknowns for \emph{JWST}.

Next, we consider our noise floor assumption for \mbox{NIRSpec.} Of course the actual performance of NIRSpec will not be known until \emph{JWST} is on orbit and fully commissioned, but there are some known issues that could make it difficult for NIRSpec to improve upon the \emph{HST} WFC3's very low noise qualities.  Specifically the \mbox{NIRSpec} image sampling is not quite as good as \emph{HST} WFC3, the detector electronics (in particular the System for Image Digitalization, Enhancement, Control, and Retrieval (SIDECAR) ASICs, used for detector readout and control)\footnote{\raggedright{\url{https://www.cosmos.esa.int/web/jwst-nirspec/nirspec-s-design}}} are not well characterized at this point, and the measured wavefront-error is acceptable, but not excellent.

NIRSpec may have more systematic noise than NIRISS because it is not spatially sampled well at wavelengths less than about 3 microns. The PSF of \emph{JWST} is on the order of $\sim$ 70 mas FWHM at 2 microns,\footnote{\raggedright{\url{https://jwst.nasa.gov/content/forScientists/faqScientists.html\#howbig}}} but the NIRSpec detectors have 100 mas pixels.\footnote{\raggedright{\url{https://jwst-docs.stsci.edu/near-infrared-spectrograph/nirspec-overview\#NIRSpecOverview-Opticalelementsanddetectors}}}  This under-sampling condition will tend to degrade resolution and add systematic noise that can generally not be completely eliminated by co-adding transits. 

``Since the NIRSpec PSF is under-sampled at most wavelengths, dithering is required to achieve nominal spectral and spatial resolution."\footnote{\raggedright{\url{https://jwst-docs.stsci.edu/near-infrared-spectrograph/nirspec-observing-modes/nirspec-fixed-slits-spectroscopy}}}

This under-sampling issue has come up with other space telescope instruments.  Discussing \emph{Spitzer}/IRAC performance, \citet{Ingalls2012} commented that ``due to the under-sampled nature of the PSF, the warm IRAC arrays show variations of as much as 8\% in sensitivity as the center of the PSF moves across a pixel due to normal spacecraft pointing wobble and drift. These intra-pixel gain variations are the largest source of correlated noise in IRAC photometry."

In addition, \citet{Grillmair2012} identified a number of \emph{Spitzer}/IRAC systematic issues ``that limit the [telescope's] attainable precision, particularly for long duration observations [e.g., transits]. These include initial pointing inaccuracies, pointing wobble, initial target drift, long-term pointing drifts, and low and high frequency jitter. Coupled with small scale, intra-pixel sensitivity variations, all of these pointing issues have the potential to produce significant, correlated photometric noise." \emph{JWST} will be affected by these issues to a degree, to be determined.

The NIRSpec wavefront-error performance is acceptable, but it is not equal to the excellent performance of \emph{HST} WFC3.  \citet{Aronstein2016} noted that ``. . . there are currently violations in the requirements for the uncertainties of 3rd order aberrations in NIRSpec."

\citet{Louie2018} argue that systematic noise can be removed by co-adding multiple transits.  In our view this may be true for some systematic noise sources, but not for all (e.g., thermal shock after slew).

\citet{Greene2016} also point out that,``Astrophysical noise (e.g., Barstow et al. 2015) and/or instrumental noise (e.g., decorrelation residuals) produce systematic noise floors that are not lowered when summing more data. . . . We do not know how well co-adding observations of multiple transits or secondary eclipses will improve our results at this time. The simulated single-transit and single-eclipse observations of our selected systems typically have total noise values only $10\% - 50\%$ larger than our adopted noise floors . . .  so systematic noise assumptions have already significantly influenced the precision of our simulated data . . . .  Given this, co-adding more data would not substantially improve the results for these very observationally favored systems with bright host stars [generally the case with \emph{TESS} targets]."

It should be mentioned that the noise floor generally has more of an impact on the more difficult detection situations, where co-adding many transits is needed for a detection.  These targets will tend to be further down the ranking list and may well fall below a time allocation cut-off line.  Reducing the noise floor would likely only have a marginal effect on the cut-off-limited ranking.

While it could be argued that our noise floor values are somewhat conservative, it would seem that for the purpose of establishing preliminary target rankings a conservative approach is appropriate.  In the end, the noise floor for any particular instrument configuration is an input value and can be changed at the user's discretion.

For the purpose of our full catalog baseline run with \texttt{JET}, we have chosen the NIRSpec instrument with the G395M disperser and the F290LP filter operating in Bright Object Time Series (BOTS) mode.  There were a number of factors that influenced this decision.  First, according to \citet{Batalha2017}, ``An observation with both NIRISS [SOSS] and NIRSpec G395M/H always yields the highest information content spectra with the tightest constraints, regardless of temperature, C/O, [M/H], cloud effects or precision.''   In addition, \citet{Bean2018} included the NIRSpec G395H as one of the consensus high priority modes for the Early Release Science Program for \emph{JWST}.  Also, \texttt{PandExo} run-times for NIRSpec are significantly shorter than for NIRISS.  We have made small variation-set runs with NIRISS, but as a practical consideration, for the full catalog baseline run we felt that NIRSpec would be a better choice.  Finally, the NIRSpec G395M wavelength coverage gives us (simulated) access to the 4.5 $\mu$m region and effects of the important atmospheric constituents CO and CO$_{2}$, which would not be true for NIRISS.

The NIRSpec simulations are performed using the S1600A1 aperture with a fixed 1.6" x 1.6" field of view.\footnote{\raggedright{\url{https://jwst-docs.stsci.edu/near-infrared-spectrograph/nirspec-observing-modes/nirspec-bright-object-time-series-spectroscopy}}}   Exposures use the SUB2048 subarray (2048x32 pixels)\footnote{\raggedright{\url{https://jwst-docs.stsci.edu/near-infrared-spectrograph/nirspec-instrumentation/nirspec-detectors/nirspec-detector-subarrays}}}  to record the full spectrum using the NRSRAPID read mode \citet{Ferruit2014}. 

Given the long PandExo run times for our baseline instrument mode (NIRSpec G395M), and given the purpose of our study (a simple detection, not a full constituent retrieval) we felt that the M disperser mode was an appropriate choice over the H mode which would have even longer run times.

It should be kept in mind that the instrument/mode chosen for the baseline run was primarily for the purpose of demonstrating the \texttt{JET} code.  The user can study other modes (e.g., NIRSpec G140M/BOTS, and G235M/BOTS, as well as NIRISS SOSS, with the GR700XD disperser) that have been implemented and tested in \texttt{JET}.

In the figures, we show the model spectrum in the background, and have re-binned the data to R $\sim$ 100 to match the resolution set in the \texttt{PandExo} simulation.  The lower (than native instrument) resolution is used to reduce overall computation time.  In the description of the analysis that follows, the re-binned model spectral values are designated $y_{\rm rebin}$.  The re-binning process is accomplished using the \texttt{Python} package \texttt{SpectRes} as described in \citet{Carnall2017}.

\subsubsection{Detecting the Transmission Spectra}\label{sec:detecting}
For a particular target/atmosphere, the questions that we need to answer are: (1) Can the spectrum be detected given the instrument's noise characteristics?, and (2) If the observation of one transit is insufficient to detect the spectrum, will the observation of additional transits pull a detectable signal out of the noise?

Specifically we want to determine, in a formal way, whether our simulated spectrum is a better fit to the model spectrum or to a flat-line spectrum.  We use a Bayesian Information Criterion ($BIC$) approach to guide this selection \citep{Wall2012, Kass1995}.  This technique will allow us to determine if the spectrum is detectable at all, and if so, how many transit observations will be needed for a strong detection.

Fortunately, we do not need to re-run the full \texttt{PandExo} simulation over and over to determine the results of observations of multiple transits.  We can compute the improvements in signal to noise, and the effects of the instrument noise floor analytically.

Our one-transit \texttt{PandExo} run, described in the previous section, generates a simulated spectrum with $n$ spectral data elements of wavelengths ($x$), spectral values ($y$), and error values ($e$).  In addition, we can extract another set of data from \texttt{PandExo} that does not include random noise.  We will call these the wavelengths ($x_{\rm 1trans}$), the spectral values ($y_{\rm 1trans}$), and the noise values ($e_{\rm 1trans}$).

We then compute the integrated multi-transit noise ($noise_{\rm multi-transit}$):
\begin{equation}\label{mt}
noise_{\rm multi-transit} = \frac{e_{\rm 1trans}}{\sqrt{ntr}}              
\end{equation}
where $ntr$ is the number of transit observations co-added.

If this value is less than the noise floor, then we reset it to the noise floor lower bound. Clearly, this multi-transit noise term will tend to smaller and smaller values as the number of transits increases (bounded by the noise floor).

Next, we compute a random noise value:
\begin{equation}\label{randomnoise}
noise_{\rm random} =  noise_{\rm multi-transit}*f_{\rm random}              
\end{equation}
where for each statistical sample, the $f_{\rm random}$ term is drawn from a standard normal distribution with mean 0 and variance 1.

Now we can recast the spectrum with random noise:
\begin{equation}\label{simspec}
x_1 = x_{\rm 1trans}             
\end{equation}

\begin{equation}\label{simspec}
spectrum_{\rm sim} = y_{\rm 1trans} + noise_{\rm random}          
\end{equation}

Adjusting for transit depth $\%$ units:
\begin{equation}\label{simspec}
y_{\rm 100} = spectrum_{\rm sim}*100         
\end{equation}

\begin{equation}\label{simspec}
e_{\rm 100} = noise_{\rm multi-transit}*100             
\end{equation}

Next, we compute the $BIC$ value for the model spectrum case.  We first determine the chi-squared statistic ($\chi_{\rm MS}^2$), and the reduced chi-squared statistic ($\chi_{\rm MS,r}^2$), for the single transit spectral data: 

\begin{equation}
\chi_{\rm MS}^2 = \sum \bigg(\frac{y_{\rm 100}-y_{\rm rebin}}{e_{\rm 100}}\bigg)^2            
\end{equation}
\bigskip

\begin{equation}
\chi_{\rm MS, r}^2 = \frac{\chi_{\rm MS}^2}{(n - n_{\rm params})}            
\end{equation}

For the model spectrum case the number of free parameters ($n_{\rm params}$) used in the calculation of the reduced chi-squared ($\chi_{\rm MS, r}^2$) is five. 
 
As discussed in Section \ref{sec:modtranspec}, \texttt{JET} uses the transmission spectroscopy code \texttt{Exo-Transmit} to generate model spectra of the target planet atmospheres.  In the \texttt{JET} implementation, \texttt{Exo-Transmit} reads in survey/catalog data for each target planet and generates a moderate resolution transmission spectrum of the planet's atmosphere for two atmosphere cases: (1) low metallicity solar with no clouds, and (2) high metallicity solar with low altitude clouds.

The \texttt{JET} code then generates a simulated observational spectrum.  Our observational simulator, \texttt{PandExo} (described in Section \ref{sec:gensimspectra}), uses the primary model as the starting point for the simulation.  It introduces noise based on the target characteristics, and the performance of the \emph{JWST} instrument under study.  

The goal of our target ranking exercise is to determine the targets for which \emph{JWST} observations could potentially be the most informative. Given that we do not know the atmospheric composition a priori, simulating transmission spectra for two drastically different atmospheres allows us to investigate which targets would be the most interesting for future atmospheric characterization via retrieval studies.  

The question is: for our calculation of the reduced chi-squared, and the model spectrum BIC, how many free parameters should we employ?   

A typical \emph{JWST} retrieval, assuming high S/N and high spectral resolution, may employ ten or even more free parameters \citep{Madhusudhan_2018}.  For our initial ranking and selection of targets, we are more interested in assessing the overall level of information content within each spectrum rather than determining detailed atmospheric abundances for individual elements. We therefore adopt a relatively low spectral resolution (R=100) and consider a broad set of potential targets, some of which may be of somewhat lower S/N.  For the model spectrum BIC analysis, we therefore consider a lower number of free parameters to correspond with the lower resolution and lower S/N of our simulated spectra. \citep{benneke2013bayesian}.  

We have chosen five representative free parameters for our model spectrum BIC calculation for each target/atmosphere case.  The parameters include:  (1) planet equilibrium temperature, (2) bulk metallicity/atmospheric composition (1x solar, 10x solar, etc.), (3) Rayleigh scattering (on/off), (4) cloud height, and (5) whether to include rainout of molecules out of the gas phase (i.e., condensation/gas models).  Of course other ``fixed" catalog values (e.g., $R_{p}$, $R_{*}$, $g_{s}$) are needed to generate the model spectra for each target, but these are not included as ``free" parameters because the planets and their host stars will have known radii, masses, and surface gravities.

To provide flexibility, the number of free parameters for the model spectrum BIC calculation is an input value and can be changed by the user.

Next, we compute an error factor to drive \mbox{$\chi_{\rm MS,r}^2$ to 1.}  \texttt{JET} runs \texttt{PandExo} for a single transit and adds a random noise value with raw error-bars (based on the single transit S/N level) to the underlying model spectrum.  Since the model spectrum is the correct model for the simulation by construction, the reduced $\chi^2$ should indicate a good fit.  In principle, a value of $\chi_{\rm MS,r}^2$ = 1 indicates that the extent of the match between the simulated observation and the model is in accord with the error variance.  The error factor ($f_{\rm error}$) is used to adjust the simulation error-bars to result in $\chi_{\rm MS,r}^2$ = 1.  These adjusted error bars are then used in the $\chi^2$ (and $BIC$) calculations for the model spectrum and the associated flat-line spectrum.

The error factor based on the raw spectrum data is:
\begin{equation}
f_{\rm error} = \sqrt{\chi_{\rm MS,r}^2}             
\end{equation}

From this we can determine the rescaled $\chi^2$ for the model atmosphere case:
\begin{equation}
\sigma_{\rm combined} = \sqrt{(f_{\rm error} * e_{\rm 100})^2}            
\end{equation}

and,
\begin{equation}
\chi_{\rm MS}^2 = \sum \bigg(\frac{y_{\rm 100}-y_{\rm rebin}}{\sigma_{\rm combined}}\bigg)^2   
\end{equation}

and now,
\begin{equation}
\chi_{\rm MS,r}^2 = \frac{\chi_{\rm MS}^2}{(n - n_{\rm params})} = 1 
\end{equation}
\bigskip

Finally, we can compute the $BIC$ value for the model spectrum case ($BIC_{\rm MS}$).  The $BIC$ is formally defined, using the traditional nomenclature, as:
\begin{equation}
BIC =\ln(n)k-2\ln({\hat {L}})
\end{equation}
where $\hat {L}$ is the maximized value of the likelihood function of the model $M$.   $\hat {L} = p(x|\hat{\theta},M)$, and $\hat {\theta }$ are the parameter values that maximize the likelihood function.  $x$ is the observed data, in our case the spectral transit depth values; $n$ is the number of data points in $x$ (in our case the wavelength points), essentially the sample size; and $k$ is the number of parameters estimated by the model (in our case the number of ``free" parameters, $n_{\rm params}$).

As discussed in \citet{Kass1995}, under the assumption that the errors are independent and follow a normal distribution, the $BIC$ can be rewritten in terms of the $\chi^2$ as: 
\begin{equation}
BIC = \chi^2 + k\ln(n)
\end{equation}

\bigskip
For the model spectrum case we have:
\begin{equation}
BIC_{\rm MS} = \chi_{\rm MS}^2 + (n_{\rm params})*\ln(n)
\end{equation}

where we have set $n_{\rm params} = 5.$
\bigskip

Next, we repeat the $BIC$ calculation for the case of the flat-line spectrum.  In this case only one number, the y-intercept of the flat line, needs to be specified.  This number can be determined by taking the median value of the simulated spectrum values.  This is consistent with the conventional approach where the model parameters are determined from the observational data.   

We can now determine the $BIC$ value for the flat-line spectrum case ($BIC_{\rm FL}$):
\begin{equation}
BIC_{\rm FL} = \chi_{\rm FL}^2 + (1)*\ln(n)
\end{equation}

For model selection we are interested in the difference in $BIC$.  When picking from multiple models, the one with the lower $BIC$ is preferred.

We can define the parameter:
\begin{equation}\label{dBIC}
dBIC = \Delta BIC = BIC_{\rm FL} - BIC_{\rm MS}
\end{equation}

For a very strong detection we need $dBIC > 10$ \citep{Kass1995}.  This loosely corresponds to a 3.6\,$\sigma$ confidence level of detection \citep{Gordon2007} for the mean value of dBIC.

Now we can loop on this analysis sequence, holding the number of transits constant and just considering new random samples of noise.  With enough samples (we use 2000 samples for each number-of-transits grid point) we build up a distribution of $dBIC$.  The distribution tends toward Gaussian from which we can determine the mean and standard deviation. 

We then move to the next higher number of transits on the grid and build up another distribution of $dBIC$, repeating the process outlined in Equations \ref{mt} through \ref{dBIC}.  We use 15 transit grid points, from 1 transit to 50 transits (with spacing increasing as the number of transits increase).  The result is a dataset of mean $dBIC$ (and standard deviation) vs number of transits.

For a particular number-of-transits grid point, if we use the mean value of $dBIC$ as our critical parameter, then we can say that we have a 50\,\% confidence level of a very strong detection if $dBIC > 10$.  Assuming a Gaussian distribution, if we reduce our $dBIC_{\rm mean}$ parameter by 1$\sigma$, then for this $dBIC_{\rm mean}$ less 1$\sigma$ we have an $\sim$ 84\,\% confidence level of a very strong detection.

For each target/atmosphere combination we first check the $dBIC_{\rm mean}$ less 1$\sigma$ for one transit.  If it is \mbox{$>$ 10} we record that detection and move on.  If not, then we check to see if the $dBIC_{\rm mean}$ less 1$\sigma$ for 50 transits is $>$ 10.  If not, then we flag this target/atmosphere as a non-detection and move on.

If neither of these conditions are met, then we need to determine the number of transits to reach the detection threshold.  We use the \texttt{Python} routine \texttt{UnivariateSpline} from the \texttt{scipy} package to interpolate the 15 element $dBIC_{\rm mean}$ less 1$\sigma$ dataset.  The smoothing parameter ($s$) is dynamic.  We set it to 2 (soft spline) if the $dBIC_{\rm mean}$ at 50 transits is less than 175; otherwise we set it to zero (spline goes through all points). Then we use the \texttt{scipy} routine \texttt{brentq} to find the crossing point of number of transits for a $dBIC_{\rm mean}$ less 1$\sigma$ of 10.

\begin{figure*}[!htbp]
	\centering
	\includegraphics[width=1.0\textwidth]{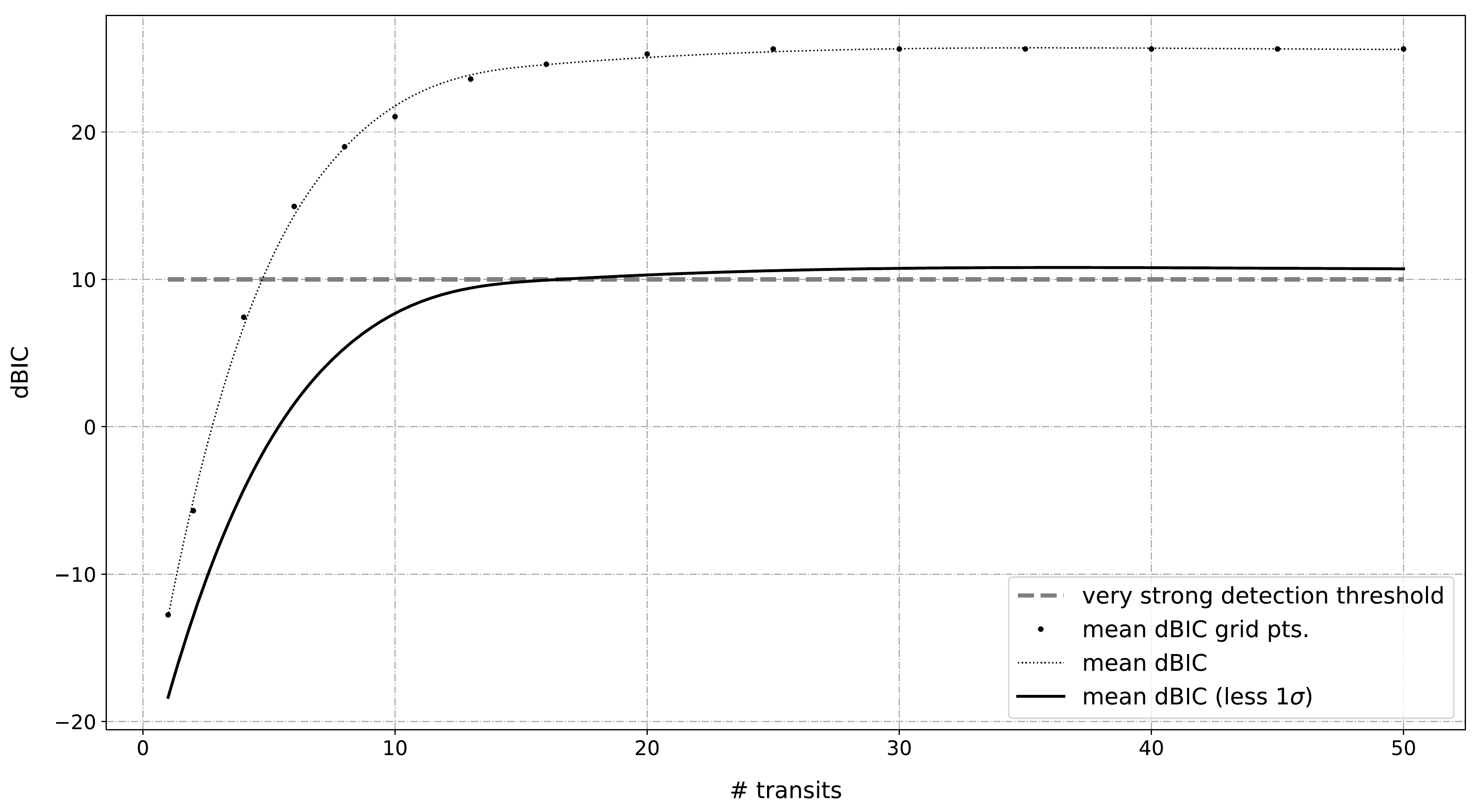}
	\setlength{\abovecaptionskip}{-10pt}
	\caption{$dBIC$ vs number of transits for Sullivan Target 1292 with a high metallicity (1000xSolar, clouds at 100\,mbar) atmosphere.  The mean $dBIC$ grid points are shown and the univariate spline fit with soft smoothing is shown as well for this mean value.  The mean less $1 \sigma$ value is also shown and is the critical line for determining detection with very high confidence. The plot shows a very strong detection for 17 transits.\label{fig:dbic1292hi}}
\end{figure*}

Figure \ref{fig:dbic1292hi} presents a plot of $dBIC$ vs number of transits for Sullivan Target 1292 with a high metallicity atmosphere.

\subsubsection{Determining the Observing Cycle Time Needed for Detection}
The critical resource is total observing time.  For a single transit observation there are a number of actions that take time.  These include: (1) slewing the spacecraft to the proper target; (2) acquiring the target; (3) allowing the detectors to settle; (4) observing the transit; (5) observing the host star out-of-transit to establish the baseline flux level; and (6) various overheads added based on experience with other space telescopes.\footnote{\raggedright{\url{https://jwst-docs.stsci.edu/jwst-observatory-functionality/jwst-observing-overheads-and-time-accounting-overview}}}  We have captured these activities and overhead factors for NIRSpec-BOTS mode and NIRISS-SOSS for an example case with a transit duration of 2.4 hours (the median transit time of the Sullivan survey) in Table \ref{tab:obstime}.

The Instrument overheads shown in the table have been generated using the \emph{JWST} Astronomer's Proposal Tool (APT).\footnote{\raggedright{\url{https://jwst-docs.stsci.edu/jwst-astronomers-proposal-tool-overview}}}  The Slew time and Instrument Overheads shown do not change for the particular instrument configuration considered over a reasonable science-time range.

So, for a given number of transits needed for detection, we can determine the total observation time needed:  
\begin{equation}\label{tT}
t_{\rm tot} = ntr*(\text{Total charged time})            
\end{equation}

\noindent where $t_{\rm tot}$ is the total observation time in hours, $ntr$ is the number of transits required for detection of the transmission spectrum.

We determine the total observation time for both atmosphere cases of each target.  Then, given the broad range of plausible atmospheric compositions we simply take an average of the two to determine our figure of merit ($tT_{\rm avg}$): 

\begin{equation}\label{tTavg}
tT_{\rm avg} = \frac{(t_{\rm tot,\,lo} + t_{\rm tot,\,hi})}{2}              
\end{equation}

\noindent where $tT_{\rm avg}$ is the average total observation time in hours.

Finally, summary data for each target is gathered by \texttt{Pdxo{\_}Master}.  Upon completion of the analysis of the target list, this summary data is written to the file: \texttt{Pdxo{\_}Out.txt} in the working directory.  The data here is in unranked target order.

\subsection{Sorting and Ranking Targets (Rank\_Master)}
\begin{figure*}[!htbp]
	\centering
	\includegraphics[width=1.0\textwidth]{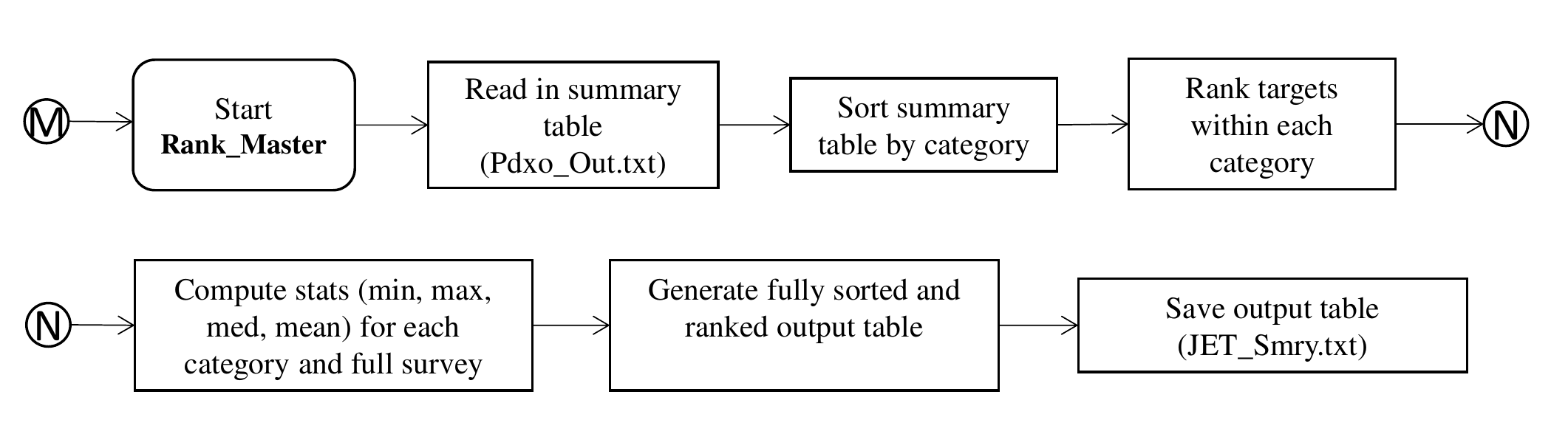}
	\setlength{\abovecaptionskip}{-15pt}
	\caption{Analysis and code flow for \texttt{JET} program element \texttt{Rank{\_}Master}.\label{fig:Rankflow}}
\end{figure*}

Once we have completed the analysis of all of the targets and computed the average total observation time required for detection of the atmosphere it is time to sort and rank the targets and provide summary statistics.  Using a two level sorting approach, we first sort the targets by category, then rank the targets within each category on the average total observing time parameter ($tT_{\rm avg}$).  Targets that have hit warning flags (e.g., detector saturation, $R_{p} > 10\rearth$, etc.) are listed at the end of the viable ranking for each category.  The sorting and ranking process is carried out by the program element \texttt{Rank{\_}Master}, shown in flowchart form in Figure \ref{fig:Rankflow}.

\clearpage
\begin{deluxetable*}{lrr}
	\tabletypesize{}
	\tablecolumns{3}
	\setlength{\tabcolsep}{10pt}
	\renewcommand{\arraystretch}{0.9}  
	\tablecaption{Observation Time Elements\label{tab:obstime}}
	\tablehead{[-0.5ex]
		\colhead{\hspace{-263pt}Time element} & \colhead{NIRSpec BOTS} & \colhead{NIRISS SOSS} \\
		\colhead{} & \colhead{\hspace{50pt}[sec]} & \colhead{\hspace{43pt}[sec]}
	}
	\startdata
	Science observing time:                                                                                       &       &  \\
	\hspace{3ex}Transit observation (e.g., $t_{\rm dur}$\tablenotemark{a} = 2.4 hrs)                              & 8640  & 8640 \\
	\hspace{3ex}Out of transit observation,\tablenotemark{b} including ``timing tax" [$t_{\rm dur}$ + 3600] (sec)  & 12240 & 12240 \\
	&       &  \\
	Slew time\tablenotemark{c,d} [(1800 + avg. slew time*4)/5] (sec):                                             & 1108  & 1108 \\
	&       &  \\
	Instrument Overheads:\tablenotemark{e}                                                                        &       &  \\
	\hspace{3ex}SAMs: small angle maneuvers                                                                       & 0     & 0 \\
	\hspace{3ex}GS Acq: guide star acquisition(s)                                                                 & 284   & 284 \\
	\hspace{3ex}Targ Acq: target acquisition, if any                                                              & 492   & 602 \\
	\hspace{3ex}Exposure Ovhd:\,\, some instruments require initial reset                                         &       &  \\
	\hspace{6ex}NIRSpec: [0.0393 * (Science) + 26]                                                                & 846   &  \\
	\hspace{6ex}NIRISS:   \hspace{1.5ex}[0.2523 * (Science) + 15]                                                 &       & 5283 \\
	\hspace{3ex}Mech: mechanism movements, including filter wheels                                                & 110   & 52 \\
	\hspace{3ex}OSS: Onboard Script System compilation                                                            & 65    & 30 \\
	\hspace{3ex}MSA: \hspace{1ex}NIRSpec MSA configuration                                                        & 0     & 0 \\
	\hspace{3ex}IRS2: \hspace{1ex}NIRSpec IRS2 Detector Readout Mode  setup                                       & 0     & 0 \\
	\hspace{3ex}Visit Ovhd: visit cleanup activities                                                              & 110   & 62 \\
	&       &  \\
	Observatory Overheads [16\% * (Science + Slew + Instr. Ovhd)]:\tablenotemark{f}                               & 3823  & 4528 \\
	&       &  \\
	Direct Scheduling Overheads:\tablenotemark{g}                                                                 & 0     & 0 \\
	&       &  \\
	Total charged time                                                                                            &       &  \\
	\hspace{3ex}[Science + Slew + Instr. Ovhd + Observ. Ovhd + DS Ovhd]:                                          & 27718 & 32829 \\
	\enddata   
	\tablenotetext{\footnotesize\text{a}}{$t_{\rm dur}$ is the transit duration}
	\tablenotetext{\footnotesize\text{b}}{to avoid systematic bias, the ``Out of transit" observation is split into two parts, pre-ingress, and post-egress.}
	\tablenotetext{\footnotesize\text{c}}{avg. slew time per target, based on program of one 53 deg. slew (of 1800 sec) followed by four 13 deg. slews (each 935 sec) repeating 95 times over course of full 10-yr fuel limited mission.  This assumes a program of approx. 8300 hr of exoplanet transmission spectroscopy with one dedicated instrument over the full mission.  If we assume roughly 7.7 hrs for each transit observed, and $\sim$ 2.3 transits for each target we could visit roughly 475 targets/transits over the course of the full mission.  Raw slew duration vs distance (angle) are given in the \emph{JWST} User Documentation: (\url{https://jwst-docs.stsci.edu/jppom/visit-overheads-timing-model/slew-times})}
	\tablenotetext{\footnotesize\text{d}}{avg. slew angle comes from the fact that there are 41,253 sq. deg. on the sky, and if our 475 targets/transits are evenly distributed, then each target is in a box of 87 sq deg (or about 9 deg. on a side).  The diagonal of this box and the avg. angular separation of the targets is $\sim$ 13 deg.}
	\tablenotetext{\footnotesize\text{e}}{instrument overhead times are based on a set of trial observing programs using the APT.}
	\tablenotetext{\footnotesize\text{f}}{this charged time supports calibration, station keeping, and momentum management activities.}
	\tablenotetext{\footnotesize\text{g}}{for very tight timing constraints or rapid turnaround targets of opportunity.}
\end{deluxetable*}
\mbox{}
\clearpage

\section{Validating the \texttt{JET} Code}\label{sec:validating}
In order to validate that the \texttt{JET} code is performing its calculations correctly, we have considered the various parameters that are determined by \texttt{JET} and compared those to manual calculations using the methods described in Section 2, and to reference data if available.  

For the model spectra and simulated instrument spectra generated by \texttt{Exo-Transmit} and \texttt{PandExo} respectively, we demonstrate that our implementation of these programs is working correctly within the \texttt{JET} framework.

\subsection{Planet Masses}\label{sec:validplanetmasses}
We first consider the planet mass estimates.  Our method for estimating planet mass based on planet radius is described in Section \ref{sec:estplanetmasses} above.  Table \ref{tab:planetaryparams5} presents basic data for the first four targets of the Sullivan catalog, for several solar system planets, and for the well known exoplanet GJ 1214b. The planet mass ($M_{p}$) estimated by \texttt{JET} is compared to a manual calculation based on the previously described methods. The \texttt{JET}-to-Manual residuals for mass are well under 1\,\% for the example cases considered (see Table \ref{tab:planetaryparams6}). 
 
\begin{table*}[!htbp]
	\centering
	\setlength{\belowcaptionskip}{0pt}
	\caption{Basic Survey Data for Validation\label{tab:planetaryparams5}} 
	\begin{center}						                                   
		\setlength{\tabcolsep}{24pt}                                       
		\renewcommand{\arraystretch}{1.2}                                  
		\begin{tabular}{r r r r r r r}                                 			
			\hline\hline                                                   
			Target        & $R_{p}$          & $P$     &  $S$             & $R_{*}$    & $\teff$     & $\jmag$    \\ 
			              & [$\rearth$]      & [days]  &  [$\searth$]     & [$\rsun$]  & [K]         & [mag]      \\ [1.0ex]  
			\hline  
			     1        & 3.31             & 9.1     & 361.7            & 1.41       & 6531        & 7.6        \\ 
			     2        & 2.19             & 14.2    & 2.1              & 0.32       & 3426        & 11.6       \\ 
			     3        & 1.74             & 5.0     & 235.0            & 0.95       & 5546        & 8.8        \\ 
			     4        & 1.48             & 2.2     & 1240.0           & 1.13       & 5984        & 7.0        \\ 
			     Earth    & 1.00             & 365.2   & 1.0              & 1.00       & 5777        & $\cdots$   \\ 
			     Mercury  & 0.38             & 87.6    & 6.7              & 1.00       & 5777        & $\cdots$   \\ 
			     Mars     & 0.53             & 686.2   & 0.4              & 1.00       & 5777        & $\cdots$   \\
			     GJ 1214b  & 2.67             & 1.6     & 17.6             & 0.22       & 3026        & 9.8  \\ [1ex] 
			\hline
			\multicolumn{7}{l}{
				\begin{minipage}{15.5cm}~\\[1.5ex]
					Notes. --- Ref Data Sources: Targets 1 through 4, Sullivan catalog;  Solar System planets, NASA Planetary Fact Sheet (\url{https://nssdc.gsfc.nasa.gov/planetary/factsheet/planet_table_ratio.html}); GJ 1214b, NASA Exoplanet Archive (\url{https://exoplanetarchive.ipac.caltech.edu/}); Sun, \citet{Carroll2006}
				\end{minipage}
			}\\
		\end{tabular}
	\end{center}		
\end{table*}

\begin{table*}[!htbp]
	\centering
	\setlength{\belowcaptionskip}{0pt}
	\caption{Validation of \texttt{JET} Calculation of Planetary Parameters\label{tab:planetaryparams6}}
	\begin{center}						                                   
		\setlength{\tabcolsep}{24pt}                                       
		\renewcommand{\arraystretch}{1.2}                                  
	\begin{tabular}{rrrrrrr}
		\hline\hline
	Target        & $a$       & $\teq$    & $g_{s}$                & $M_{p}$      & $t_{\rm dur}$ & $nt_{\rm 10yr}$ \\ 
	              & [$\rsun$] &     [K]   & [$\text{m s}^{-2}$]    & [$\mearth$]  & [hrs]         &                 \\ [1.0ex]
	\hline                                                                      
	\multicolumn{7}{l} {\texttt{JET} computed values:} \\
	1                                                                        & 20.4  & 1215 & 9.8   & 11.0  & 4.9  & 163      \\
	2                                                                        & 17.0  & 333  & 11.1  & 5.4   & 2.2  & 137      \\
	3                                                                        & 12.3  & 1091 & 11.9  & 3.7   & 3.0  & 280      \\
	4                                                                        & 7.4   & 1653 & 12.5  & 2.8   & 2.5  & 514      \\
	Earth                                                                    & 215.1 & 279  & 9.5   & 1.0   & 13.1 & $\cdots$ \\
	Mercury                                                                  & 83.3  & 448  & 2.1   & 0.0   & 8.1  & $\cdots$ \\
	Mars                                                                     & 328.0 & 226  & 3.5   & 0.1   & 16.1 & $\cdots$ \\
	GJ 1214b                                                                 & 3.0   & 570  & 10.5  & 7.6   & 1.0  & $\cdots$ \\
	                                                                         &       &      &       &       &      &        \\[-1.0ex]
	\hline
	\multicolumn{7}{l} {Manual calculation (by methods of Section \ref{sec:analysisframework}):} \\
	1                                                                        & 20.5  & 1213 & 9.8   & 11.0  & 4.9  & 163      \\
	2                                                                        & 17.0  & 333  & 11.1  & 5.4   & 2.2  & 136      \\
	3                                                                        & 12.4  & 1089 & 11.9  & 3.7   & 3.0  & 282      \\
	4                                                                        & 7.4   & 1651 & 12.5  & 2.8   & 2.6  & 510      \\
	Earth                                                                    & 215.5 & 278  & 9.5   & 1.0   & 13.1 & $\cdots$ \\
	Mercury                                                                  & 83.4  & 447  & 2.1   & 0.0   & 8.1  & $\cdots$ \\
	Mars                                                                     & 328.7 & 225  & 3.5   & 0.1   & 16.1 & $\cdots$ \\
	GJ 1214b                                                                 & 3.0   & 570  & 10.5  & 7.6   & 1.0  & $\cdots$ \\
	                                                                         &       &      &       &       &      &        \\[-1.0ex]
	\hline
	\multicolumn{7}{l} {\texttt{JET} to Manual residual (\%):}   \\
	1                                                                        & 0.2   & -0.1 & -0.2  & 0.0   & 0.3  & 0.0      \\
	2                                                                        & 0.2   & -0.1 & -0.2  & 0.0   & 0.3  & -0.7     \\
	3                                                                        & 0.2   & -0.1 & -0.2  & 0.0   & 0.3  & 0.7      \\
	4                                                                        & 0.2   & -0.1 & -0.2  & 0.0   & 0.3  & -0.7     \\
	Earth                                                                    & 0.2   & -0.1 & -0.2  & 0.0   & 0.3  & $\cdots$ \\
	Mercury                                                                  & 0.2   & -0.1 & -0.2  & 0.0   & 0.3  & $\cdots$ \\
	Mars                                                                     & 0.2   & -0.1 & -0.2  & 0.0   & 0.3  & $\cdots$ \\
	GJ 1214b                                                                 & 0.2   & -0.1 & 0.1   & 0.3   & 0.3  & $\cdots$ \\
	                                                                         &       &      &       &       &      &        \\[-1.0ex]
	\hline                                                                         
	\multicolumn{7}{l} {Reference parameters:} \\
	Earth                                                                    & 215.5 & 279  & 9.8   & 1.0   & 13.0 & $\cdots$ \\
	Mercury                                                                  & 83.4  & 449  & 3.7   & 0.1   & 8.1  & $\cdots$ \\
	Mars                                                                     & 327.6 & 218  & 3.7   & 0.1   & 16.0 & $\cdots$ \\
	GJ 1214b                                                                 & 3.0   & 604  & 8.8   & 6.6   & 0.9  & $\cdots$ \\
	                                                                         &       &      &       &       &      &        \\[-1.0ex]
	\hline                                                                         
	\multicolumn{7}{l} {\texttt{JET} to Reference residual (\%):}  \\
	Earth                                                                    & -0.2  & -0.2 & -2.9  & -2.9  & 0.6  & $\cdots$ \\
	Mercury                                                                  & -0.1  & -0.3 & -44.4 & -44.9 & -0.5 & $\cdots$ \\
	Mars                                                                     & 0.1   & 3.5  & -6.1  & -6.7  & 0.4  & $\cdots$ \\
	GJ 1214b                                                                 & 0.3   & -5.5 & 18.8  & 16.1  & 9.9  & $\cdots$ \\[2.0ex]
	\hline
	\multicolumn{7}{l}{
		\begin{minipage}{15.5cm}~\\[1.5ex]
			Notes. --- Ref Data Sources: Targets 1 through 4, Sullivan catalog;  Solar System planets, NASA Planetary Fact Sheet (\url{https://nssdc.gsfc.nasa.gov/planetary/factsheet/planet_table_ratio.html}); NASA About Transits (\url{https://www.nasa.gov/kepler/overview/abouttransits}); GJ 1214b, NASA Exoplanet Archive (\url{https://exoplanetarchive.ipac.caltech.edu/}); Sun, \citep{Carroll2006}
		\end{minipage}
	}\\
	\end{tabular}
\end{center}
\end{table*}

For the solar system planets and for GJ 1214b, we compared the \texttt{JET} mass estimates to reference masses.  For the solar system planets the \texttt{JET} mass estimates are within 7\,\% of reference, except for Mercury, which with its very small radius, has a mass that is not well approximated by the methods of \citet{0004-637X-834-1-17}.  Mercury has a radius of roughly $0.38\rearth$, whereas the smallest planet radius in the Sullivan catalog is roughly $0.70\rearth$. The mass estimate for GJ 1214b is off by about 16\,\%. This is still within the acceptable bounds of uncertainty of our estimation approach.

\subsection{System Parameters, and Transits Observable}
In a similar manner, we validate the \texttt{JET} calculations for the other planetary system parameters ($a$, $\teq$, $g_{s}$, $t_{\rm dur}$, and $ntr_{\rm 10yr}$).  Again, Table \ref{tab:planetaryparams6} presents the results of this validation study.  For all of the targets the \texttt{JET}-to-Manual calculation residuals are well under 1\,\%.  The very small errors are due to minor roundoff and constant differences between the manual calculations and those carried out by \texttt{JET}.  

For the \texttt{JET}-to-Reference comparison we see some bigger differences. The simple \texttt{JET} estimate of $\teq$ is slightly off for Mars and GJ 1214b, probably due to our assumption of zero albedo and other small differences.  The reference $\teq$ for Earth is shown for zero albedo.  Finally, the \texttt{JET}-to-Reference residual for the GJ 1214b transit duration is off by roughly 10\,\%.  This could be due to many factors, including differences in the assumptions for eccentricity (value $< 0.14$), inclination (values between 87.63 and 90 deg.), host star radius (values between 0.204 and 0.228$\rsun$), or planet radius (values between 2.19 and 3.05$\rearth$).\footnote{\url{https://exoplanetarchive.ipac.caltech.edu/}}

\subsection{Model Spectra}
To validate the calculation of model transmission spectra we have taken a benchmark spectrum generated by \texttt{Exo-Transmit} for the exoplanet GJ 1214b (obtained by private communication with the \texttt{Exo-Transmit} code's lead author, E. Kempton, and described in Section 2.2 of \citet{1538-3873-129-974-044402}) and compared it with a model spectrum produced by the \texttt{JET} \texttt{ExoT{\_}Master} subprogram. The benchmark \texttt{Exo-Transmit} spectral data is based on planetary system parameters from the \texttt{exoplanet.eu} database.\footnote{\url{http://exoplanet.eu/catalog/gj_1214_b/}}  For the comparison \texttt{JET} spectrum we have used \texttt{exoplanet.eu} for the basic catalog-like input data and relied on the \texttt{JET} computed values for the other system parameters. 

\begin{figure*}[!htbp]
	\centering
	\includegraphics[width=0.95\textwidth]{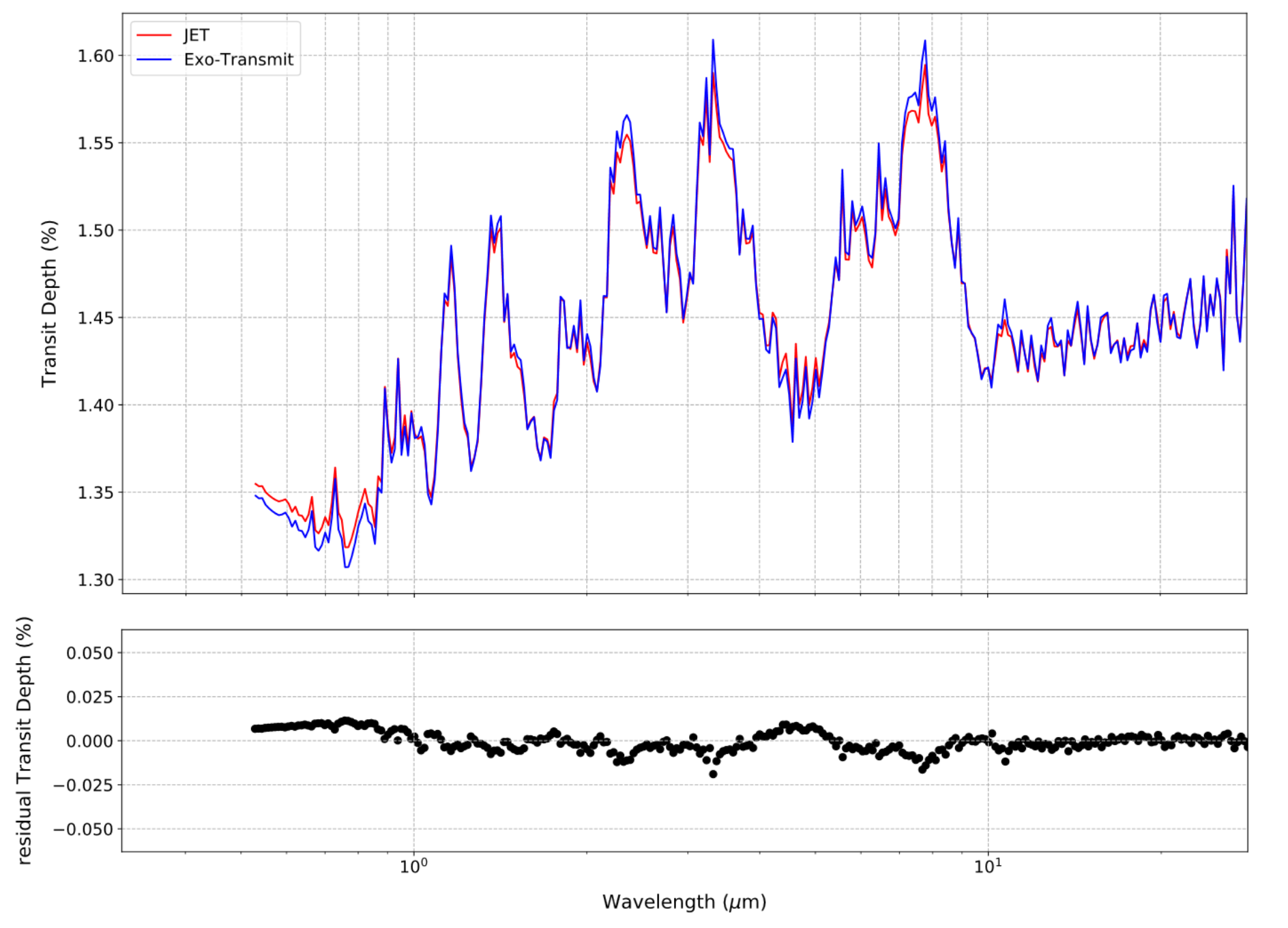}
	\includegraphics[width=0.95\textwidth]{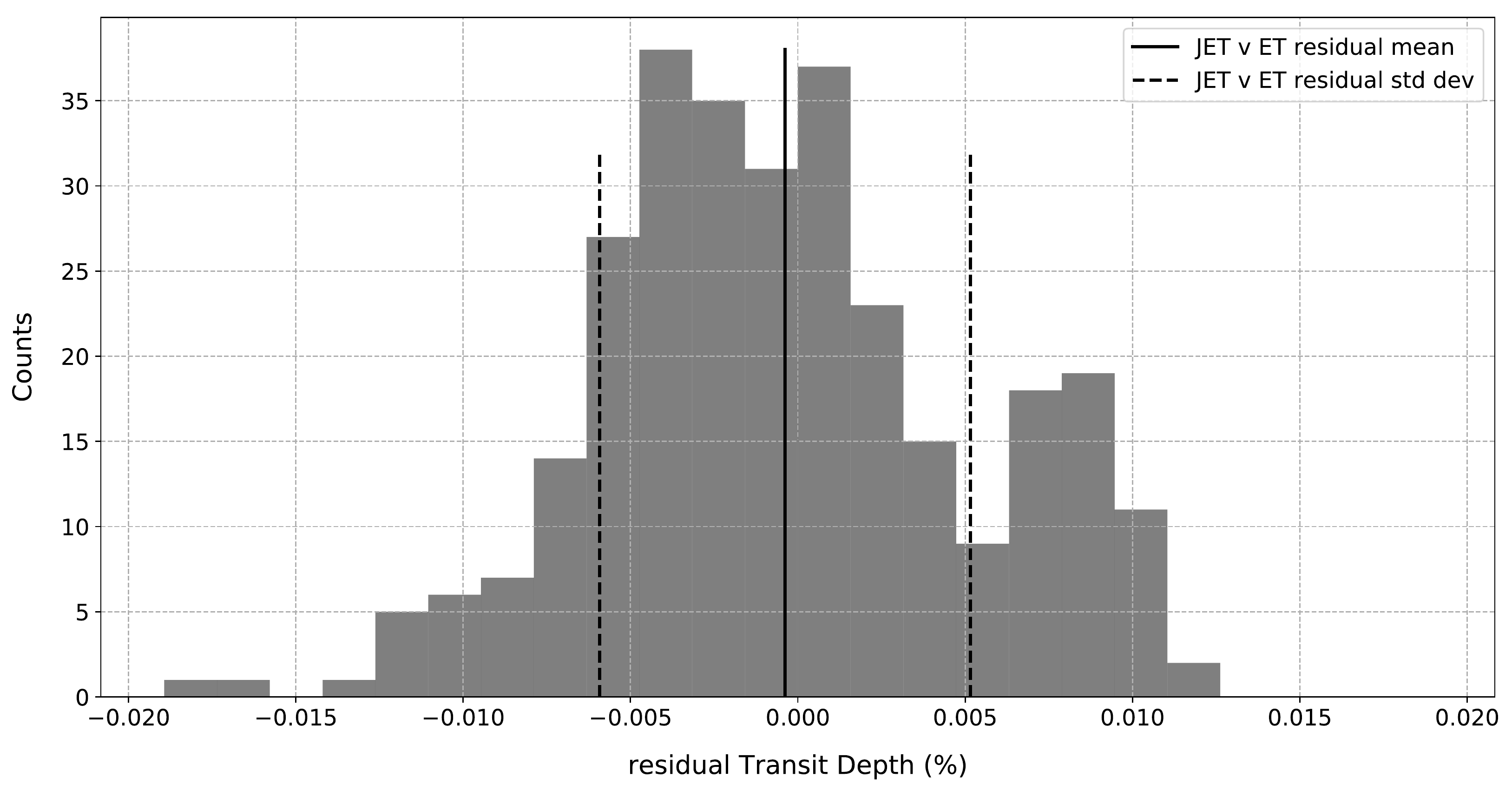}
	\setlength{\abovecaptionskip}{10pt}
	\caption{\emph{Top:} Comparison of a test transmission spectrum for GJ 1214b generated by our \texttt{JET} code with a benchmark spectrum generated by \texttt{Exo-Transmit} (E. M.-R. Kempton, personal communication, September 14, 2018) shows excellent agreement across a broad spectral range. \emph{Center:} Residuals are shown at the same scale. \emph{Bottom:} The \texttt{JET}-to-\texttt{Exo-Transmit} benchmark residuals are small and show no systematic bias. The mean of the residuals is approximately - 4\,ppm, with a standard deviation of approximately 55\,ppm.\label{fig:JETvETresidual}}
\end{figure*}

The two spectra shown in the upper panel of Figure \ref{fig:JETvETresidual} are a very close match.  The small differences seen are likely due to differences in the re-binning and smoothing algorithms used, and to minor differences in input parameters for the planet-host star system.  The center panel shows the \texttt{JET}-to-\texttt{Exo-Transmit} residuals at the same scale. The lower panel shows the distribution of the \texttt{JET}-to-\texttt{Exo-Transmit} residuals. The mean of the residuals is approximately - 4\,ppm, with a standard deviation of approximately 55\,ppm.

As shown in Figure \ref{fig:JETvETresidual}, \texttt{JET} has properly implemented the underlying \texttt{Exo-Transmit} code and is producing consistent transmission spectra.

\subsection{Simulated Spectra}
Our approach to validation of our calculation of simulated instrument spectra is essentially the same as that of the previous section.  We have taken a benchmark spectrum generated by \texttt{PandExo} for the exoplanet GJ 1214b (obtained by private communication with the \texttt{PandExo} code's lead author, Natasha Batalha, and generally described in \citet{1538-3873-129-976-064501}) and compared it with a simulated spectrum produced by the \texttt{JET} \texttt{Pdxo{\_}Master} subprogram.

\begin{figure*}[!htbp]
	\centering
	\includegraphics[width=0.95\textwidth]{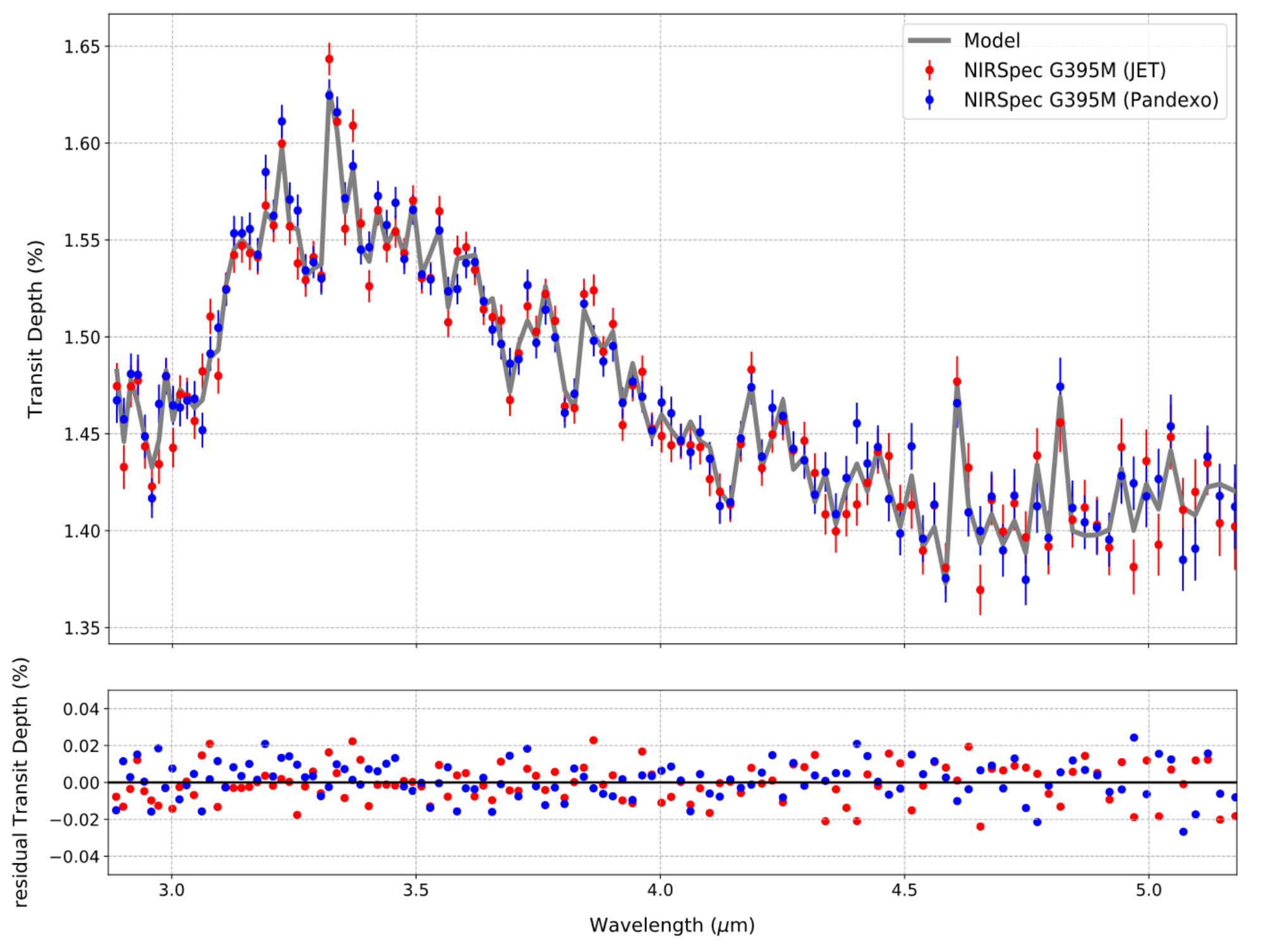}
	\includegraphics[width=0.95\textwidth]{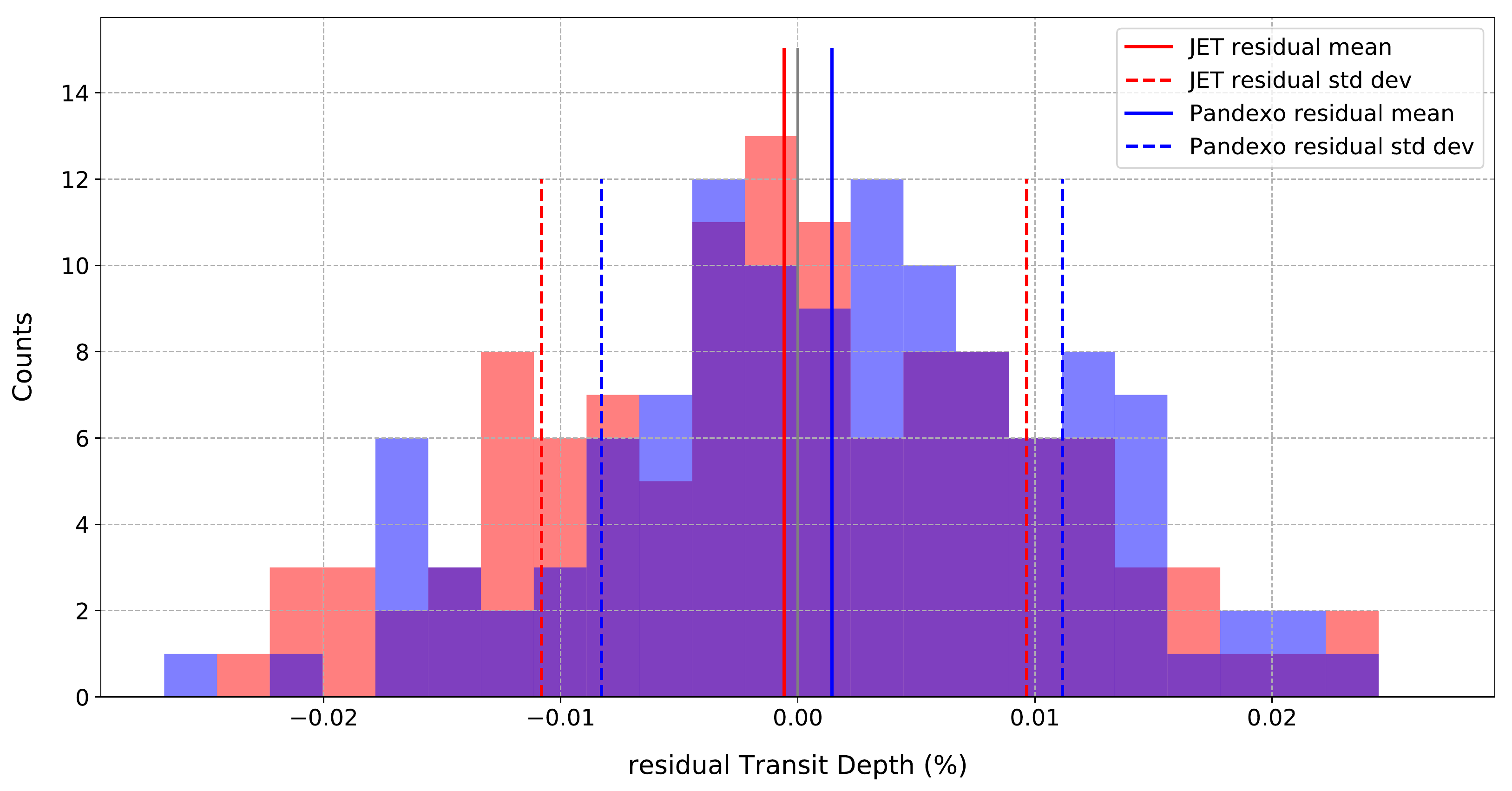}
	\setlength{\abovecaptionskip}{5pt}
	\caption{\emph{Top:} Comparison of a simulated spectrum for GJ 1214b generated by our \texttt{JET} code with a benchmark spectrum generated by \texttt{PandExo} (Natasha Batalha, personal communication, October 22, 2018) shows good agreement across the bandpass. \emph{Center:} Residuals are shown at the same scale. \emph{Bottom:} Here we overlay the \texttt{JET}-to-model residual histogram on the \texttt{PandExo}-to-model residual histogram. There is good agreement. The difference in the mean of the residuals is approximately 20\,ppm, with the difference in standard deviation of the residuals approximately 5\,ppm.\label{fig:JETvPdxoresidual}}
\end{figure*}

The two spectra (in this case the data points representing the instrument spectra for a single transit with random noise) shown in the upper panel of Figure \ref{fig:JETvPdxoresidual} are a close match.  The small differences seen are likely due to the fact that each case is a different random noise instance and would only agree in overall distribution. The center panel shows the \texttt{JET}-to-model residuals and the \texttt{PandExo}-to-model residuals. The lower panel shows the distributions of the two residual plots (\texttt{JET}-to-model, and \texttt{PandExo}-to-model) overlaid.  They are a close match.  The difference in the mean of the residuals is approximately 20\,ppm, with the difference in standard deviation of the residuals approximately 5\,ppm.

Figure \ref{fig:JETvPdxoresidual} demonstrates that \texttt{JET} has properly implemented the underlying \texttt{PandExo} code and is producing consistent simulated instrument spectra.

\section{Results/Discussion}\label{sec:resultsdiscuss}
Now that we have described our analysis framework and validated the calculations in the \texttt{JET} code, we present the results of our full baseline run.  In addition, we present limited case studies of the effects of variations of some of the driving parameters of our analysis, including the noise floor, the atmospheric equation of state, the detection threshold, and the instrument/mode selected.  

\subsection{Baseline (Full Survey)}
The input parameters for the baseline run of the 1984 target Sullivan catalog were shown previously in Table \ref{tab:inputparams}.  Excerpts of the summary output table for the baseline run are included in Table \ref{tab:BaselineRunCat}. A machine-readable version of the full output table is available, however it does not include the summary statistic sections. 

We now have a list of (simulated \emph{TESS} catalog) targets that is categorized and ranked within each category on the average total observation time to detect an atmosphere ($tT_{\rm avg}$).  

The overall statistics for the baseline run are presented in Table \ref{tab:baselinestats1}.  Out of the total 1984 targets in the catalog, we show a full unambiguous detection (a detection with less than 50 transits observed for both low and high metallicity atmosphere cases) of 1070 targets.  Some of these detections may well be unrealistic, given the large number of transit observations (and very long observation times) required.  The Table shows how certain factors reduce the overall detection numbers.  For example, 36 targets are eliminated because their host star is too bright, leading to a detector saturation condition for this instrument/mode.  Of course the most significant screening factor is the lack of a strong detection ($dBIC < 10$) of the high metallicity atmospheres for 828 targets.  A number of these targets may show a detection of the low metallicity atmosphere, but given that the high metallicity atmospheres are undetectable, we consider these to be ambiguous detections and eliminate them from further consideration.

We can see that there are 49 targets with $R_{p} > 10\rearth$, properly falling into Category 7 (sub-Jovians), but that are outside the range of our analysis.  Also, we note (somewhat surprisingly) that there is only one target where the number of transits needed for detection is greater than the number observable within the 10-year fuel life of the spacecraft.

In Table \ref{tab:baselinestats2} we slice the data a different way.  Columns three and four show numbers of targets that meet certain cutoff values for $nt_{\rm hi}$, and $tT_{\rm avg}$.  Column five shows accumulated observing hours by each category.  We see that for $nt_{\rm hi} < 10$ and $tT_{\rm avg} < 35$ hours, we detect roughly 600 targets with approximately 11,000 hours of observing time.  Likewise, the next three columns show that for $nt_{\rm hi} < 6$ and $tT_{\rm avg} < 20$ hours, we detect roughly 300 targets with approximately 4,000 hours of observing time.

From Table \ref{tab:baselinestats2}, we can also see, for example, that for Category 4 (Cool Neptunes) there are 37 targets where a high metallicity atmosphere (the difficult case) can be detected with less than 6 transit observations, and 45 targets where an average atmosphere can be detected in under 20 hours of observation for each target.

In Table \ref{tab:besttargets} we show the best targets for an 8300 hr (10\,\% of the anticipated 10-year mission total observing time) program by category for (1) the full Sullivan catalog, (2) the \texttt{JET} baseline ranking with all viable targets for Cat 1 (Cool Terrestrials), and the remaining hours applied evenly for other categories, (3) the \texttt{JET} baseline ranking with all viable targets for Cat 1, and a fixed number of targets for other categories, and (4) planets with actual atmospheric characterization by transmission spectroscopy to date, based on the NASA Exoplanet Archive (and associated references).

\begin{figure*}[!htbp]
	\centering
	\begin{picture}(800, 275)
	\includegraphics[width=1.02\textwidth]{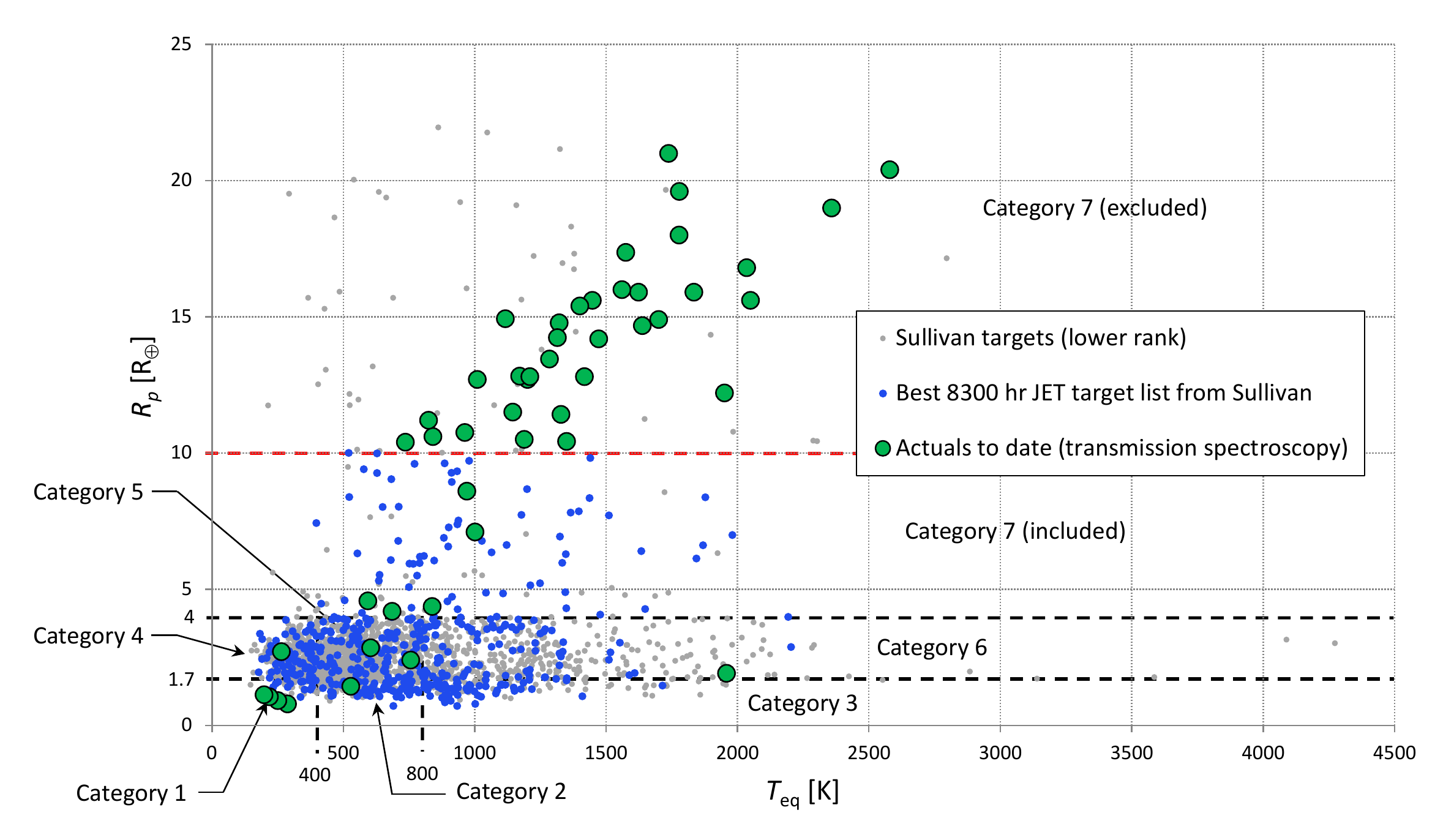}
	\end{picture}
	\begin{picture}(497, 305)
	\includegraphics[width=1.01\textwidth]{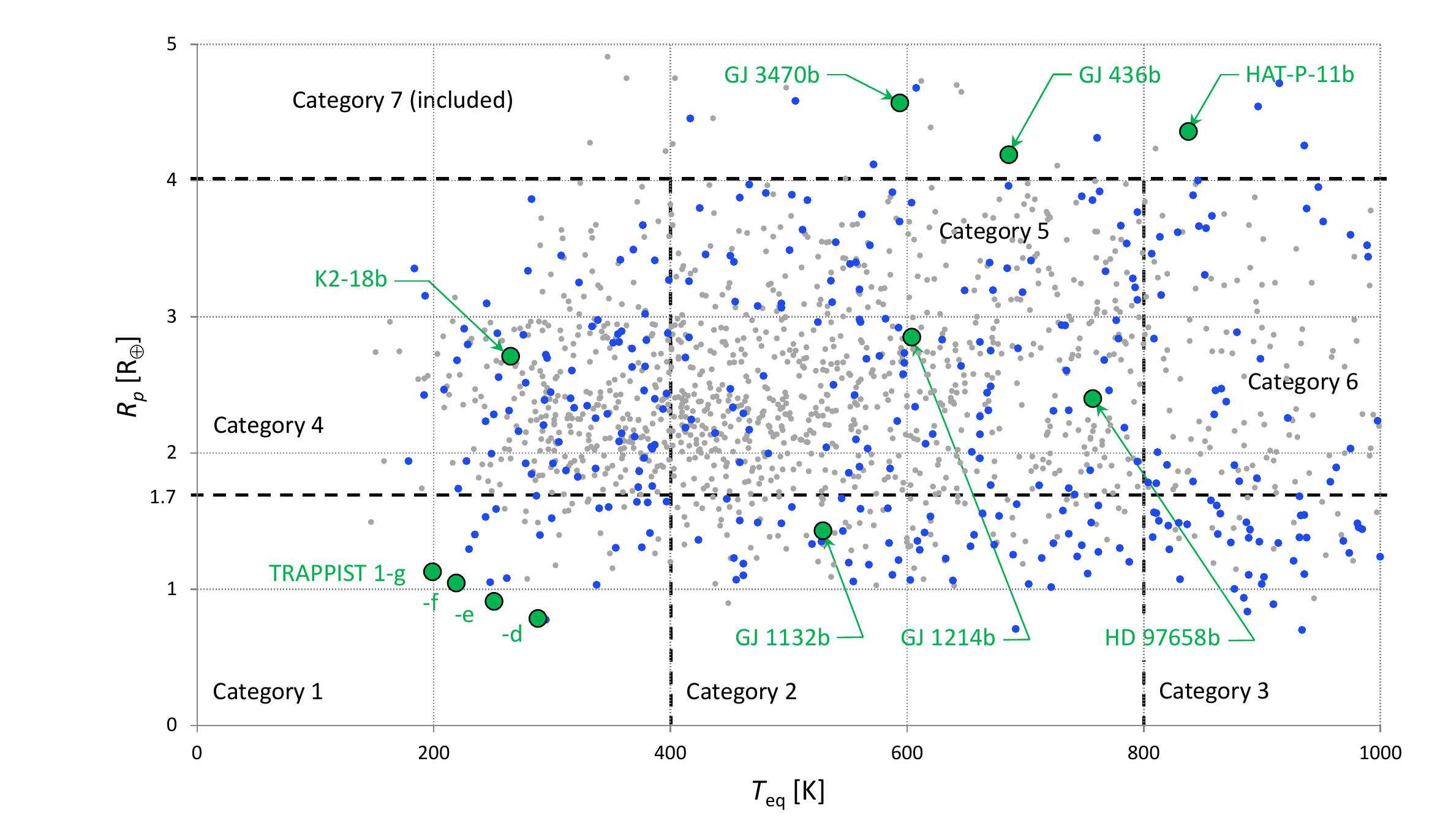}
	\end{picture}
	\setlength{\abovecaptionskip}{-10pt}
	\caption{Target planets plotted by category on a radius ($R_{p}$) vs equilibrium temperature ($\teq$) grid. \emph{Top:} Full range of Sullivan catalog (1984 targets), the best 8300 hr \texttt{JET} target list from Sullivan (462 targets), and actual transmission spectroscopy targets to date (48 targets). \emph{Bottom:} Same data focused on the densest region of the plot, with planet names labeled.\label{fig:demog}}
\end{figure*}

We show the target planets plotted by category on a radius ($R_{p}$) vs equilibrium temperature ($\teq$) grid in Figure \ref{fig:demog}. In this figure we combine a plot of the full Sullivan catalog (1984 targets), with the best 8300 hr \texttt{JET} target list from the Sullivan catalog (462 targets; as shown in Col 4 of Table \ref{tab:besttargets}), and with planets with actual atmospheric characterization (by transmission spectroscopy) to date (48 targets).\footnote{\url{https://exoplanetarchive.ipac.caltech.edu/cgi-bin/TblView/nph-tblView?app=ExoTbls\&config=transitspec}} There is some ambiguity regarding which planetary atmospheres have actually been ``characterized."  Many of the spectroscopic observations obtained so far are of very low resolution and in some cases only multi-band photometry has been employed.

As we mentioned in Section \ref{sec:Targetranking}, a recent paper by \citet{Kempton2018a} ranked the Sullivan targets for transmission spectroscopy using an analytical metric (the transmission spectroscopy metric, TSM).  In Table \ref{tab:BaselineRunCompare} we show the top habitable zone transmission spectroscopy targets taken from Table 2 of their paper, with the \texttt{JET} target ranking side-by-side.  There is general agreement, but differences are evident.

The TSM is an algebraic function of the planetary radius, mass, equilibrium temperature, and host star radius, that is proportional to the expected S/N of the observation.  In our analysis we use a total observation time figure of merit ($tT_{\rm avg}$) for ranking that is based on a statistical analysis of the fit of simulated spectra to model spectra for each target.  Our ranking figure of merit is also related to the expected S/N of the observation, but in a more indirect and complex way.  The imperfect agreement between the two rankings is reasonable given the \mbox{significantly different analytical approaches taken}.

\begin{figure*}[!htbp]
	\centering
	\hspace{0.10em}\includegraphics[width=0.488\textwidth]{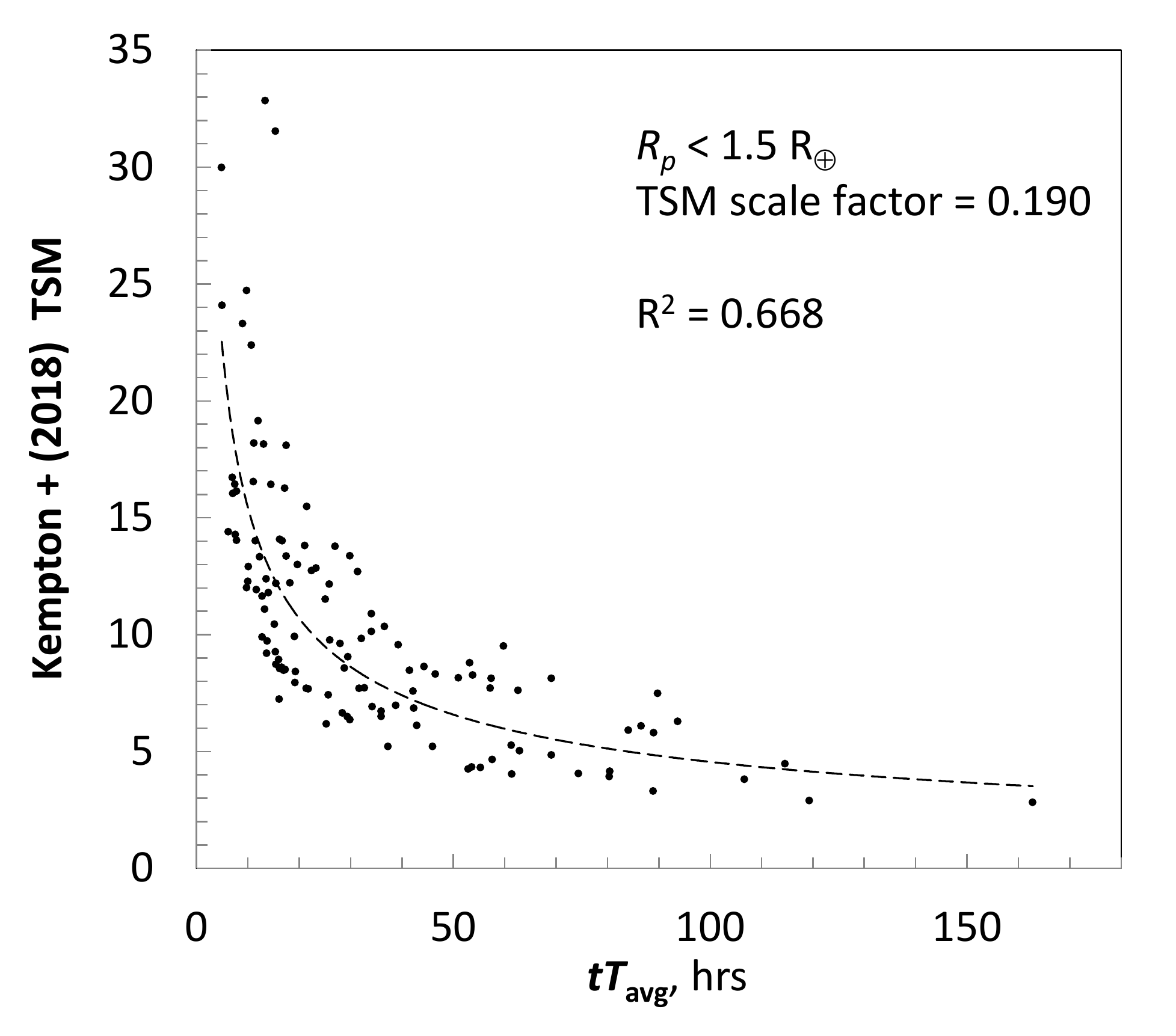}
	\includegraphics[width=0.501\textwidth]{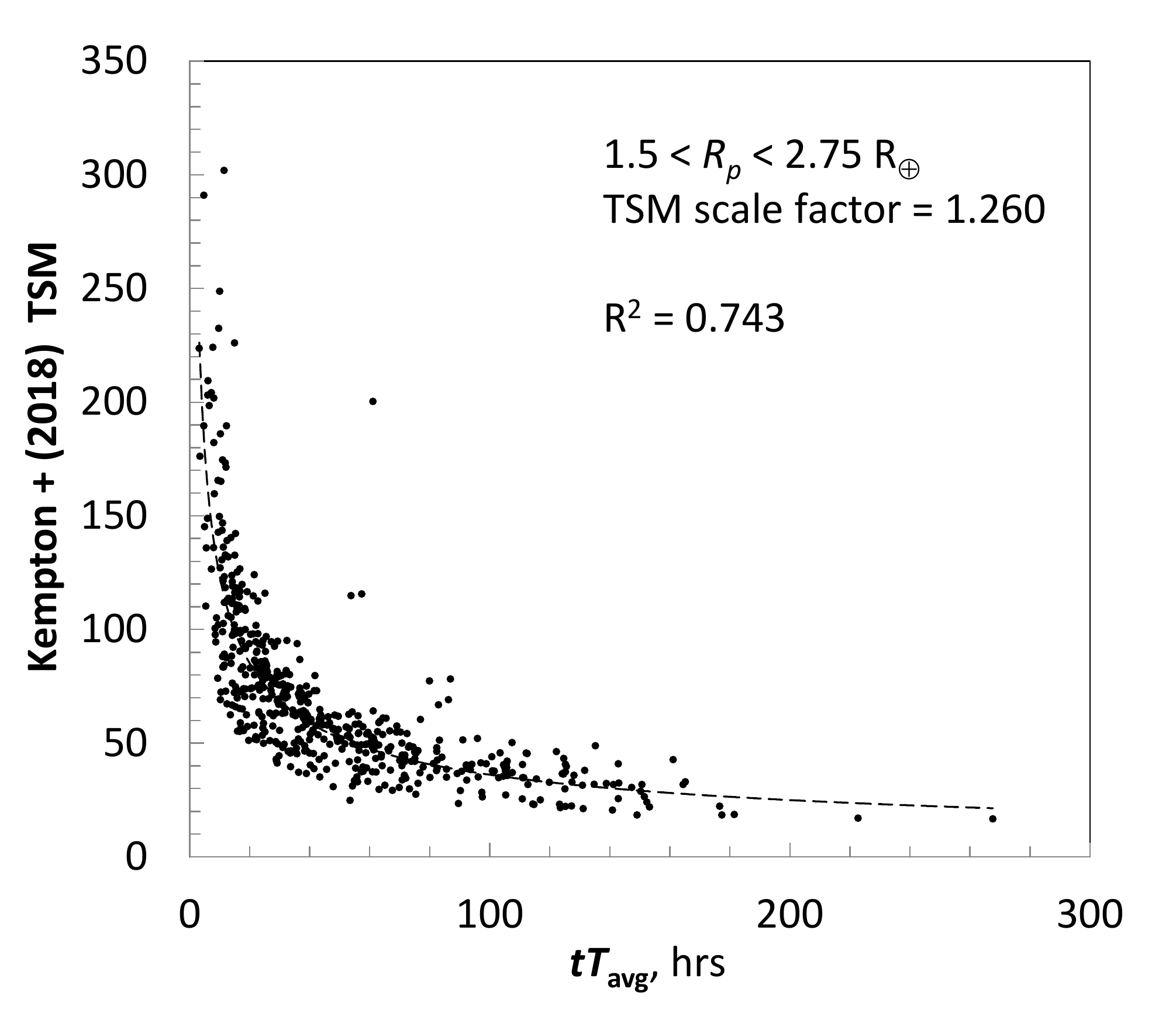}\\
	\flushleft
	\includegraphics[width=0.494\textwidth]{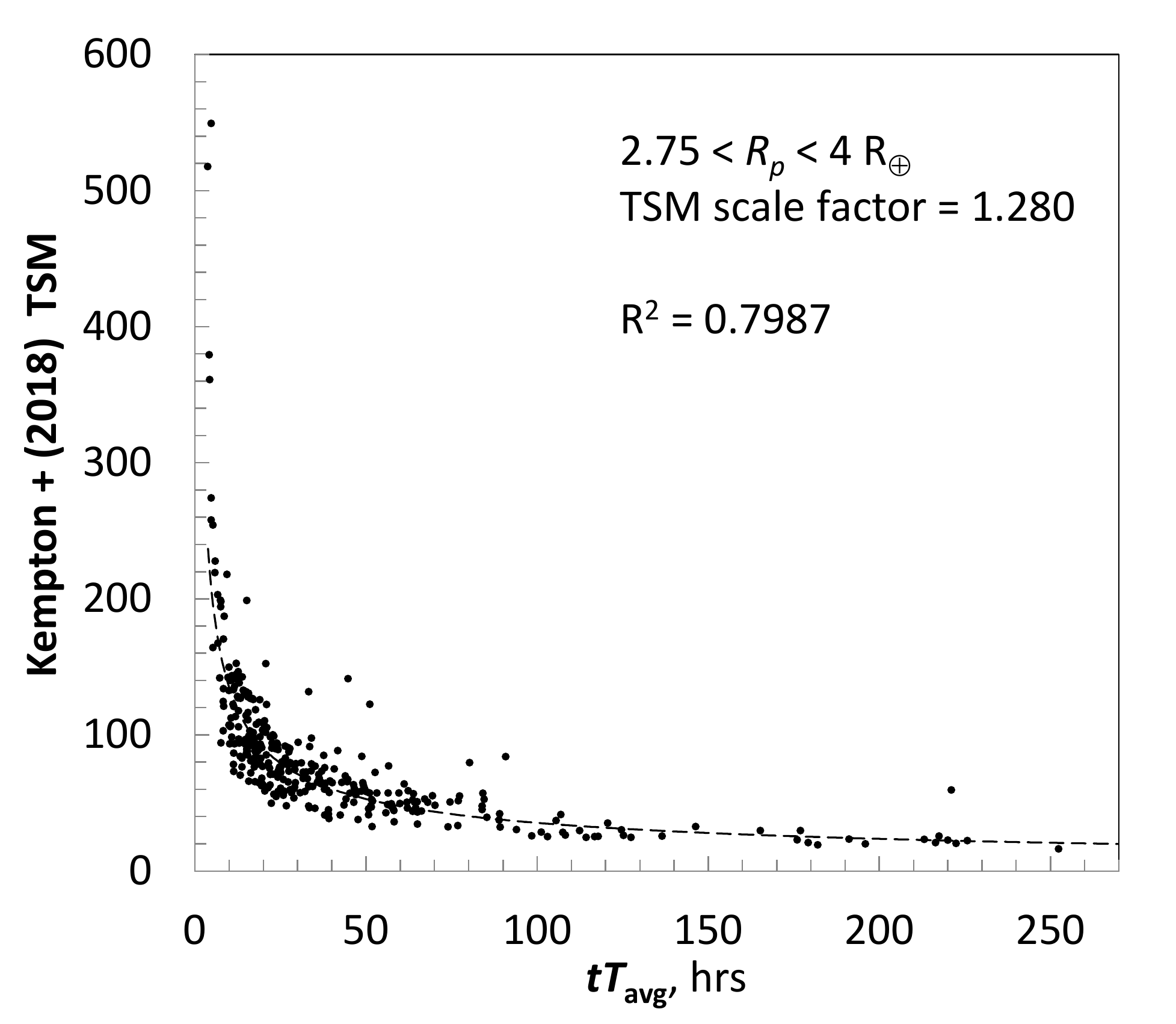}
	\includegraphics[width=0.487\textwidth]{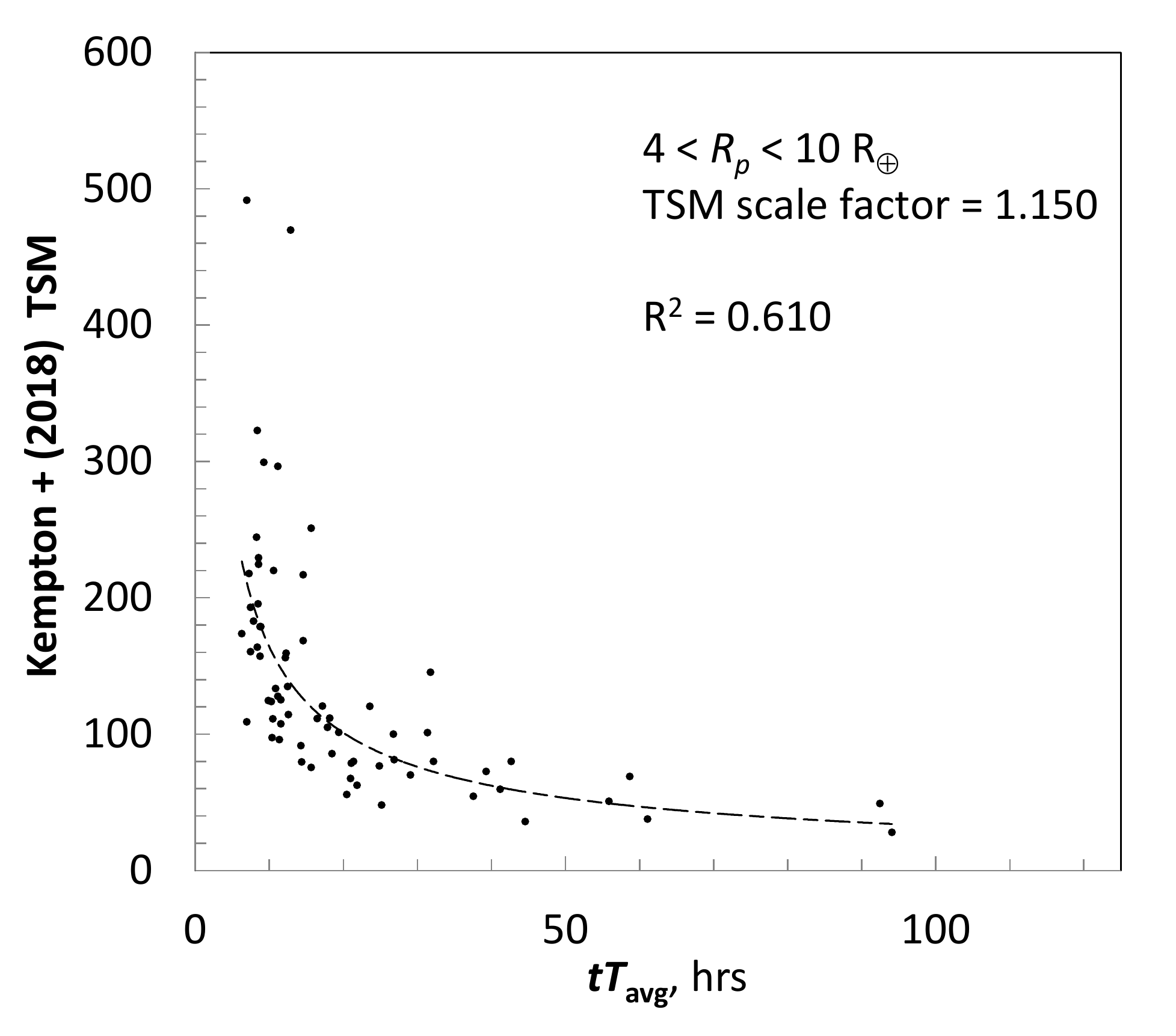}
	\caption{Comparison of the \texttt{JET} ranking metric, $tT_{\rm avg}$, to the transmission spectroscopy metric, TSM, from Equation 1 of \citet{Kempton2018a} for the full baseline \texttt{JET} run (1070 targets with unambiguous detection). Here we are using the Kempton demographic categories and scale factors for their ``statistical" sample. The black dashed lines mark a power law least squares fit to the data, with $\rm R^2$ values shown in each case. The Kempton ranking tracks the \texttt{JET} ranking reasonably well in all cases, and particularly well for the $2.75 < R_{p} < 4\rearth$ demographic. \label{fig:Kempt_JET1}} 
\end{figure*}

We compare our ranking metric, $tT_{\rm avg}$, to the TSM in Figure \ref{fig:Kempt_JET1}.  This is for their ``statistical" sample. We see a good general correspondence between the two approaches, but also some dispersion between these two ranking schemes. 

\begin{figure}[!htbp]
	\centering
	\hspace{0.10em}\includegraphics[width=0.495\textwidth]{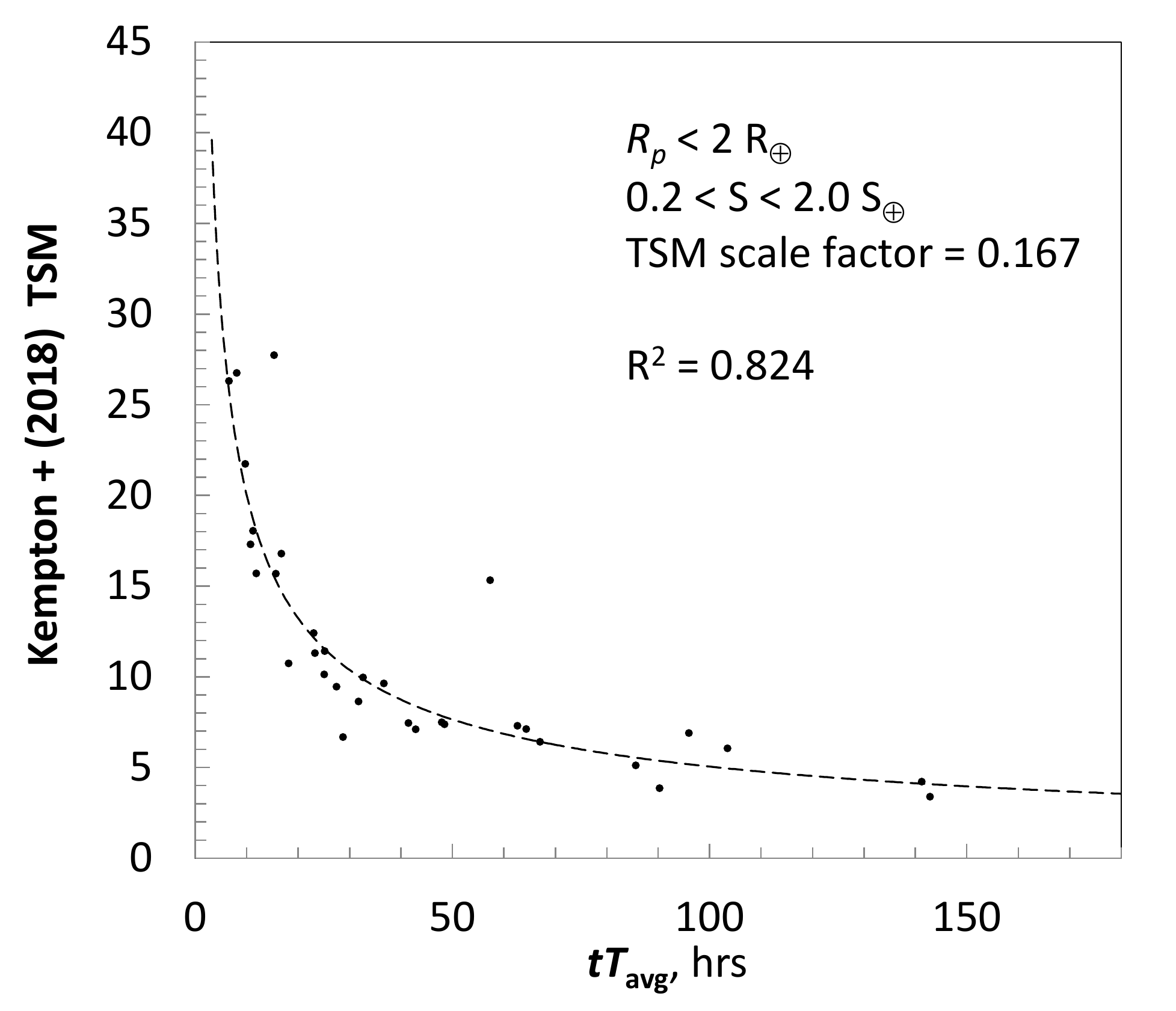}
	\caption{Same as the previous figure, but now showing only the ``small temperate" target sample. The Kempton ranking tracks the \texttt{JET} ranking very well for this demographic.\label{fig:Kempt_JET5}}  
\end{figure}

In Figure \ref{fig:Kempt_JET5} we show the comparison for their ``small temperate" target sample.  Again, we see good correspondence with only a small amount of dispersion.

\subsection{Noise Floor Variation}\label{sec:noisefloorvar}
\begin{figure}[!htbp]
	\centering
	\centerline{\includegraphics[width=0.50\textwidth]{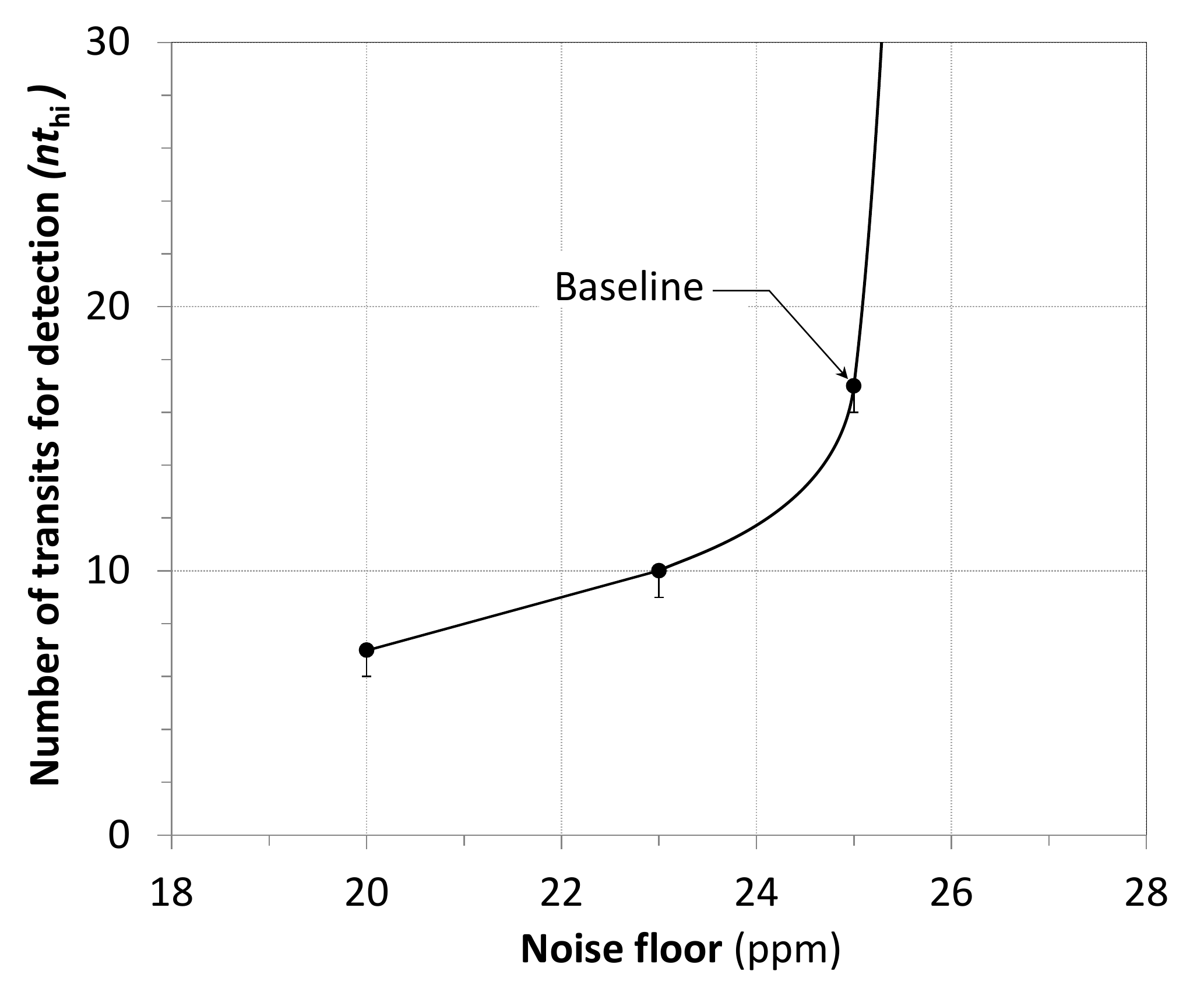}}	
	\setlength{\abovecaptionskip}{0pt}
	\caption{Effect of noise floor variation on number of transits needed for detection, for Sullivan Target 1292, with a high metallicity atmosphere (1000xSolar), clouds at 100\,mbar, a detection threshold ($dBIC$) of 10, and using NIRSpec G395M. There is no detection for a noise floor above approximately 25\,ppm. The number of transits are rounded to the next highest whole number of transits.\label{fig:nfloorvar}}
\end{figure}

Variation of the instrument noise floor has a powerful effect on the number of transits needed for detection for the more difficult targets (those with weak spectral features and bright host stars).  Figure \ref{fig:nfloorvar} presents the results of our noise floor variation study for Sullivan Target 1292.  As discussed in Section \ref{sec:gensimspectra}, the noise floor for NIRSpec G395M is estimated to be approximately 25\,ppm.  Of course, the true value of the noise floor will not be known until \emph{JWST} is launched and has completed detailed commissioning procedures. 

Our baseline run for this target with a noise floor of 25\,ppm shows that we need 17 transits to detect the high-metallicity atmosphere.  This falls to 7 transits for a noise floor of 20\,ppm.  Remarkably, if the noise floor turns out to be only 1\,ppm higher than the baseline (or 26\,ppm) there will be no detection of the atmosphere no matter how many transits are observed.  Clearly the target ranking can be affected significantly by minor variations in the estimate of the instrument noise floor.

It should be pointed out that in the figure the error-bars do not indicate uncertainty (which could be significant), but are meant to indicate that the number of transits has been rounded up to the next highest whole transit.

\clearpage
\begin{deluxetable*}{rrrrrrrrrrrrrrrrrr}
	\tablecolumns{18}
	\tabletypesize{\footnotesize}
	\setlength{\tabcolsep}{4.8pt}
	\renewcommand{\arraystretch}{1.3}
	\tablecaption{Excerpt of Baseline Run (top 10 targets for each category shown)\label{tab:BaselineRunCat}}
	\tablehead{
		\colhead{Cat} &
		\colhead{Rank} &
		\colhead{Targ} &
		\colhead{$R_{p}$} &
		\colhead{$P$} &
		\colhead{$S$} &
		\colhead{$R_{*}$} &
		\colhead{$\teff$} &
		\colhead{$\jmag$} &
		\colhead{$a$} &
		\colhead{$\teq$} &
		\colhead{$g_{s}$} &
		\colhead{$M_{p}$} &
		\colhead{$t_{\rm dur}$} &
		\colhead{$nt_{\rm 10yr}$} &
		\colhead{$nt_{\rm lo}$} &
		\colhead{$nt_{\rm hi}$} &
		\colhead{$tT_{\rm avg}$} \\ 
		&
		&
		&
		\colhead{[$\rearth$]} &
		\colhead{[days]} &
		\colhead{[$\searth$]} &
		\colhead{[$\rsun$]} &
		\colhead{[K]} &
		\colhead{[mag]} &
		\colhead{[$\rsun$]} &
		\colhead{[K]} &
		\colhead{[$\text{m s}^{-2}$]} &
		\colhead{[$\mearth$]} &
		\colhead{[hrs]} &
		&
		&
		&
		\colhead{[hrs]}
	}
	\startdata
	1 & 1  & 922 & 1.40 & 4.39 & 1.17 & 0.12 & 2729 & 11.3 & 5.1 & 290 & 12.72 & 2.53 & 0.8 & 300 & 1 & 4 & 9.8 \\
	1 & 2  & 98 & 1.64 & 5.61 & 4.14 & 0.23 & 3325 & 8.6 & 8.0 & 397 & 12.11 & 3.33 & 1.3 & 173 & 1 & 3 & 10.1 \\
	1 & 3  & 105 & 1.68 & 7.58 & 1.12 & 0.14 & 3130 & 11.3 & 8.5 & 287 & 12.02 & 3.48 & 1.1 & 256 & 1 & 4 & 11.3 \\
	1 & 4  & 139 & 1.60 & 8.11 & 2.45 & 0.21 & 3170 & 8.1 & 8.8 & 348 & 12.20 & 3.20 & 1.6 & 145 & 1 & 3 & 11.5 \\
	1 & 5  & 1248 & 1.31 & 7.07 & 3.31 & 0.24 & 3351 & 7.9 & 9.5 & 376 & 12.98 & 2.26 & 1.4 & 275 & 1 & 4 & 13.4 \\
	1 & 6  & 204 & 0.77 & 14.32 & 1.27 & 0.24 & 3345 & 7.9 & 15.1 & 295 & 6.35 & 0.39 & 1.8 & 73 & 1 & 4 & 15.4 \\
	1 & 7  & 1247 & 1.41 & 4.24 & 3.59 & 0.16 & 3224 & 11.3 & 5.7 & 383 & 12.68 & 2.57 & 1.0 & 458 & 1 & 7 & 17.2 \\
	1 & 8  & 1919 & 1.40 & 9.41 & 0.51 & 0.12 & 2838 & 12.2 & 8.7 & 235 & 12.71 & 2.54 & 1.1 & 173 & 1 & 7 & 18.2 \\
	1 & 9  & 1878 & 1.08 & 9.96 & 0.78 & 0.14 & 3128 & 11.4 & 10.2 & 262 & 10.74 & 1.28 & 1.1 & 97 & 1 & 9 & 23.3 \\
	1 & 10 & 45 & 1.05 & 15.96 & 0.63 & 0.16 & 3228 & 10.9 & 14.0 & 248 & 10.27 & 1.15 & 1.5 & 70 & 1 & 8 & 25.1 \\
	\\
	\hline
	2 & 1  & 1622 & 0.71 & 0.77 & 38.12 & 0.17 & 3231 & 11.4 & 1.9 & 692 & 5.48 & 0.28 & 0.6 & 2531 & 1 & 2 & 4.9 \\
	2 & 2  & 1856 & 1.55 & 0.55 & 32.21 & 0.13 & 3055 & 13.2 & 1.3 & 664 & 12.31 & 3.03 & 0.4 & 2077 & 1 & 3 & 6.0 \\
	2 & 3  & 17 & 1.11 & 0.79 & 53.39 & 0.22 & 3304 & 11.1 & 2.2 & 753 & 11.26 & 1.42 & 0.7 & 1446 & 1 & 3 & 7.0 \\
	2 & 4  & 857 & 1.50 & 2.27 & 7.37 & 0.16 & 3141 & 11.3 & 3.8 & 459 & 12.44 & 2.86 & 0.8 & 886 & 1 & 4 & 9.6 \\
	2 & 5  & 1255 & 1.22 & 0.59 & 26.65 & 0.14 & 2871 & 12.0 & 1.4 & 633 & 13.11 & 2.00 & 0.5 & 3223 & 1 & 6 & 10.7 \\
	2 & 6  & 882 & 1.06 & 1.61 & 15.80 & 0.19 & 3242 & 11.1 & 3.2 & 555 & 10.36 & 1.18 & 0.8 & 1187 & 1 & 5 & 11.2 \\
	2 & 7  & 893 & 1.07 & 2.56 & 7.17 & 0.17 & 3155 & 11.0 & 4.1 & 456 & 10.55 & 1.23 & 0.9 & 773 & 1 & 5 & 12.0 \\
	2 & 8  & 805 & 1.67 & 1.82 & 14.63 & 0.13 & 3758 & 12.5 & 3.1 & 545 & 12.06 & 3.41 & 0.7 & 571 & 1 & 6 & 12.2 \\
	2 & 9  & 1533 & 1.30 & 0.56 & 61.35 & 0.18 & 3231 & 12.9 & 1.6 & 780 & 13.00 & 2.24 & 0.5 & 3935 & 2 & 6 & 12.8 \\
	2 & 10 & 1013 & 1.59 & 0.88 & 19.35 & 0.14 & 3006 & 13.0 & 1.9 & 584 & 12.22 & 3.16 & 0.6 & 2176 & 1 & 7 & 13.0 \\
	\\
	\hline
	3 & 1 & 1745 & 1.09 & 0.51 & 110.20 & 0.24 & 3337 & 10.4 & 1.7 & 902 & 10.88 & 1.31 & 0.6 & 3781 & 1 & 2 & 5.0 \\
	3 & 2 & 281 & 0.70 & 0.85 & 126.50 & 0.39 & 3557 & 10.1 & 2.8 & 934 & 5.39 & 0.27 & 0.9 & 2697 & 1 & 2 & 6.2 \\
	3 & 3 & 1421 & 1.40 & 0.63 & 92.11 & 0.27 & 3359 & 10.9 & 2.1 & 863 & 12.70 & 2.55 & 0.7 & 3226 & 1 & 3 & 7.1 \\
	3 & 4 & 1057 & 1.04 & 0.70 & 108.80 & 0.31 & 3425 & 10.2 & 2.3 & 900 & 10.07 & 1.10 & 0.8 & 1385 & 1 & 3 & 7.5 \\
	3 & 5 & 378 & 1.04 & 0.59 & 184.10 & 0.37 & 3511 & 10.1 & 2.1 & 1026 & 10.13 & 1.12 & 0.8 & 3290 & 1 & 3 & 7.6 \\
	3 & 6 & 146 & 0.89 & 0.76 & 113.70 & 0.34 & 3476 & 10.3 & 2.5 & 910 & 7.89 & 0.64 & 0.8 & 2550 & 1 & 3 & 7.8 \\
	3 & 7 & 419 & 1.09 & 0.64 & 174.90 & 0.38 & 3532 & 9.6 & 2.3 & 1013 & 10.94 & 1.33 & 0.8 & 2971 & 1 & 3 & 7.8 \\
	3 & 8 & 13 & 1.56 & 0.62 & 71.16 & 0.22 & 3283 & 12.1 & 1.8 & 809 & 12.29 & 3.06 & 0.6 & 3127 & 1 & 4 & 8.5 \\
	3 & 9 & 1855 & 1.35 & 1.12 & 108.00 & 0.42 & 3640 & 8.1 & 3.5 & 898 & 12.86 & 2.38 & 1.1 & 1148 & 1 & 3 & 9.0 \\
	3 & 10 & 917 & 1.65 & 1.35 & 89.71 & 0.43 & 3670 & 9.9 & 4.0 & 857 & 12.09 & 3.36 & 1.2 & 720 & 1 & 3 & 9.4 \\
	\\
	\hline
	4 & 1 & 8 & 2.81 & 5.62 & 2.56 & 0.17 & 3228 & 11.1 & 7.0 & 352 & 10.30 & 8.28 & 1.2 & 173 & 1 & 1 & 4.8 \\
	4 & 2 & 921 & 2.68 & 9.98 & 0.39 & 0.12 & 2729 & 11.3 & 8.9 & 220 & 10.45 & 7.65 & 1.2 & 132 & 1 & 1 & 4.8 \\
	4 & 3 & 1810 & 2.83 & 3.84 & 3.46 & 0.16 & 2844 & 9.8 & 4.6 & 380 & 10.28 & 8.38 & 1.2 & 253 & 1 & 1 & 4.8 \\
	4 & 4 & 1444 & 2.89 & 6.90 & 2.75 & 0.21 & 3284 & 10.3 & 8.8 & 359 & 10.21 & 8.71 & 1.4 & 297 & 1 & 1 & 5.3 \\
	4 & 5 & 1864 & 2.56 & 4.82 & 0.71 & 0.11 & 2529 & 12.5 & 5.3 & 255 & 10.60 & 7.06 & 0.9 & 201 & 1 & 2 & 6.2 \\
	4 & 6 & 1804 & 1.94 & 17.79 & 0.45 & 0.16 & 2844 & 9.8 & 12.7 & 228 & 11.52 & 4.41 & 1.9 & 55 & 1 & 1 & 6.6 \\
	4 & 7 & 1296 & 1.92 & 11.56 & 1.00 & 0.17 & 3236 & 10.0 & 11.4 & 278 & 11.55 & 4.35 & 1.4 & 168 & 1 & 2 & 8.1 \\
	4 & 8 & 822 & 2.93 & 9.05 & 2.07 & 0.22 & 3300 & 10.6 & 10.8 & 334 & 10.17 & 8.89 & 1.6 & 134 & 1 & 2 & 8.6 \\
	4 & 9 & 783 & 2.29 & 6.20 & 2.42 & 0.19 & 3225 & 10.1 & 8.3 & 347 & 10.96 & 5.84 & 1.2 & 319 & 1 & 3 & 9.7 \\
	4 & 10 & 113 & 3.02 & 4.62 & 3.41 & 0.18 & 3006 & 12.5 & 5.8 & 379 & 10.07 & 9.37 & 1.3 & 413 & 1 & 3 & 10.0 \\
	\\
	\hline
	5 & 1 & 182 & 2.27 & 0.57 & 31.90 & 0.14 & 2871 & 12.8 & 1.3 & 662 & 10.99 & 5.76 & 0.5 & 1778 & 1 & 1 & 3.2 \\
	5 & 2 & 1930 & 3.46 & 1.81 & 5.66 & 0.13 & 2877 & 11.6 & 3.0 & 430 & 9.67 & 11.79 & 0.8 & 613 & 1 & 1 & 3.8 \\
	5 & 3 & 934 & 3.64 & 2.16 & 11.39 & 0.19 & 3253 & 11.2 & 3.9 & 512 & 9.52 & 12.85 & 1.0 & 883 & 1 & 1 & 4.2 \\
	5 & 4 & 1260 & 3.91 & 1.91 & 19.91 & 0.25 & 3342 & 10.7 & 4.0 & 588 & 9.32 & 14.55 & 1.0 & 1015 & 1 & 1 & 4.4 \\
	5 & 5 & 901 & 3.80 & 2.81 & 5.40 & 0.16 & 2793 & 12.3 & 3.6 & 425 & 9.40 & 13.83 & 1.2 & 346 & 1 & 1 & 4.8 \\
	5 & 6 & 1130 & 2.01 & 0.66 & 30.50 & 0.15 & 2999 & 12.6 & 1.6 & 655 & 11.40 & 4.68 & 0.5 & 1747 & 1 & 2 & 4.8 \\
	5 & 7 & 60 & 3.19 & 2.46 & 34.13 & 0.41 & 3587 & 10.0 & 5.8 & 673 & 9.91 & 10.30 & 1.4 & 677 & 1 & 1 & 5.3 \\
	5 & 8 & 183 & 1.94 & 0.82 & 66.23 & 0.26 & 3363 & 11.2 & 2.4 & 795 & 11.52 & 4.40 & 0.7 & 1179 & 1 & 2 & 5.6 \\
	5 & 9 & 1050 & 3.91 & 1.94 & 8.92 & 0.15 & 3206 & 13.5 & 3.3 & 481 & 9.32 & 14.52 & 0.8 & 999 & 1 & 2 & 5.9 \\
	5 & 10 & 33 & 3.85 & 1.85 & 11.79 & 0.17 & 3225 & 13.0 & 3.4 & 516 & 9.36 & 14.18 & 0.9 & 1048 & 1 & 2 & 6.0 \\
	\\
	\hline
	6 & 1 & 1956 & 2.38 & 0.60 & 95.05 & 0.25 & 3336 & 11.6 & 1.9 & 870 & 10.83 & 6.24 & 0.7 & 3258 & 1 & 1 & 3.5 \\
	6 & 2 & 993 & 1.91 & 0.52 & 98.50 & 0.23 & 3300 & 11.9 & 1.6 & 877 & 11.57 & 4.30 & 0.6 & 1885 & 1 & 2 & 5.0 \\
	6 & 3 & 1903 & 2.03 & 0.52 & 150.20 & 0.30 & 3407 & 11.7 & 1.8 & 975 & 11.36 & 4.78 & 0.7 & 1871 & 1 & 2 & 5.4 \\
	6 & 4 & 945 & 3.87 & 3.22 & 307.70 & 0.82 & 5383 & 8.3 & 8.8 & 1167 & 9.34 & 14.30 & 2.4 & 539 & 1 & 1 & 7.7 \\
	6 & 5 & 115 & 1.76 & 1.19 & 431.20 & 0.64 & 4583 & 7.5 & 4.2 & 1269 & 11.87 & 3.74 & 1.4 & 813 & 1 & 2 & 8.1 \\
	6 & 6 & 231 & 2.88 & 1.87 & 99.23 & 0.53 & 3918 & 10.2 & 5.3 & 879 & 10.21 & 8.67 & 1.5 & 612 & 1 & 2 & 8.3 \\
	6 & 7 & 1682 & 3.98 & 4.27 & 230.10 & 0.86 & 5470 & 8.3 & 10.9 & 1085 & 9.27 & 14.98 & 2.7 & 447 & 1 & 1 & 8.4 \\
	6 & 8 & 866 & 3.59 & 2.13 & 72.91 & 0.51 & 3824 & 10.4 & 5.6 & 814 & 9.56 & 12.54 & 1.6 & 521 & 1 & 2 & 8.5 \\
	6 & 9 & 1516 & 1.77 & 0.67 & 307.80 & 0.49 & 3804 & 9.8 & 2.6 & 1167 & 11.83 & 3.79 & 1.0 & 5152 & 1 & 3 & 8.6 \\
	6 & 10 & 207 & 1.79 & 0.64 & 83.65 & 0.25 & 3330 & 12.2 & 1.9 & 842 & 11.80 & 3.85 & 0.7 & 2965 & 1 & 4 & 8.8 \\
	\\
	\hline
	7 & 1 & 79 & 7.51 & 2.07 & 129.10 & 0.60 & 4147 & 11.1 & 5.8 & 939 & 7.65 & 43.99 & 1.8 & 955 & 1 & 1 & 6.3 \\
	7 & 2 & 104 & 5.07 & 3.86 & 52.77 & 0.58 & 4122 & 10.7 & 8.8 & 751 & 8.61 & 22.58 & 2.1 & 252 & 1 & 1 & 7.0 \\
	7 & 3 & 860 & 7.70 & 1.66 & 868.70 & 0.86 & 5470 & 8.4 & 5.6 & 1512 & 7.59 & 45.94 & 2.1 & 629 & 1 & 1 & 7.0 \\
	7 & 4 & 905 & 6.61 & 2.82 & 264.50 & 0.74 & 5152 & 9.7 & 7.8 & 1123 & 7.95 & 35.45 & 2.2 & 626 & 1 & 1 & 7.3 \\
	7 & 5 & 352 & 8.93 & 3.56 & 116.10 & 0.67 & 4673 & 10.8 & 8.8 & 914 & 7.26 & 59.02 & 2.3 & 546 & 1 & 1 & 7.5 \\
	7 & 6 & 1898 & 6.77 & 4.63 & 42.24 & 0.59 & 4134 & 10.5 & 9.9 & 710 & 7.89 & 36.87 & 2.3 & 412 & 1 & 1 & 7.5 \\
	7 & 7 & 1784 & 6.88 & 4.46 & 101.40 & 0.69 & 4835 & 10.0 & 10.3 & 884 & 7.86 & 37.91 & 2.5 & 226 & 1 & 1 & 7.9 \\
	7 & 8 & 168 & 9.26 & 6.53 & 26.21 & 0.58 & 4144 & 10.3 & 12.5 & 630 & 7.18 & 62.77 & 2.6 & 149 & 1 & 1 & 8.3 \\
	7 & 9 & 608 & 9.71 & 4.15 & 153.20 & 0.79 & 5000 & 10.9 & 10.2 & 980 & 7.08 & 68.04 & 2.7 & 485 & 1 & 1 & 8.4 \\
	7 & 10 & 1213 & 8.34 & 2.69 & 710.60 & 1.00 & 5834 & 8.7 & 8.3 & 1438 & 7.41 & 52.55 & 2.7 & 839 & 1 & 1 & 8.4 \\
	\enddata 
	\tablecomments{ This table is available in its entirety in machine-readable form.}   
\end{deluxetable*}
\mbox{}
\clearpage

\begin{deluxetable*}{clccccccc}
	\tabletypesize{}
	\tablecolumns{9}
	\setlength{\tabcolsep}{5pt}
	\tablecaption{Baseline Run Statistics 1\label{tab:baselinestats1}}
	\tablehead{[-2.0ex]
		\colhead{} &
		\colhead{} &
		\colhead{} &
		\multicolumn{5}{c}{---------------------------------non-detection --------------------------------} &
		\colhead{} \\[-0.5ex]
		\colhead{} &
		\colhead{} &
		\colhead{} &
		\colhead{} &
		\colhead{} &
		\colhead{$dBIC < $\,10} &
		\colhead{$dBIC < $\,10} &
		\colhead{nt req'd} &
		\colhead{Full} \\
		\colhead{} &
		\colhead{} &
		\colhead{} &
		\colhead{$\jmag < $\,lim} &
		\colhead{$R_{p} > $\,10$\rearth$} &
		\colhead{for $nt_{\rm lo} < $\,50} &
		\colhead{for $nt_{\rm hi} < $\,50} &
		\colhead{$> nt_{\rm 10yr}$} &
		\colhead{Detection} \\
		\colhead{Cat} &
		\colhead{Description} &
		\colhead{targets} &
		\colhead{targets} &
		\colhead{targets} &
		\colhead{targets} &
		\colhead{targets} &
		\colhead{targets} &
		\colhead{targets}
	}
	\startdata
	1 & Cool Terrestrials & 41   & 1  & $\cdots$ & $\cdots$ & 21  & $\cdots$ & 19   \\
	2 & Warm Terrestrials & 134  & 2  & $\cdots$ & 2        & 53  & $\cdots$ & 79   \\
	3 & Hot Terrestrials  & 119  & 6  & $\cdots$ & 5        & 43  & $\cdots$ & 70   \\
	4 & Cool Neptunes     & 371  & 1  & $\cdots$ & $\cdots$ & 128 & 1        & 241  \\
	5 & Warm Neptunes     & 768  & 7  & $\cdots$ & 17       & 343 & $\cdots$ & 418  \\
	6 & Hot Neptunes      & 400  & 12 & $\cdots$ & 27       & 214 & $\cdots$ & 174  \\
	7 & sub-Jovians       & 151  & 7  & 49       & 1        & 26  & $\cdots$ & 69   \\
	\hline
	Total &               & 1984 & 36 & 49       & 52       & 828 & 1        & 1070 \\
	\enddata 
	\tablecomments{Some overlap in non-detection conditions (``Total" line will not sum across row).}               
\end{deluxetable*}

\clearpage
\begin{deluxetable*}{clcccccc}
	\tabletypesize{}
	\tablecolumns{8}
	\setlength{\tabcolsep}{6pt}
	\tablecaption{Baseline Run Statistics 2\label{tab:baselinestats2}}
	\tablehead{
		\colhead{} &
		\colhead{} &
		\colhead{$nt_{\rm hi} < $\,10} &
		\colhead{$tT_{\rm avg} < $\,35 hrs} &
		\colhead{accum hrs} &
		\colhead{for $nt_{\rm hi} < $\,6} &
		\colhead{for $tT_{\rm avg} < $\,20 hrs} &
		\colhead{accum hrs} \\
		\colhead{Cat} &
		\colhead{Description} &
		\colhead{targets} &
		\colhead{targets} &
		\colhead{by Cat\tablenotemark{a}} &
		\colhead{targets} &
		\colhead{targets} &
		\colhead{by Cat\tablenotemark{b}} 
	}
	\startdata
	1 & Cool Terrestrials &   10  &   9   &   167   &   4  &    4  &    50 \\
	2 & Warm Terrestrials &   29  &   37  &   768   &   4  &   20  &   295 \\
	3 & Hot Terrestrials  &   50  &   51  &   898   &  21  &   35  &   468 \\
	4 & Cool Neptunes     &  107  &  101  &  2165   &  37  &   45  &   611 \\
	5 & Warm Neptunes     &  211  &  207  &  4077   &  96  &  108  &  1443 \\
	6 & Hot Neptunes      &  131  &  108  &  2165   &  75  &   57  &   809 \\
	7 & sub-Jovians       &   59  &   52  &   791   &  51  &   39  &   458 \\
	\hline
	Total  &              &  597  &  565  &  11032  &  288 &  308  &  4133
	\enddata   
	\tablecomments{Columns 3, 4, 6, and 7 are numbers of targets; columns 5, and 8 are accumulated observing hrs.}
	\tablenotetext{\footnotesize\text{a}}{based on $tT_{avg} < 35$ hrs column}
	\tablenotetext{\footnotesize\text{b}}{based on $tT_{avg} < 20$ hrs column}         
\end{deluxetable*}
\mbox{}
\clearpage

\begin{deluxetable*}{clcccc}
	\tablecolumns{6}
	\setlength{\tabcolsep}{12pt}
	\renewcommand{\arraystretch}{1.2}
	\tablecaption{Best Targets for 8300 hr \emph{JWST} Transmission Spectroscopy Program\label{tab:besttargets}}
	\tablehead{
		\colhead{} &
		\colhead{} &
		\colhead{} &
		\colhead{\texttt{JET} targets} &
		\colhead{\texttt{JET} targets} &
		\colhead{Actual atms} \\
		\colhead{Cat} &
		\colhead{Description} &
		\colhead{Sullivan (full)\tablenotemark{a}} &
		\colhead{even hrs\tablenotemark{b}} &
		\colhead{even targets\tablenotemark{c}} &
		\colhead{characterized to date\tablenotemark{d}}
	}
	\startdata
	1 & Cool Terrestrials & 41   & 19  & 19  & 4        \\
	2 & Warm Terrestrials & 134  & 52  & 67  & 1        \\
	3 & Hot Terrestrials  & 119  & 62  & 67  & $\cdots$ \\
	4 & Cool Neptunes     & 371  & 74  & 67  & 1        \\
	5 & Warm Neptunes     & 768  & 105 & 67  & 2        \\
	6 & Hot Neptunes      & 400  & 83  & 67  & 1        \\
	7 & sub-Jovians       & 151  & 67  & 67  & 39       \\
	\hline
	Total &               & 1984 & 462 & 421 & 48       \\
	\enddata 
	\tablenotetext{\footnotesize\text{a}}{From the simulated \citet{Sullivan2015} \emph{TESS} Catalog}             
	\tablenotetext{\footnotesize\text{b}}{\texttt{JET} baseline ranking with all viable targets for Cat 1 (Cool Terrestrials), and remaining hours applied evenly for other categories (8300 hr program total)}  
	\tablenotetext{\footnotesize\text{c}}{\texttt{JET} baseline ranking with all viable targets for Cat 1, and a fixed number of targets for other categories (8300 hr program total)}
	\tablenotetext{\footnotesize\text{d}}{Planets with actual atmospheric characterization by transmission spectroscopy to date, based on the NASA Exoplanet Archive}               
\end{deluxetable*}
 
\clearpage
\begin{deluxetable*}{rrrrrrrrrrcr}
	\tablecolumns{12}
	\setlength{\tabcolsep}{8.7pt}
	\renewcommand{\arraystretch}{1.2}
	\tablecaption{Top Habitable Zone Transmission Spectroscopy \mbox{Targets Ranked by \citet{Kempton2018a} and by \texttt{JET}}\label{tab:BaselineRunCompare}}
	\tablehead{
		\multicolumn{8}{c}{\citet{Kempton2018a} ranking} & & \multicolumn{3}{c}{\texttt{JET} ranking}\\
		\cline{1-8}
		\cline{10-12}
		\colhead{Targ\tablenotemark{a}} &
		\colhead{TSM\tablenotemark{b}} &
		\colhead{$M_{p}$} &
		\colhead{$R_{p}$} &
		\colhead{$M_{*}$} &
		\colhead{$R_{*}$} &
		\colhead{$S$} &
		\colhead{$\jmag$} &
		\colhead{} &
		\colhead{Cat} &
		\colhead{Rank} &
		\colhead{$tT_{\rm avg}$} \\ 
		&
		&
		\colhead{[$\mearth$]} &
		\colhead{[$\rearth$]} &
		\colhead{[$\msun$]} &
		\colhead{[$\rsun$]} &
		\colhead{[$\searth$]} &
		\colhead{[mag]} &
		\colhead{} &
		\colhead{} &
		\colhead{[in Cat]} &
		\colhead{[hrs]}
	}
	\startdata
	204 & 27.9 & 0.39 & 0.77 & 0.2 & 0.24 & 1.27 & 7.91 &  & 1 & 6 & 15.4 \\
	1296 & 26.8 & 4.35 & 1.92 & 0.12 & 0.17 & 1 & 10 &  & 4 & 7 & 8.1 \\
	1804 & 26.5 & 4.42 & 1.94 & 0.07 & 0.16 & 0.45 & 9.78 &  & 4 & 6 & 6.6 \\
	1308 & 23.2 & 2.83 & 1.49 & 0.12 & 0.25 & 1.15 & 7.97 &  & 1 & $no\,\,detection$\tablenotemark{c} & $\cdots$ \\
	922 & 21.6 & 2.53 & 1.39 & 0.06 & 0.12 & 1.17 & 11.26 &  & 1 & 1 & 9.8 \\
	405 & 19.4 & 3.11 & 1.58 & 0.26 & 0.38 & 1.61 & 6.85 &  & 1 & $saturated$\tablenotemark{d} & $\cdots$ \\
	105 & 17.9 & 3.48 & 1.68 & 0.1 & 0.14 & 1.12 & 11.27 &  & 1 & 3 & 11.3 \\
	48 & 17.3 & 4.64 & 1.99 & 0.12 & 0.16 & 0.64 & 11.1 &  & 4 & 12 & 10.8 \\
	1244 & 16.8 & 3.99 & 1.82 & 0.11 & 0.16 & 1.79 & 11.34 &  & 4 & 42 & 16.8 \\
	991 & 15.8 & 3.67 & 1.74 & 0.1 & 0.16 & 0.39 & 10.53 &  & 4 & 19 & 11.9 \\ [-2ex]
	\enddata
	\tablenotetext{\footnotesize\text{a}}{Target planet numbers from the simulated \citet{Sullivan2015} \emph{TESS} Catalog}             
	\tablenotetext{\footnotesize\text{b}}{Transmission Spectroscopy Metric, for Scale factor = 0.167, calculated for small temperate sample.}
	\tablenotetext{\footnotesize\text{c}}{No detection, $dBIC < 10$ for 50 transits with hi-metal atm.}
	\tablenotetext{\footnotesize\text{d}}{Host star magnitude exceeds brightness limit of detector.}
	   
\end{deluxetable*}

\subsection{Atmospheric Equation of State Variation}
\begin{figure}[!htbp]
	\centering
	\centerline{\includegraphics[width=0.49\textwidth]{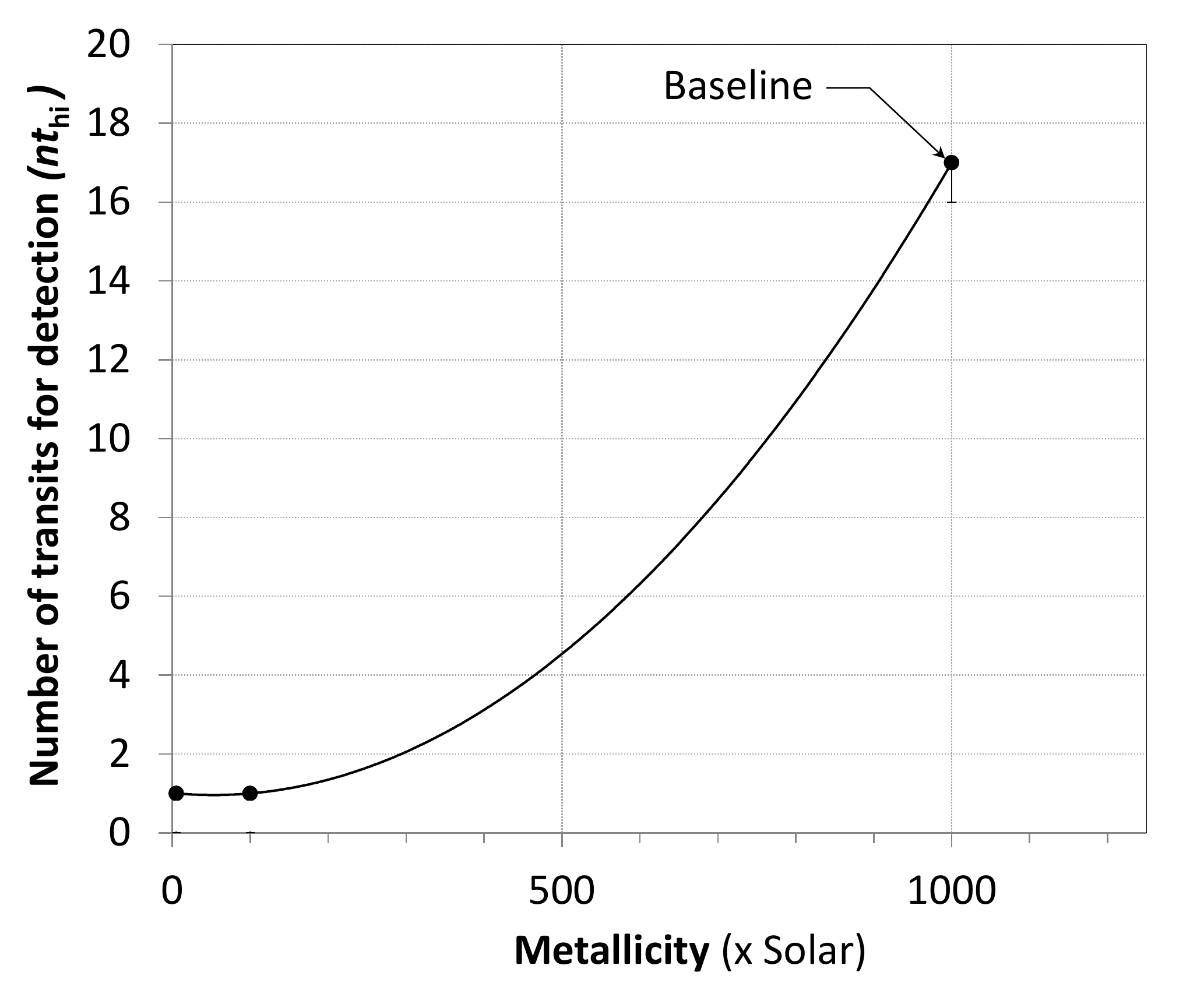}}
	\flushleft		
	\setlength{\abovecaptionskip}{0pt}
	\caption{Effect of atmospheric metallicity on number of transits needed for detection for Sullivan Target 1292, a detection threshold ($dBIC$) of 10, and using NIRSpec G395M with a noise floor of 25\,ppm, and no clouds. This behavior is consistent with Figure 5 of \citet{1538-3873-129-974-044402}.  We were limited in the choice of metallicity levels by the equation of state files available in the \texttt{Exo-Transmit} installation.\label{fig:metalvar}}
\end{figure}

Variation of the atmospheric metallicity has a very significant effect on our results.  As we can see from Figure 5 of \citet{1538-3873-129-974-044402}, spectral line strength is strongly driven by metallicity.  Again, for Sullivan Target 1292, in Figure \ref{fig:metalvar} we can see that the number of transits needed for detection falls almost by an order of magnitude between the 1000xSolar and the 50xSolar metallicity levels.  Unfortunately, the only intermediate level that we could examine was a 100xSolar metallicity.  The equation of state files available in the \texttt{Exo-Transmit} installation do not include files in the regime between 100x and 1000xSolar.  It was beyond the scope of this project to construct a new equation of state file to fill in the gap.  It is unlikely, however, that any intermediate data points would significantly change our conclusions about the number of transits needed for detection here. 

We did not directly examine the variation in our results due to changes in cloud levels.  Again, Figure 5 of \citet{1538-3873-129-974-044402} guides what we would expect to see.  Since we were trying to bound the atmospheric cases, our low metallicity (5xSolar) case with no clouds would seem to be a good choice for a best-on-best case.  The high metallicity (1000xSolar) case with clouds at 100\,mbar would seem to be a nearly worst-on-worst case.  As the figure indicates, the effect of clouds (at any altitude) on the strength of spectral features at this high metallicity is not very strong. 

\subsection{Detection Threshold Variation}
\begin{figure}[!htbp]
	\centering
	\centerline{\includegraphics[width=0.50\textwidth]{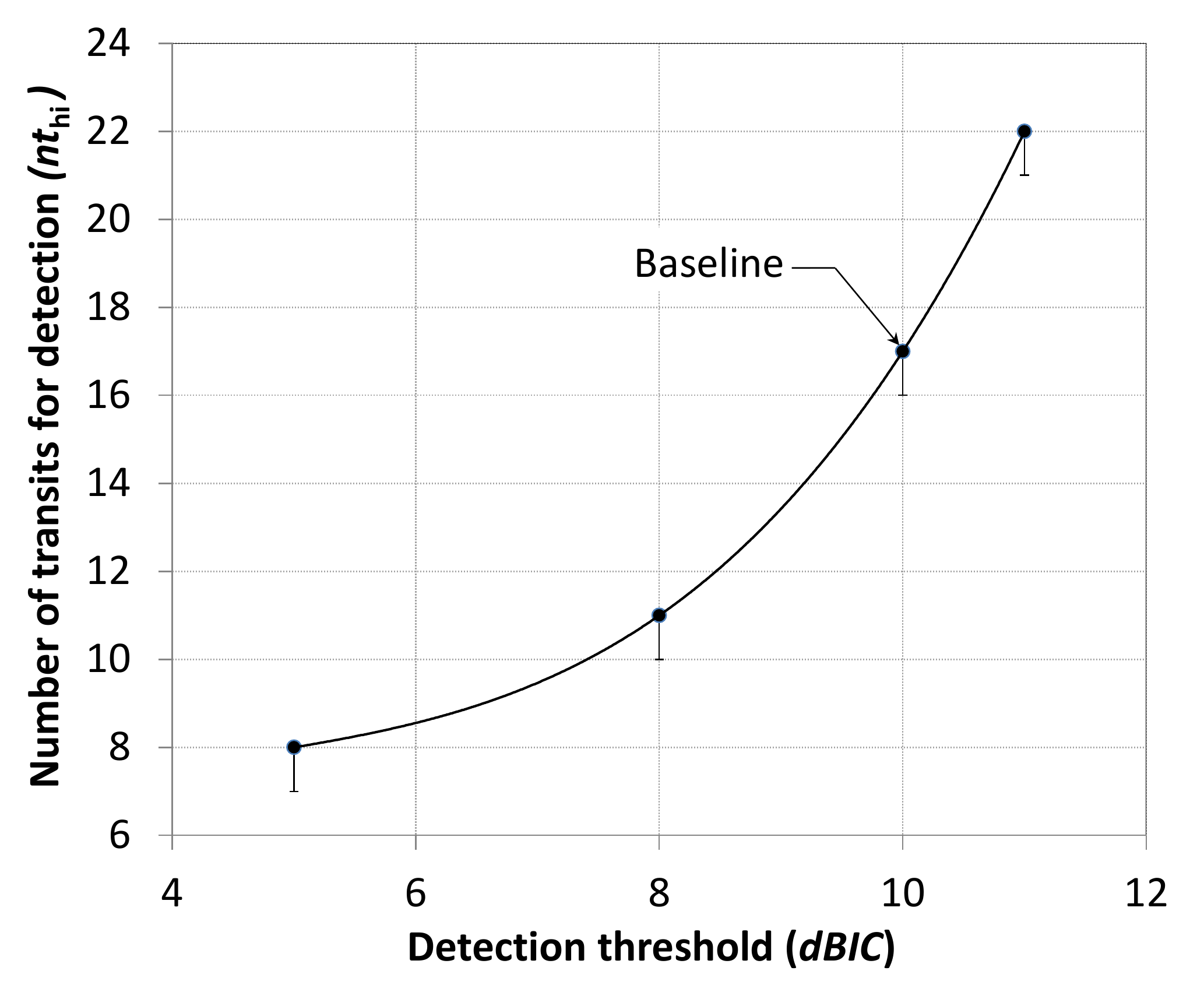}}		
	\setlength{\abovecaptionskip}{0pt}
	\caption{Effect of lowering the detection threshold ($dBIC$) on the number of transits needed for detection for Sullivan Target 1292, a high metallicity atmosphere (1000xSolar), with clouds at 100\,mbar, and using NIRSpec G395M with a noise floor at 25\,ppm. A $dBIC$ of 10 is considered a very strong detection, while a $dBIC$ of 6 is still considered a strong detection.\label{fig:threshvar}}
\end{figure}

The number of transits needed for detection is significantly effected by our choice of $dBIC$ detection threshold.  \citet{Kass1995} suggest that a $dBIC$ of 6 can be considered a strong detection, while a $dBIC$ of 10 is very strong.  Not surprisingly, our Figure \ref{fig:threshvar} shows that for Sullivan Target 1292 with a high metallicity atmosphere, as we lower the detection threshold from 10 to 5, we see a drop in the number of transits required for detection from 17 to 8. 

Given that we are using the dBIC mean less 1$\,\sigma$ values for detection (the bold line shown in Figure \ref{fig:dbic1292hi} in Section \ref{sec:detecting}), the threshold of 10 gives us a confidence level for a very strong detection of approximately 84\,\%.

\subsection{Instrument Variation}
\begin{figure*}[!htbp]
	\centering
	\includegraphics[width=1.0\textwidth]{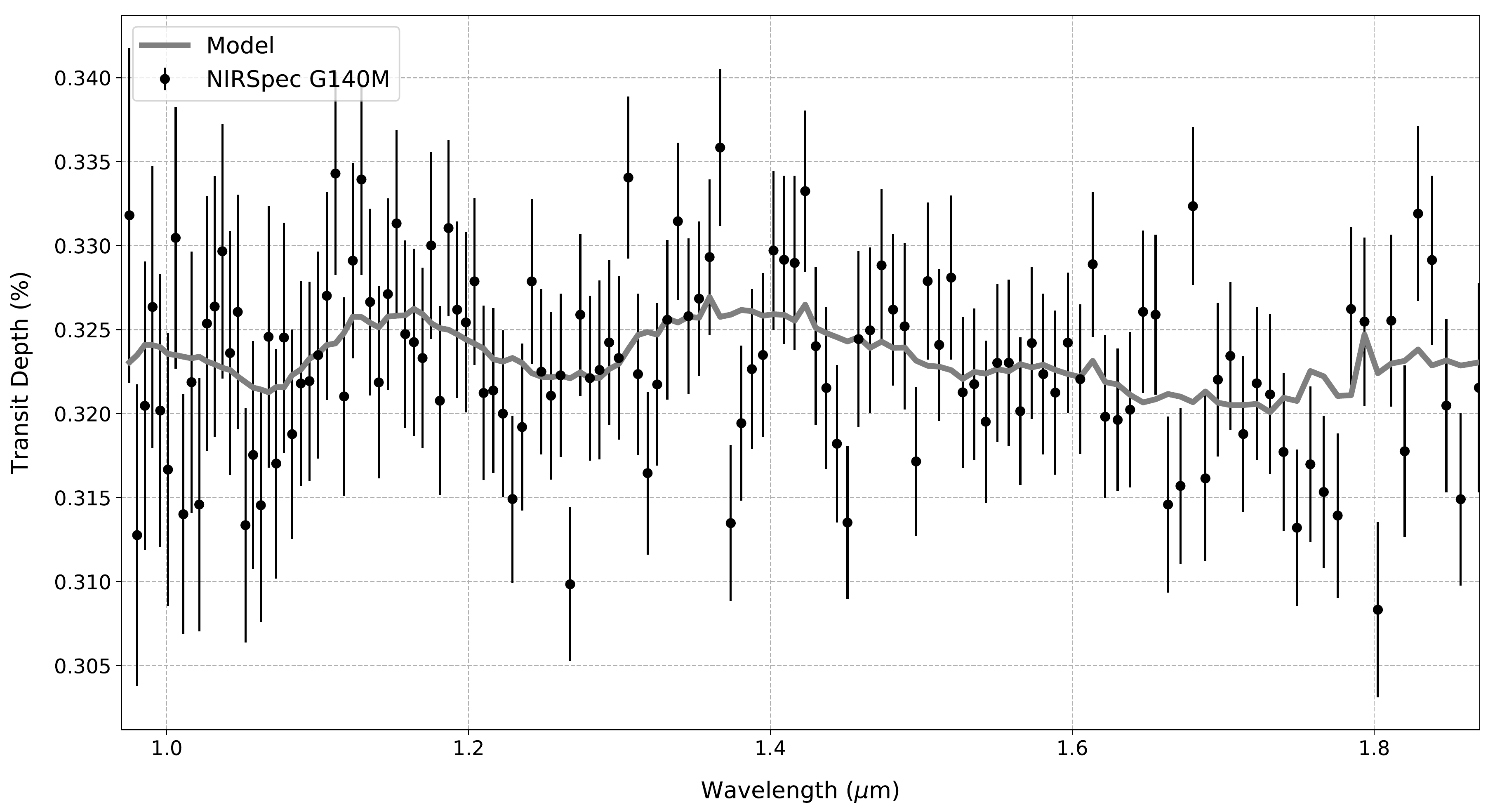}
	\setlength{\abovecaptionskip}{0pt}
	\caption{Simulated spectrum for a single transit of Sullivan Target 1292 with a high-metallicity atmosphere and clouds at 100\,mbar, using NIRSpec G140M with a wavelength range of $0.97\text{  - 1.87 } \mu \text{m}$.  The noise floor for this instrument is approximately 25\,ppm.  The model spectrum shown as the gray background line has been binned down to a resolution (R $\sim$ 100) consistent with the simulated data. A $dBIC$ = 10 detection takes observation of four transits.\label{fig:tsp1292hinirspecg140m}}
\end{figure*}

\begin{figure*}[!htbp]
	\centering
	\includegraphics[width=1.0\textwidth]{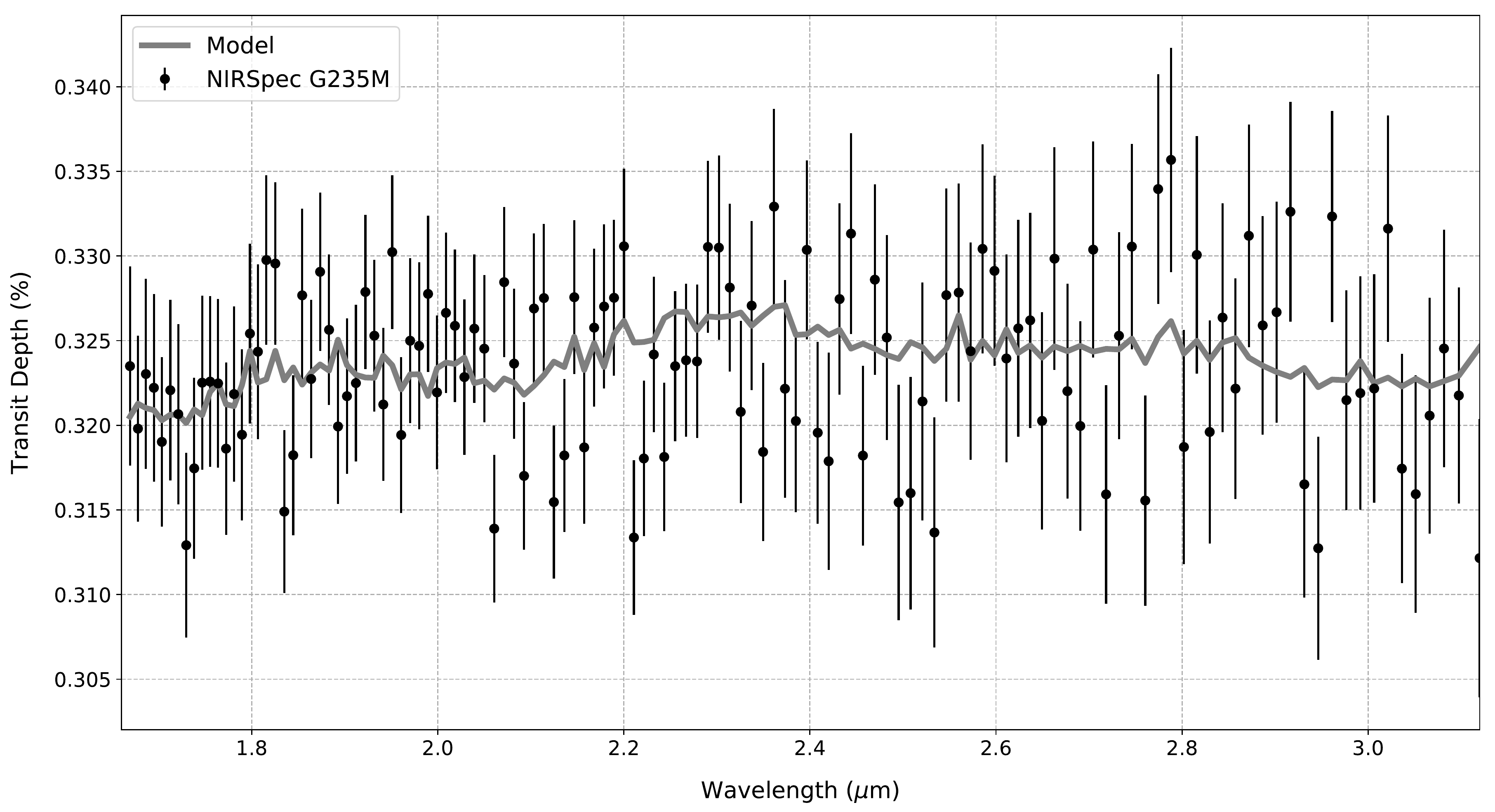}
	\setlength{\abovecaptionskip}{-10pt}
	\caption{Similar to Figure \ref{fig:tsp1292hinirspecg140m}, but for the G235M disperser and a wavelength range of $1.66\text{  - 3.12 } \mu \text{m}$.  A $dBIC$ = 10 detection takes observation of five transits.\label{fig:tsp1292hinirspecg235m}}
\end{figure*}

\begin{figure*}[!htbp]
	\centering
	\includegraphics[width=1.0\textwidth]{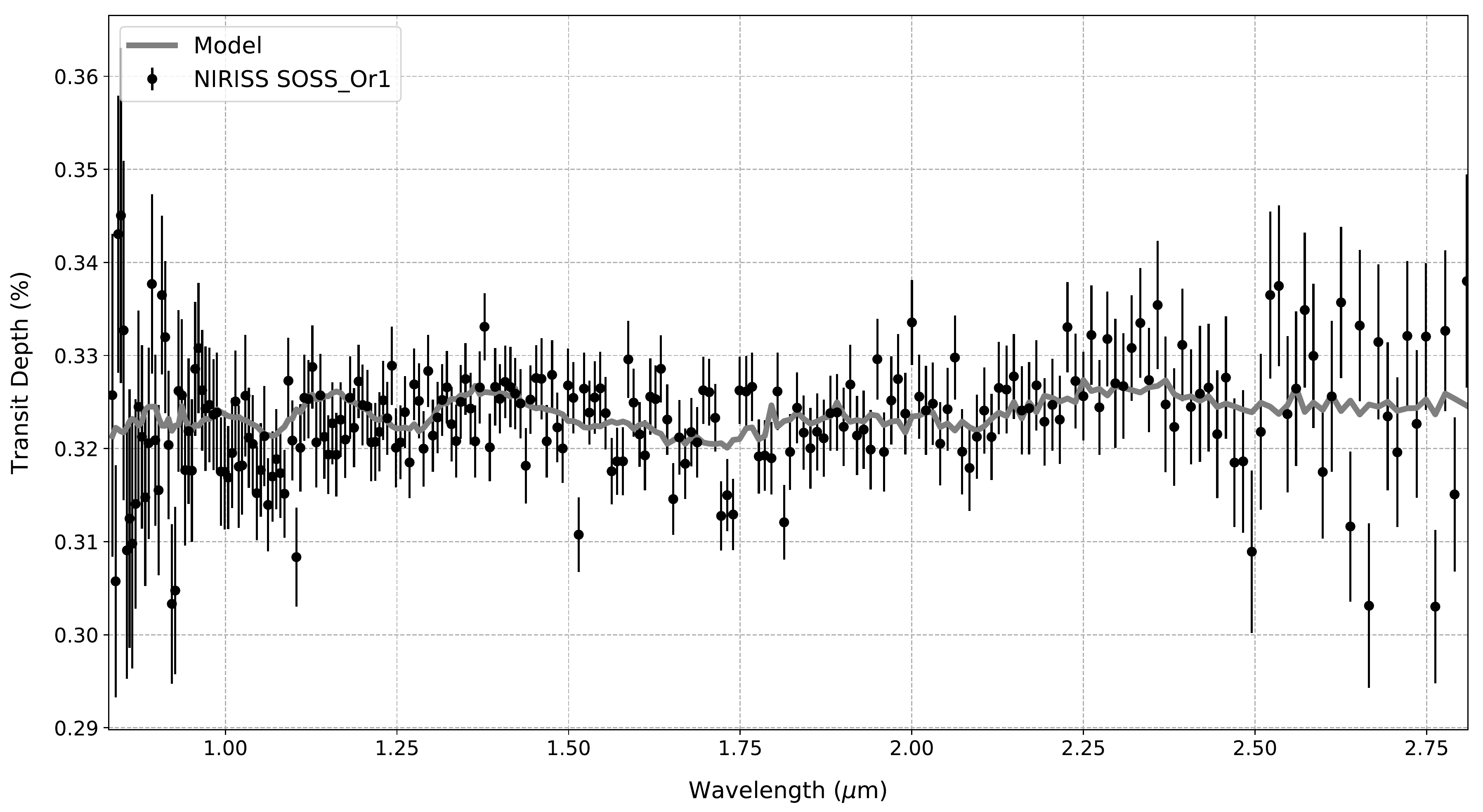}
	\setlength{\abovecaptionskip}{-10pt}
	\caption{Simulated spectrum for a single transit of Sullivan Target 1292 with a high-metallicity atmosphere using NIRISS SOSS (Order 1) with a wavelength range of $0.83\text{  - 2.82 } \mu \text{m}$.  The noise floor for this instrument/mode is approximately 20\,ppm.  The model spectrum shown as the gray background line has been binned down to a resolution (R $\sim$ 100) consistent with the simulated data. A $dBIC$ = 10 detection takes observation of two transits.\label{fig:tsp1292hinirisssoss}}
\end{figure*}

\emph{JWST} has four main instruments, each with multiple observing modes.  All four (NIRCam, NIRSpec, NIRISS, and MIRI) will to varying degrees be used for exoplanet transmission spectroscopy.  We have been able to implement four of these instrument/modes with \texttt{JET}.  We previously presented single-transit spectra for NIRSpec G395M in Figure \ref{fig:tsp1292lonirspecg395m}. 
We now present additional single-transit spectra for each of the other instrument/modes in Figures \ref{fig:tsp1292hinirspecg140m} through \ref{fig:tsp1292hinirisssoss}. 

We have summarized results of variation of the \emph{JWST} instrument/mode on the detection of a high metallicity atmosphere for Sullivan Target 1292 in Table \ref{tab:instrvar}.

We see that only two transits are needed to detect the target with NIRISS SOSS, while 17 transits are needed using NIRSpec G395M.  This does not necessarily imply that the former instrument is better than the latter.  The wavelength ranges are different, and the capabilities of the instruments are complementary rather than competing.  It has been suggested \citep{Batalha2017} that these two instruments be used in tandem, since there is little overlap in their wavelength coverage, and both have relatively high precision.
\bigskip

\begin{table*}[!htbp]
	\centering
	\setlength{\belowcaptionskip}{-5pt}
	\caption{Effect of Instrument Choice on Detection (Sullivan Target 1292)\label{tab:instrvar}}   
	\begin{center}                                                                                  
		\setlength{\tabcolsep}{20pt}                                                                 
		\renewcommand{\arraystretch}{1.4}                                                            
		\begin{tabular}{l c c c c}                                                                   
			\hline\hline
			Instrument/Mode          &  R   &   nfloor  & $\lambda$ range   & $nt_{\rm hi}$  \\[-1.00ex]  
			&      &   (ppm)   & ($\mu$m)          &            \\[ 0.30ex]  
			\hline                                            
			NIRSpec G395M & 100 & 25 & 2.87 \,-\, 5.18 & 17 \\
			NIRSpec G235M & 100 & 25 & 1.66 \,-\, 3.12 & 5 \\
			NIRSpec G140M & 100 & 25 & 0.97 \,-\, 1.87  & 4 \\
			NIRISS SOSS{\_}Or1 & 100 & 20 & 0.83 \,-\, 2.81 & 2 \\   [0.5ex]                       
			\hline
			\multicolumn{5}{l}{
				\begin{minipage}{13cm}~\\[1.5ex]
					Notes. --- noise floor values (nfloor) are pre-launch estimates only. The R values shown are not the native resolving power of the instrument, but are the re-binned values for consistency between the model spectra and simulated spectra in our analysis.
				\end{minipage}
			}\\
		\end{tabular}
	\end{center}
\end{table*}

\section{Conclusions/Further Study}\label{sec:conclusions}

\subsection{Conclusions}
We have developed an analytical framework and associated computer code that can assist the community in determining the best exoplanet targets for atmospheric characterization by \emph{JWST}.  The tools can also be used to prioritize targets that would be worthy of follow up with RV observations to better determine the planet masses.  

We have demonstrated that a target catalog can be categorized and ranked for minimum observation time to detect an atmosphere, even though we have no direct knowledge of the atmospheric properties of the target planets. 

Our use of the Bayesian Information Criterion for model selection in our target atmosphere detection algorithm is one of the strengths of this work. In addition, any target prioritization tool must take into consideration all of the observing time overheads for a transit observation.  Even a very short transit can be expensive in terms of the overall observing time requirement.

In the Sullivan (simulated \emph{TESS} detection) catalog we only saw one target where the number of transits observable during the fuel life of the spacecraft was less than the number needed for an atmospheric detection.  The observing constraints imposed by the spacecraft's orbital location and pointing limitations do not appear to be significant, at least for the shorter period planets that are the focus of surveys like those coming from \emph{TESS}.

The instrument noise floor is a critical parameter.  Particularly for difficult targets (small transit depth and reduced feature prominence), a very small change in the assumption for the noise floor can change the number of transits needed for detection significantly.  This, of course, could substantially change the overall target rankings.  Once the instruments have been better characterized on orbit (during commissioning), the prioritization analysis will need to be repeated to reflect the updated information on instrument precision.
 
Perhaps the weakest element in our analytical approach is our assumption that each target effectively has an atmosphere of average metallicity.  This may be true if we consider the entire dataset, but it is almost certainly not true for any particular target.  In reality any given target may be skewed away from average metallicity, either high or low.  This, of course, would affect the detectability of the atmosphere for that target, and the ranking. Our detection model provides a reasonable basis for ranking, but we must accept that there are uncertainties.

To date there have been less than 50 exoplanet atmospheres that have been characterized by transmission spectroscopy. Most of these studies have been at relatively low resolution and some only consist of a few data points gathered by multi-band photometry.  \emph{JWST} has the potential to increase the number of atmospheres characterized by at least an order of magnitude.

Our baseline run of the Sullivan catalog showed detection of over 1100 planet targets (atmospheres). This figure significantly overstates the realistic/practical target set that could be observed.  The fraction of the 10-year \emph{JWST} mission devoted to exoplanet transmission spectroscopy is estimated to be on the order of 10\,\%.  This translates into roughly 8300 hours overall.  Our analysis indicates that between 400 and 500 target/atmospheres could realistically be detected and studied during the mission using NIRSpec G395M. Observations with other instrument/modes (e.g., NIRISS SOSS) could well add to this target/atmosphere characterization total. 

It should also be noted, as we discussed in Section \ref{sec:syntargetsurveys}, that there have been several new studies of potential \emph{TESS} planet yields since the Sullivan paper was published. The categorization and ranking results could well be different using these updated catalogs.

\emph{JWST} will provide access to many important atmospheric spectral features in the infrared, including water vapor, methane, carbon dioxide, carbon monoxide, ammonia, sodium, potassium, and others. In addition, this access will be over a much wider wavelength range and at much higher resolution than what has been done previously \citep{Madhusudhan2019}. This wealth of new spectral data will dramatically enhance our knowledge of many exoplanet physical processes, including planetary formation, geology, climate, environment, and potential habitability \citep{Kalirai2018a}.

With \emph{JWST} we will be in a position to begin to assess the demographics of planetary atmospheres and the relationship with the properties of the planet's host stars. In particular, determining planetary atmosphere metallicity can provide insight into formation scenarios. The wide wavelength range of \emph{JWST} will allow us to determine overall atmospheric spectral shapes and to probe vertical energy redistribution.  The effects of inversions and other temperature distributions can potentially be studied.  We also expect to be able to determine atmospheric properties of smaller rocky planets orbiting M-dwarf host stars. It may in certain cases be possible to detect atmospheric biosignatures, but this will be challenging for \emph{JWST} \citep{KrissansenTotton2018, LustigYaeger2019}.  Unfortunately the characterization of the atmosphere of an Earth-like planet around a Sun-like star will be beyond the reach of \emph{JWST}.  For that we will have to wait for the next generation of space observatories \citep{LUVOIRTeam2019, Gaudi2019, Gaudi2019a, Roberge2019, Roberge2019a}. 

\subsection{Opportunities for Further Study}
We have identified a number of opportunities to refine the analysis that we have described in this work and to consider other lines of investigation.

As we have mentioned, the \texttt{JET} code is slow when run on a typical home computer.  An effort to make the code more efficient would be helpful.  In particular, consideration of processing multiple targets with a parallel processing architecture would seem to be a worthwhile effort. In addition, a new open-source \texttt{Python} package, PLanetary Atmospheric Transmission for Observer Noobs (\texttt{PLATON}), described by \citet{Zhang2018}, may reduce the computation time necessary for generating model transmission spectra. Unfortunately, this code was not available when we made the decision to use \texttt{Exo-Transmit}. 

As we discussed, we have been able to implement the NIRSpec G395M and NIRISS SOSS (Order 1) with \texttt{JET}.  Doing a full catalog run with NIRISS SOSS and then combining the output with our existing NIRSpec G395M baseline would cover a wide wavelength range ($0.83\text{  - 5.18 } \mu \text{m}$), and could yield interesting results.

In addition, for the sake of completeness, it would be helpful to expand the list of instrument/modes that \texttt{JET} could address.  This would include the higher resolution modes for NIRSpec, the NIRCam grism (with F322W2 and/or F444W filters), the second order NIRISS SOSS mode, and MIRI LRS.

In Section \ref{sec:resultsdiscuss} we discussed single target parameter variations of noise floor, detection threshold, etc.  We could consider full runs with small parameter variations.  This would provide a more thorough study of how small parameter variations could change overall target rankings.

Of course the most useful test of the analysis framework and code would be to run it on actual \emph{TESS} (or other precursor) catalog data.  \emph{TESS} is delivering datasets periodically over its two year mission.  It would be helpful to take these early datasets and format them (similar to the Sullivan catalog) or otherwise prepare them to be used as input to the \texttt{JET} code.

With further study we may find ways to refine our estimates of planet atmosphere metallicity that are better than simply taking the average of high and low bounds.  This will allow us to prepare target rankings with less uncertainty in the results. 

Our efforts so far have been focused on transmission spectroscopy, but exploring emission spectroscopy with our analysis approach would seem to be an interesting area for further study.  This would be a major excursion from what we have done so far.  The \texttt{Exo-Transmit} code is aimed at transmission spectroscopy, but there are other codes that have the capability to model atmospheric emission spectra as well as transmission; these include: \texttt{PyDisort} \citep{Stamnes:88}, \texttt{NEMESIS} \citep{Irwin2008a}, \texttt{CHIMERA} \citep{Line2013}, \texttt{ATMO} \citep{Tremblin2015}, \texttt{TauRex} \citep{Waldmann2015}, \texttt{HELIOS} \citep{Malik2017}, \texttt{PLATON}, and \texttt{petitRADTRANS} \citep{Molliere2019}.  \texttt{PandExo} can be used to simulate \emph{JWST} emission (occultation) spectra.

There have been a number of delays in the \emph{JWST} launch schedule; however, there is a positive aspect to the latest delay, in that there should be time to refine the approach to target optimization presented here, or by others, as well as time to incorporate actual \emph{TESS} survey results into the analysis.

\begin{acknowledgments}
This paper grew out of C. Fortenbach's Master's thesis project at San Francisco State University.  C.F. would like to acknowledge the support of Prof. Mohsen Janatpour (College of San Mateo), and Profs. Joseph Barranco (SF State Univ.), Andisheh Mahdavi (SF State Univ.), and Stephen Kane (UC Riverside).  We also appreciate the assistance of Josh Lamstein, Shervin Sahba, Dirk Kessler, Paul Seawell, Craig Schuler, and Arjun Savel who helped with various aspects of the code development and preparation of the manuscript.  We are grateful to Prof. Eliza Kempton (Univ. of Maryland) for her guidance on installation and usage of the \texttt{Exo-Transmit} code.  We would like to thank Dr. Tom Greene (NASA Ames) for his insight and guidance on a number of issues.  We would also like to acknowledge the assistance of Dr. Natasha Batalha (NASA Ames) the lead author of the \texttt{PandExo} code. This work has made use of the NASA Exoplanet Archive, which is operated by the California Institute of Technology, under contract with the National Aeronautics and Space Administration under the Exoplanet Exploration Program. C.D.D. acknowledges support from the Alfred P. Sloan Foundation, the David and Lucile Packard Foundation, and the Hellman Fellows Fund.   
\end{acknowledgments}

\facilities{\emph{JWST}, \emph{TESS}, {Exoplanet Archive}}.
\software{\texttt{astropy} \citep{astropy_et_al2018}, \texttt{Exo-Transmit} \citep{1538-3873-129-974-044402}, \texttt{Forecaster} \citep{0004-637X-834-1-17}, \texttt{matplotlib} \citep{hunter2007}, \texttt{numpy} \citep{oliphant2015}, \texttt{PandExo} \citep{1538-3873-129-976-064501}, \texttt{scipy} \citep{jones_et_al2001}, \texttt{SpectRes} \citep{Carnall2017}}.

\bibliography{jwst_cf_biblio_Jun2019}

\clearpage
\enddocument